\documentclass[
 reprint,
 amsmath,amssymb,
 pra, aps,
floatfix,
longbibliography, showkeys
]{revtex4-2}

\usepackage{placeins}
\usepackage[colorlinks,citecolor=blue,urlcolor=blue]{hyperref}
\usepackage{dcolumn}
\usepackage{bm}
\usepackage{color}
\usepackage{marvosym}
\usepackage{booktabs}
\usepackage{siunitx,url}
\usepackage{mathtools}
\usepackage{graphicx,currfile-abspath}
\usepackage{xcolor}
\usepackage[caption=false]{subfig}

\newcommand{\kT}{\ensuremath k_\text{B}T}
\newcommand{\gammaN}{\ensuremath \gamma_\text{N}}
\newcommand{\gammaS}{\ensuremath \gamma_\text{S}}
\newcommand{\gammaD}{\ensuremath \gamma_\text{D}}
\newcommand{\phiN}{\ensuremath \phi^\text{N}}
\newcommand{\phiD}{\ensuremath \phi^\text{D}}
\newcommand{\omegaS}{\ensuremath \omega_\text{S}}
\newcommand{\nuN}{\ensuremath \nu_\text{N}}
\newcommand{\nuS}{\ensuremath \nu_\text{S}}
\newcommand{\LambdaDN}{\ensuremath \Lambda_\text{DN}}
\newcommand{\LambdaDS}{\ensuremath \Lambda_\text{DS}}
\newcommand{\LambdaSN}{\ensuremath \Lambda_\text{SN}}
\newcommand{\gammaDN}{\ensuremath \gamma^\text{DN}}
\newcommand{\gammaDS}{\ensuremath \gamma^\text{DS}}
\newcommand{\gammaSN}{\ensuremath \gamma^\text{SN}}
\newcommand{\SDN}{\ensuremath S^\text{DN}}

\newcommand{\NS}{\ensuremath N_\text{S}}
\newcommand{\VN}{\ensuremath V_\text{N}}
\newcommand{\cN}{\ensuremath c_\text{N}}
\newcommand{\cD}{\ensuremath c_\text{D}}
\newcommand{\mN}{\ensuremath m^\text{N}}
\newcommand{\mS}{\ensuremath m^\text{S}}
\newcommand{\mD}{\ensuremath m^\text{D}}

\newcommand\dif[2]{\frac{\mathrm{d}#1}{\mathrm{d}#2}}

\newcommand{\Lint}{\ensuremath L_\text{int}}
\newcommand{\LintNorm}{\ensuremath L_\text{int}^\text{norm}}
\newcommand{\meanN}{\ensuremath \bar{N}}
\newcommand{\Athaliana}{\textit{A. thaliana}}
\newcommand{\Celegans}{\textit{C. elegans}}

\newcommand{\zyp}{\textit{zyp1}}

\newcommand{\diff}{\mathrm d}
\newcommand{\Eqref}[1]{\mbox{Eq.\hspace{0.25em}\eqref{#1}}}
\newcommand{\Eqsref}[1]{\mbox{Eqs.\hspace{0.25em}\eqref{#1}}}
\newcommand{\figref}[1]{\mbox{Fig.\hspace{0.25em}\ref{#1}}}
\newcommand{\Figref}[1]{\mbox{Fig.\hspace{0.25em}\ref{#1}}}

\newcommand{\secref}[1]{\mbox{section\hspace{0.25em}\ref{#1}}}
\newcommand{\Secref}[1]{\mbox{Section\hspace{0.25em}\ref{#1}}}
\newcommand{\refcite}[1]{\mbox{reference\hspace{0.25em}\cite{#1}}}
\newcommand{\appref}[1]{\mbox{App.\hspace{0.25em}\ref{#1}}}

\begin{document}
\preprint{APS/123-QED}

\title{Coarsening model of chromosomal crossover placement}

\author{Marcel Ernst}
\email[]{marcel.ernst@ds.mpg.de}
\author{Riccardo Rossetto}
\author{David Zwicker}
\email[]{david.zwicker@ds.mpg.de}
\affiliation{Max Planck Institute for Dynamics and Self-Organization, Am Fa\ss{}berg 17, 37077 G\"ottingen, Germany}

\date{\today}

\begin{abstract}
    Chromosomal crossovers play a crucial role in meiotic cell division, as they ensure proper chromosome segregation and increase genetic variability. Experiments have consistently revealed two key observations across species: (i) the number of crossovers per chromosome is typically small, but at least one, and (ii) crossovers on the same chromosome are subject to interference, i.e., they are more separated than expected by chance. These observations can be explained by a recently proposed coarsening model, where the dynamics of droplets associated with chromosomes designate crossovers. We provide a comprehensive analysis of the coarsening model, which we also extend by including material exchanges between droplets, the synaptonemal complex, and the nucleoplasm. We derive scaling laws for the crossover count, which allows us to analyze data across species. Moreover, our model provides a coherent explanation of experimental data across mutants, including the wild-type and \zyp{}-mutant of \Athaliana{}. Consequently, the extended coarsening model provides a solid framework for investigating the underlying mechanisms of crossover placement.
\end{abstract}


\maketitle

\section{Introduction}\label{sec:introduction}
Chromosomal crossovers (COs) are crucial for successful meiosis and thus sexual reproduction.
COs promote reciprocal genetic exchange increasing genetic variability, and they create a physical connection between homologous chromosomes ensuring correct segregation~\cite{Berchowitz:2010aa,globus2012joy,Anderson:2014aa,Wang2015,Zickler:2016aa,Otto2019,smith2020new}. 
The latter aspect is established by \emph{CO assurance}, which describes the observation that essentially all chromosomes contain at least one CO~\cite{jones1984control,borner2004crossover,Diezmann:2021aa,Lloyd2022,Girard2023}.
However, the number of COs rarely exceeds a few per chromosome~\cite{Cole:2012aa,Zickler:2015aa,Fernandes:2018aa}, and their placement is subject to \emph{CO interference}, implying COs on the same chromosome are more separated than expected by chance~\cite{Sturtevant:1913aa,Sturtevant:1915aa,Muller:1916aa,Kleckner:2003aa, Diezmann:2021aa,Lloyd2022,Girard2023}.

CO formation is an evolutionarily conserved process:
During prophase I of meiosis I, compacted homologous chromosomes pair up and are connected along their axes by the synaptonemal complex (SC)~\cite{carpenter1975electron,Zickler:1999aa,page2004genetics,Capilla-Perez2021}, which is highly ordered, but permits lateral diffusion~\cite{Stauffer:2019aa,Rog:2017aa}, akin to a liquid crystal~\cite{Rog:2017aa}.
Class~I COs, which are subject to CO interference, form from pre-cursors known as recombination nodules (RNs) that associate with the SC~\cite{carpenter1975electron,page2004genetics,Berchowitz:2010aa}, whereas less prevalent class~II COs result from a separate mechanism and appear to be non-interfering~\cite{Otto2019, Girard2023, Lloyd2022}.
In early pachytene, many RNs form, typically around DNA double-strand breaks (DSBs), but the RN count decreases over time~\cite{Berchowitz:2010aa}, so that eventually only a few late RNs remain, which typically lead to COs~\cite{carpenter1975electron,Zickler:1999aa,neyton2004association,baudat2007regulating,Hollingsworth:2004aa,
Zhang:2014ab,Diezmann:2021aa}.
Despite significant progress in the last decades, the mechanisms designating which RNs become COs remains highly debated~\cite{Wang2015,Zickler:2016aa,Otto2019,smith2020new,Diezmann:2021aa,Lloyd2022,Girard2023}.
In particular, there is no comprehensive model that explains CO placement across species and mutants.

\begin{figure}[b]
\centering
\includegraphics[width=\columnwidth]{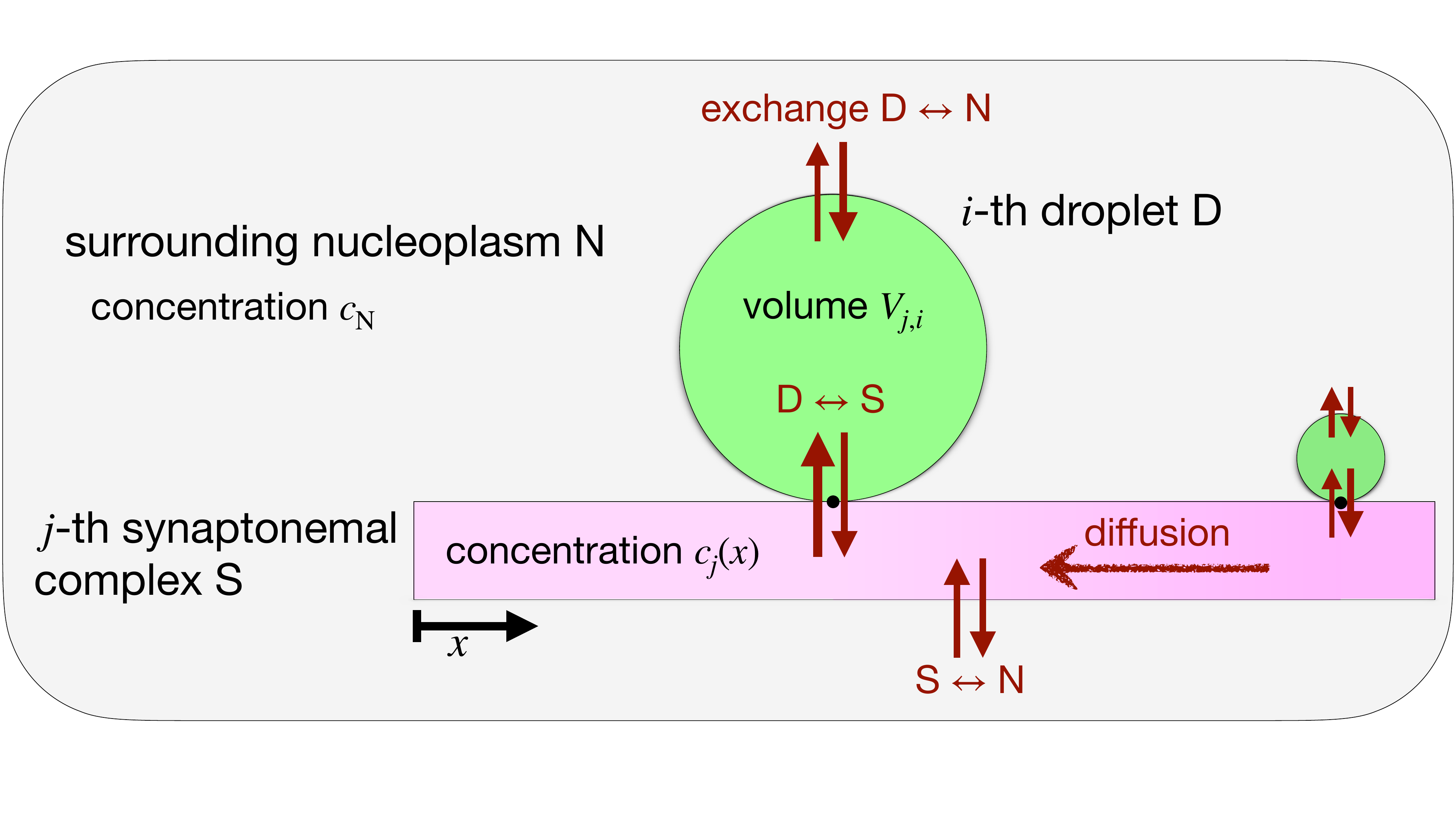}\vspace{-0.7cm}
\caption{
\textbf{Schematic of the extended coarsening model of crossover placement},
showing crossover precursor droplet (green) associated with a synaptonemal complex (SC, pink) embedded in nucleoplasm.
All compartments exchange material (red arrows) and the key droplpet component diffuses along the SC.
The system is characterized by the average concentration $\cN(t)$ in nucleoplasm, the concentration $c_j(x, t)$ along the SC, and droplet volumes $V_{j,i}(t)$.
}
\label{fig:Figure1_sketch}
\end{figure}

Models of CO placement can be broadly classified as mechanical models and diffusion-based models:
In mechanical models, such as the beam-film-model~\cite{Kleckner:2004aa,Zhang:2014ab}, COs are designated by physical constraints on chromosomes.
In contrast, diffusion-based models use the localization of signaling molecules to promote COs~\cite{King:1990aa,zhang2018compartmentalized,Diezmann:2021aa}.
In particular, the recently proposed coarsening model~\cite{Durand2022,Morgan2021,Zhang:2025aa} can explain key features of CO placement in wild-type \Athaliana{}~\cite{Durand2022,Morgan2021,Girard2023} and \Celegans{}~\cite{Zhang:2025aa}.
The model describes RNs as droplets that exchange material by diffusion along the SC, so droplets compete for material.
Although different proteins are involved (particularly HEI10 in plants~\cite{ziolkowski2017natural,Morgan2021} and ZHP-3/4 in \Celegans{}~\cite{zhang2018compartmentalized}), it is suggested that the general mechanism is conserved across species~\cite{Girard2023}.
However, it is currently unclear how the coarsening model could explain other conditions, particularly mutants lacking SCs, like the \zyp{}-mutant of \Athaliana{}, where diffusion likely takes place via the nucleoplasm~\cite{SYM1994283, Capilla-Perez2021,Durand2022,Fozard:2023aa}.
\citet{Fozard:2023aa} proposed separate models for wild type and the \zyp{}-mutant, but a comprehensive, quantitative description is still lacking.

We here present a thermodynamically consistent version of the coarsening model that includes nucleoplasmic exchange and coherently explains experimental data across mutants.
We use \Athaliana{} as an example to develop the model, but generalize results to other species later.
In particular, we discuss CO assurance and CO interference quantitatively and we derive asymptotic scaling relations for CO counts, which allow us to constrain model parameters.
We also investigate details of CO placement, like CO homeostasis (changing number of DSBs hardly affects the CO count~\cite{Berchowitz:2010aa, Martini:2006aa,Cole:2012aa,Wang2015}) and heterochiasmy in wild type (male and female exhibit different CO counts~\cite{Durand2022,Fozard:2023aa,Girard2023}).
The augmented coarsening model thus enables a quantitative analysis of CO placement, and the role of nucleoplasmic exchange, across species and mutants.

\section{Coarsening model of crossover placement}
\subsection{Thermodynamically-consistent coarsening model including nucleoplasm}
\label{sub:thermodynamically_consistent_coarsening_model}

To model placement of class I COs, we describe RNs as droplets that grow and compete for material.
For simplicity, we assume that RNs existing after a fixed time $t$ become COs, so we are essentially asking which droplets survive.
To do this, we model the coupled systems of nucleoplasm (N), synaptonemal complex (S) linking homologous chromosomes, and droplets (D) (\Figref{fig:Figure1_sketch}).
We consider a nucleoplasm with constant volume $\VN$, and assume that diffusive transport of HEI10
in the nucleoplasm is fast compared to diffusion along the SC. 
Consequently, we assume that spatial variations of the HEI10 concentration are negligible in the nucleoplasm, and we characterize it by an average concentration $\cN(t)$. 
The $\NS$ SCs in the nucleoplasm have physical lengths $L_j$ for $j = 1,\ldots,\NS$, and a constant effective cross section $a^2$, which defines our fundamental length scale $a$.
We parameterize the HEI10 concentration along SC by $c_j(x, t)$, where $x\in[0,L_j]$ parametrizes the corresponding arc length.
The $j$-th SC possesses $N_j$ HEI10 droplets placed along the arc at positions $x_{j,i}$ for $i=1,\ldots,N_j$.
For simplicity, we assume constant HEI10 concentration $\cD$ in droplets, so they are fully described by their volume $V_{j,i}$.
The system's state is thus characterized by the nuclear concentration $\cN$, the concentrations $c_j(x)$ on the  SCs, and the droplet volumes $V_{j,i}$.

We derive a simple dynamical description based on the idea that the system relaxes toward equilibrium.
Even though active processes might play a role in reality~\cite{Zhang:2025aa}, we find that they are not necessary to explain the data.
We base our description on the chemical potentials of HEI10, $\mu_\alpha = \kT \ln \left(\gamma_\alpha c_\alpha/\cD\right)$, in the three compartments $\alpha=\mathrm{N},\mathrm{S},\mathrm{D}$, where $\kT$ is the energy scale, we use $\cD$ as a reference concentration, and $\gamma_\alpha$ are activity coefficients accounting for interactions.

The biophysical mechanism of underlying HEI10 accumulation is currently unclear.
HEI10 can phase separate \textit{in vitro}~\cite{wang2023hei10}, whereas \textit{in vivo}  active processes, such as post-translational modification, might also be important~\cite{kim2024control,Zhang:2025aa}.
For simplicity, we describe HEI10 as an ideal solution in the nucleoplasm and SC, implying that both activity coefficients are constant.
In contrast, we assume that the affinity of HEI10 toward droplets increases with their volume, which we capture by a phenomenological expression of $\gamma_\mathrm{D}$.
Taken together,
\begin{subequations}
\label{eqn:affinities}
\begin{align}
    \gammaN&=1\;, \\
    \gammaS &=\exp(-\omegaS/\kT) \;, \\
    \label{eqn:droplet_affinity}
    \gammaD(V_{j,i})&= \gamma_0 \left(\frac{V_{j,i}}{a^3}\right)^{-\nu}  \;,
\end{align}
\end{subequations}
where we chose the nucleoplasm as reference, $\omegaS$ accounts for the affinity of HEI10 molecules toward  SC, and $\gamma_0$ is related to the affinity of HEI10 to a droplet of reference volume $a^3$.
The size-dependence is captured by the exponent $\nu>0$, with $\nu=\frac13$ corresponding to typical surface tension effects~\cite{Weber2019, zwicker2025physics}.
These affinities govern the long-term behavior of the system when it tends toward thermodynamic equilibrium.
At that point, all chemical potentials are balanced, implying
\begin{align}
    \label{eqn:equilibrium}
    \cN^*
    = \gammaS  c_j^*
    = \gamma_0 \cD \left(\dfrac{V^*_{j,i}}{a^3}\right)^{-\nu}
    \;.
\end{align}
In particular, the form of \Eqref{eqn:droplet_affinity} implies that at most a single droplet will remain in the system after a very long time, but to understand droplet placement at finite times, we need to study the dynamics.

The dynamics of the system comprise diffusion of HEI10 along the SC and its exchange between all compartments (\Figref{fig:Figure1_sketch}).
While we assume simple diffusion, describing the exchange dynamics requires more care to avoid spurious circular fluxes between the compartments.
We thus base the exchange on a thermodynamically-consistent description, which obeys detailed balance.
In particular, we apply transition-state theory~\cite{Zwicker:2022aa,kirschbaum2022chemical} to describe the net flux~$S^{\alpha\beta}$ from compartment $\alpha$ to $\beta$,
\begin{align}
S^{\alpha\beta} 
= \tilde\Lambda_{\alpha\beta} \left[\exp{\left(\frac{\mu_\alpha}{\kT}\right)} - \exp{\left(\frac{\mu_\beta}{\kT}\right)}\right]
	\;,
\end{align}
where $\tilde\Lambda_{\alpha\beta}$ sets the overall kinetic rate of the particular exchange and the square bracket captures the thermodynamic force involving the affinities defined in \Eqsref{eqn:affinities}.
Since the exchange between droplets and the surrounding may involve various processes, the integrated exchange rate generally depends on droplet size.
We capture this by generic power laws,
\begin{subequations}
\begin{align}
    S^{\mathrm{DS}}_{j,i} 
    &= \LambdaDS \left(\dfrac{V_{j,i}}{a^3}\right)^{\nuS}
            \left[\gammaS c_j - \gamma_0 \cD\left(\dfrac{V_{j,i}}{a^3}\right)^{-\nu}\right]
\;,\label{eq:SDS}
\\[5pt]
    S^{\mathrm{DN}}_{j,i}
    &= \LambdaDN \left(\dfrac{V_{j,i}}{a^3}\right)^{\nuN} \left[\cN - \gamma_0 \cD \left(\dfrac{V_{j,i}}{a^3}\right)^{-\nu}\right]
\;, \label{eq:SDN} 
\end{align}
where $\Lambda_{\mathrm{D}\beta}$ are base kinetic rates, whereas \mbox{$0 \leq \nu_\beta \leq 1$} captures corresponding size sensitivities, for $\beta~=~\mathrm{S},~\mathrm{N}$.
Particular cases include size-independent exchange ($\nu_\beta~=~0$), diffusion-limited exchange ($\nu_\beta=\frac13$~\cite{Lifshitz1961,Wagner:1961aa}), conversion-limited exchange ($\nu_\beta=\frac23$~\cite{Wagner:1961aa}), and volume-limited exchange ($\nu_\beta=1$).
In contrast, the exchange between nucleoplasm and SC reads
\begin{align}
    s^{\mathrm{SN}}_{j} 
    &= \frac{\LambdaSN}{a^3}\bigl[\cN - \gammaS c_j\bigr]\;,
 \label{eq:SSN}
\end{align}
\end{subequations}
where $\LambdaSN$ denotes the kinetic rate.
Taken together, the dynamics are described by 
\begin{subequations}
\label{eqn:dynamical_equations}
\begin{align}
    \cD \frac{\diff V_{j,i}}{\diff t} &= S^{\mathrm{DS}}_{j,i} + S^{\mathrm{DN}}_{j,i}
    \;, \label{eq:dMDdt} 
 \\
\frac{\partial c_{j}}{\partial t} &= D \partial_x^2c_{j} + s^{\mathrm{SN}}_{j} - \frac{1}{a^2}\sum_{i=1}^{N_j}  S^{\mathrm{DS}}_{j,i} \delta{(x-x_{j,i})}
	\;,	\label{eq:dlambdaSdt}
\\ 
    \VN\frac{\diff \cN}{\diff t} &= -\sum_{j=1}^{\NS} \biggl[  \sum_{i=1}^{N_j}  \SDN_{j,i} + a^2 \int_0^{L_j} s^{\mathrm{SN}}_{j} \diff x \biggr] 
    \;,\label{eq:dMNdt}
\end{align}
\end{subequations}
where $D$ is the diffusivity of HEI10 along the SC, and we impose no-flux boundary conditions at both ends.
These equations conserve the total amount $M$ of HEI10,
\begin{align}
	\label{eqn:conservation}
    M = \VN\cN + a^2 \sum_{j=1}^{\NS} \int_0^{L_j} c_j \diff x +  \cD\sum_{i=1}^{N_j} V_{j,i}
    \;.
\end{align}
Consequently, the system's state is fully specified by $c_j$ and $V_{j,i}$, whereas $\cN$ can be obtained from \Eqref{eqn:conservation}.
Given an initial condition, the dynamics of the system follow from Eqs.~\eqref{eq:dMDdt}--\eqref{eq:dlambdaSdt}.
In the following, we discuss four relevant cases of the model:
(i)~initial loading of HEI10 onto the SC without considering droplets,
(ii)~coarsening of droplets along a single SC without nucleoplasm,
(iii)~growth and coarsening of droplets without SC, and
(iv)~the combined case.

\subsection{Exchange rate determines SC loading rate and accelerates diffusion along SC}
\label{sec:depletion_accelerates_loading}
To study the initial loading of HEI10 onto already assembled SCs before droplets form, we consider the exchange of material with  nucleoplasm.
A mode analysis (\appref{app:relaxation_no_droplets}) shows that the overall dynamics relax toward equilibrium with rate
\begin{align}
	\label{eqn:loading_rate}
   k_\mathrm{load} &=\frac{\LambdaSN}{a^3} \left(\gammaS + \frac{V_\mathrm{S}}{\VN}\right)
     \;,
\end{align}
where $V_\mathrm{S} = a^2L$ denotes the volume of all SCs with total length $L=\sum_j L_j$.
This expression shows that loading onto SCs is faster for larger exchange rates~$\LambdaSN$ and if SCs contains fewer molecules, e.g., for small cross-sections~$a^2$ or lower affinity (higher $\gammaS$).
The second factor in \Eqref{eqn:loading_rate} accounts for the depletion of the nucleoplasm, which accelerates the relaxation.

\label{sec:nucleoplasm_accelerates_diffusion}

We next study the dynamics of heterogeneous HEI10 profiles by investigating diffusive relaxation along an SC.
A mode analysis (\appref{app:relaxation_no_droplets}) shows that deviations of wave length $\ell$ relax with rate
\begin{align}
    k_\mathrm{diff}(\ell) = \frac{4\pi^2 D}{\ell^2} + \frac{\gammaS \LambdaSN}{a^3}
    \,.
\end{align}
This expression emphasizes two contributions to spatial transport:
The first term describes diffusive transport along the SC, which is slower for longer distances~$\ell$.
The second term describes alternative transport by unbinding, fast diffusion through nucleoplasm, and rebinding, which additionally facilitates transport between different SCs.
This term becomes more important for larger exchange rate~$\LambdaSN$, but it does not depend on $\ell$ since we assumed fast diffusion in nucleoplasm.
In particular, the rate of transport via nucleoplasm is similar to \Eqref{eqn:loading_rate} in the typical situation where $\gammaS$ is comparable or larger than $V_\mathrm{S}/\VN$.
Consequently, faster loading of the SC implies stronger transport via nucleoplasm and between SCs, which will affect droplet coarsening.
To study coarsening, we first focus on the simpler case without nucleoplasmic exchange before considering it again in \secref{sec:no_SC}.

\subsection{Droplet coarsening via the SC implies linear scaling of droplet count with SC length}
\label{sec:coarsening_SC}

\begin{figure*}[ht]
    \centering
\subfloat{
\raisebox{200pt}[0pt]{\makebox[-5pt][l]{\textsf{\textbf{A}}}}
\includegraphics[width=0.50\textwidth]{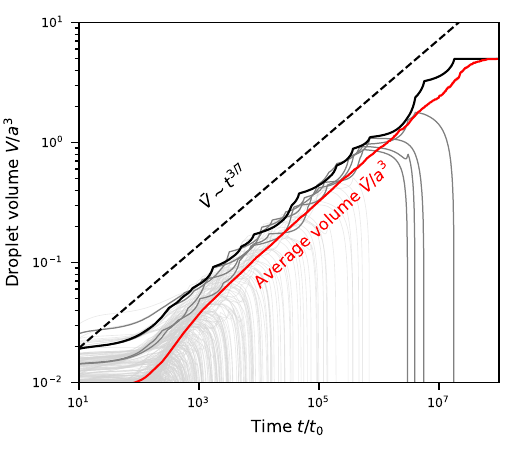}}
\subfloat{
\raisebox{200pt}[0pt]{\makebox[-5pt][l]{\textsf{\textbf{B}}}}
\includegraphics[width=0.50\textwidth]{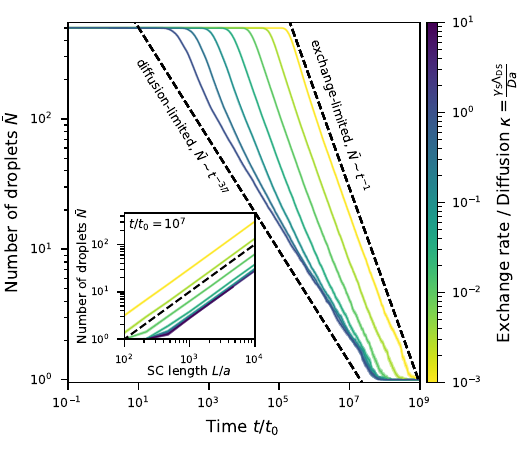} }
\caption{\textbf{Numerical investigation of the coarsening model without nucleoplasmic exchange.} (A) Time development of droplet sizes $V/a^3$ as a function of time $t/t_0$ for an exemplaric system with $\kappa=1$ without nucleoplasm, where $t_0 = a^2/D$.
The grey lines represent the volumes of the individual droplets over time, the thick grey lines the five larges droplet, whereas while the black line represents the droplet that remains until the end. 
The red line gives the average volume $\bar V(t/t_0)$ (red line) of the remaining droplets, whereas the dashed line gives the theoretical scaling for the diffusion-limited scenario for large $\kappa = \gammaS \LambdaDS/(D a)$ which is consistent with the average scaling in this scenario.
(B) Average droplet count $\meanN$ as a function of non-dimensional time $t/t_0$ for varying $\kappa$.
The inset shows the number of droplets $\meanN$ as a function of SC length $L/a$ for varying $\kappa$ with $\Delta x_\mathrm{grid}/a = 0.25$ and $N_\mathrm{samples} \geq 160$.
(A--B): Additional parameters used in the simulations: 
$\nu=\frac13$, $\nuS=0$, $\gamma_0/\gammaS=10^{-4}$, $L = 500\,a$, $N^\mathrm{init} a/L=1$, $M = 10^{-2}\cD L a^2$, $\bar{V}^\mathrm{init} = 10^{-3}a^3$, $\sigma_V^\mathrm{init} = 10^{-1}\bar{V}^\mathrm{init}$, $\Delta x_\mathrm{grid}= 0.5\,a$, and $N_\mathrm{samples} = 100$.
}
\label{fig:Figure2}
\end{figure*} 

To study the simplest case of droplet coarsening, we focus on the dynamics of droplets on a single SC, neglecting  nucleoplasm.
For simplicity, we consider the case where the SC has been loaded homogeneously with HEI10, implying $c(x, 0) = \bar c$, and that multiple small HEI10 droplets already exist.
We distribute the droplets uniformly along the SC and draw their volumes from a Gaussian distribution around a mean volume that is large enough to ensure that the initial droplets do not dissolve spontaneously.
We then simulate Eqs.~\eqref{eqn:dynamical_equations} to obtain droplet volumes as a function of time.
\Figref{fig:Figure2}A shows that most droplets vanish in a short time, while some survive longer.
This coarsening process is characterized by a growing average droplet volume $\bar V(t)$ (red line), and it ceases when only a single droplet remains, which then obeys \Eqref{eqn:equilibrium}.
We quantify the coarsening dynamics by measuring the average droplet volume $\bar V(t)$, which exhibits a power-law scaling of $\bar V(t) \propto t^{3/7}$ (\Figref{fig:Figure2}A).
This power-law scaling deviates from the seminal Lifshitz--Slyozov--Wagner scaling of $t^1$~\cite{Lifshitz1961,Wagner:1961aa}, even though we chose the corresponding volume sensitivity $\nu=\frac13$.
This observation suggest that restriction to one-dimensional diffusion slows down coarsening.

Additional simulations (\Figref{fig:Figure2}B) show that coarsening depends on the ratio of the exchange rate $\LambdaDS$ to the diffusivity $D$, which we quantify by the non-dimensional ratio $\kappa=\gammaS \LambdaDS/(D a)$.
In particular, we find less steep slopes when the exchange between SC and droplets is fast compared to diffusion of HEI10 along the SC ($\kappa\rightarrow\infty$).
In that scenario, HEI10 concentrations on the SC at positions $x_i$ of droplets are dictated by  local equilibrium, $\gammaS c(x_i) = \cD \gamma_0 (V_i/a^3)^{-\nu}$, which depends on the droplet volume~$V_i$.
Since droplets of different size result in different concentration on the SC, the dynamics are governed by  relatively slow diffusion between neighboring droplets.
An asymptotic analysis inspired by Lifshitz, Slyozov, and Wagner~\cite{Lifshitz1961, Wagner:1961aa} predicts the number~$N$ of droplets remaining on the SC (\appref{ssec:scaling_ds}),
\begin{align}
	\label{eq:scaling_DS_diffusion}
    N(t) 
    &\approx K_0(\nu) \frac{L}{a} \left(\frac{\lambda}{\cD a^2}\right)^{\frac{1+\nu}{2+\nu}} 
    \left(\frac{\gammaS}{\gamma_0} \frac{a^2}{D\,t}\right)^{\frac{1}{2+\nu}}
    \;,
\end{align}
where $K_0(\nu)\approx 1$ collects constants that only depend on $\nu$ (\appref{ssec:scaling_ds}), and $\lambda$ quantifies the initial line density of HEI10 along the SC including  HEI10 contained in  initial droplets, which sets the total amount~$M = \lambda L$.

In the opposing case of fast diffusion ($\kappa\rightarrow 0 $), concentrations are equilibrated along the entire SC, so slow exchange determines the dynamics of individual droplets.
The corresponding asymptotic analysis predicts (\appref{ssec:scaling_exchange_limited})
\begin{align}
	\label{eq:scaling_DS_exchange}
    N(t) &\approx K_1(\nu, \nuS)  \frac{\lambda L }{\cD a^3}
    \left(\frac{1}{\gamma_0} \frac{a^3}{\LambdaDS\,t}\right)^{\frac{1}{1+\nu-\nuS}}
    \;
\end{align}
for $0<\nu, \nuS < 1$, where $K_1(\nu, \nuS)$ again collects the constants and only depends on $\nu$ and $\nuS$ (\appref{ssec:scaling_exchange_limited}).
\Figref{fig:Figure2}B shows that both scaling laws are consistent with numerical data in respective limiting cases.
We show that our predictions in both limits also captures the corresponding distribution of droplet sizes (\Figref{fig:Figure_theory_distributions} and {\figref{fig:Figure_numerics_distributions}}).
Taken together, these scaling laws allow us to make quantitative predictions for large and small $\kappa$, and intermediate values interpolate between the extremes.

\begin{figure*}[t]
    \centering
\subfloat{
\raisebox{135pt}[0pt]{\makebox[-5pt][l]{\textsf{\textbf{A}}}}
\includegraphics[width=0.333\textwidth]{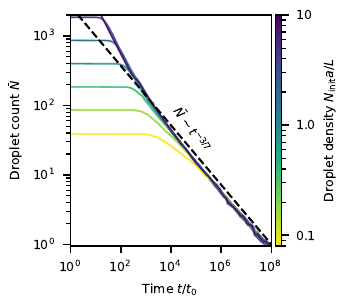}}
\subfloat{
\raisebox{135pt}[0pt]{\makebox[-5pt][l]{\textsf{\textbf{B}}}}
\includegraphics[width=0.333\textwidth]{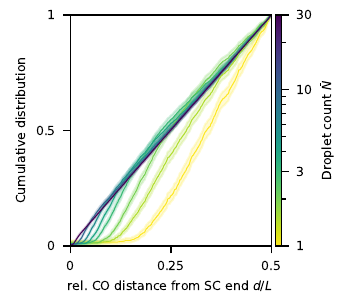}}
\subfloat{
\raisebox{135pt}[0pt]{\makebox[-5pt][l]{\textsf{\textbf{C}}}}
\includegraphics[width=0.333\textwidth]{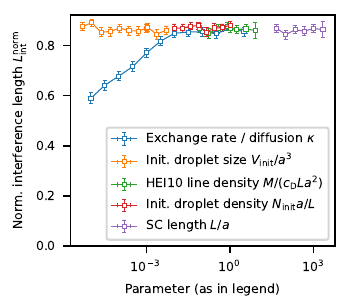}}
\caption{\textbf{Numerical investigation of the coarsening model without nucleoplasmic exchange.}
(A) Time development of the average number of droplets $\meanN$ as a function of time $t/t_0$  for varying initial droplet density $N_\mathrm{init}a/L$.
The overall HEI10 line density $M/(\cD L a^2)$ per unit length along the SC is kept constant.
The dashed line gives the theoretical asymptotic scaling for the diffusion-limited scenario with $\bar{N}\sim t^{-\frac{1}{2+\nu}} \hat{=} t^{-3/7}$.
(B) Cumulative distribution of relative distance $d/L$ of droplets from the SC end as a function of the average droplet count $\meanN$ with $\kappa=1$, $N^\mathrm{samples}=5000$ and $\Delta x_\mathrm{grid} =a$.
(C) Normalized interference length $\LintNorm$ for different values as a function of varying values of initial droplet density $N_\mathrm{init} a/L$, SC length $L/a$, initial droplet size $V_\mathrm{init} / a^3$, HEI10 line density $M/(\cD a^2 L)$ and varying ratio of exchange rate and diffusion $\kappa = \gammaS \LambdaDS/(D a)$ for a fixed number of crossover $\meanN = 4$.
(A--C) Additional parameters from \Figref{fig:Figure2}.
}
\label{fig:Figure3}
\end{figure*} 
The analytical predictions in the limiting cases of $\kappa$ both imply that the number of droplets, $N$, scales linearly with SC length~$L$.
Numerical simulations show that this linear scaling is preserved for all values of $\kappa$ (inset in \Figref{fig:Figure2}B).
Consequently, doubling $L$ also doubles the droplet count, independent  of whether coarsening is limited by exchange or diffusion.
In contrast, $\kappa$ affects the coarsening dynamics (\Figref{fig:Figure2}B).
In the limit of $\kappa\rightarrow 0$, we obtain the seminal Lifshitz-Slyozov scaling $N\sim t^{-1}$ for $\nu=\nuS=\frac13$~\cite{Lifshitz1961, Wagner:1961aa}, whereas the diffusion-limited case predicts $N \sim t^{-\frac37}$, reflecting that diffusion is restricted to one dimension.
Both scaling laws also predict that only the initial line density~$\lambda$ is relevant for the late-stage dynamics, whereas other details of the initial condition are negligible.
This suggests that droplet patterning is insensitive to details of the initial conditions.

\subsection{Details of initial conditions do not affect droplet patterning}
\label{sec:droplet_patterning_only_SC}

To study droplet patterning, we return to detailed numerical simulations.
We first vary the number~$N_\mathrm{init}$ of initial droplets while reducing the concentration~$\bar c$ on the SC to keep the total amount~$M$ constant.
\Figref{fig:Figure3}A shows that $N_\mathrm{init}$ affects the average CO count~$\meanN$ only early, whereas $\meanN$ is independent of $N_\mathrm{init}$ in late stages and instead follows the scaling relation~\eqref{eq:scaling_DS_diffusion}.
Similarly, we find that changing SC length~$L$, the volume of initial droplets $V_\mathrm{init}$, or the HEI10 line density $\lambda=M/L$ does not affect the asymptotic behavior, when accounting for the scaling relation (\Figref{fig:Figure_numerics_without_nucleoplasm} in \appref{ssec:no_nucleoplasm}).

We next investigate droplet positions in detail.
When few droplets remain, our model predicts they are less likely to be near the ends of the SC (\Figref{fig:Figure3}B), even though we started with a homogeneous distribution of initial droplets.
This behavior follows from the competition of droplets for material: Droplets toward the end can only receive material from one side, so they tend to grow less and get outcompeted by more centrally located droplets.
This boundary effect becomes more severe for smaller droplet count~$\meanN$ (\Figref{fig:Figure3}B) since coarsening has progressed more.

Our numerical simulations also exhibit CO interference, i.e., small inter-CO distances are inhibited (\Figref{fig:Figure_numerics_distributions}B).
To quantify CO interference, we use the interference length $\Lint$, which estimates the size of the region around a droplet where interference takes place~\cite{Ernst:2024aa}.
This region increases during coarsening, and theoretical models predict that $\Lint$ scales with $\meanN^{-1}$~\cite{Ernst:2024aa}.
We thus consider the normalized interference length, $\LintNorm~=~\meanN \Lint / L $, where $L$ is the length of the single SC we consider.
\Figref{fig:Figure3}C shows that CO interference, as measured by $\LintNorm$, is not affected by details of the initial condition (i.e., initial droplet density $N_\mathrm{init}/L$, initial droplet size $V_\mathrm{init}$, HEI10 line density $\lambda =M/L$, SC length $L$), as long as we compare data for the same average droplet count $\meanN$.
In contrast, decreasing $\kappa$, i.e., decreasing the exchange rate $\LambdaDS$ relative to diffusion $D$ along the SC, reduces $\LintNorm$, suggesting weaker CO interference.
Note that $\LintNorm$ does not vanish, even for arbitrarily small $\kappa$, because coarsening leads to a narrow distribution of CO counts.
This contradicts the null hypothesis that COs are placed independently along all SCs, implying a Poisson distribution of CO counts and thus finite $\LintNorm$ (\appref{ssec:interference_length}).
However, CO positions are uncorrelated, consistent with the asymptotic limit of fast diffusion ($\kappa\rightarrow 0$), where the SC is homogeneous and CO positions become irrelevant for  material exchange.
A similar behavior emerges when the SC is completely absent. 
Taken together, the average droplet count $\meanN$ determines CO patterning, and small $\kappa$ reduces CO interference.

\subsection{Droplets in mutants without SC first grow and then coarsen}
\label{sec:no_SC}

\begin{figure*}[t]
    \centering
\subfloat{
\raisebox{140pt}[0pt]{\makebox[-5pt][l]{\textsf{\textbf{A}}}}
\includegraphics[width=0.3830\textwidth]{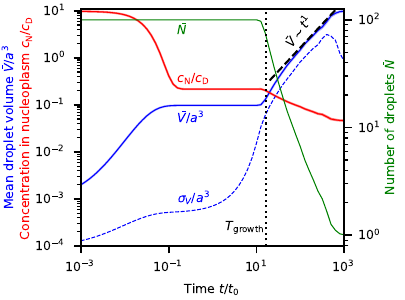}}
\subfloat{
\raisebox{140pt}[0pt]{\makebox[-5pt][l]{\textsf{\textbf{B}}}}
\includegraphics[width=0.2837 \textwidth]{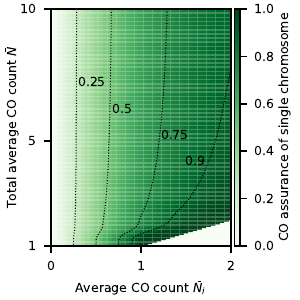}}
\subfloat{
\raisebox{140pt}[0pt]{\makebox[-5pt][l]{\textsf{\textbf{C}}}}
\includegraphics[width=0.3333\textwidth]{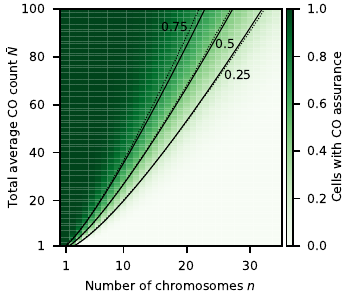}}
\caption{\textbf{Numerical investigation of the coarsening model without synaptonemal complex.} 
(A) Shown is an exemplary time-development of the average droplet volume $\bar{V}/a^3$ and average variance of droplets sizes $\sigma_V/a^3$, as well as the  HEI10 concentration in nucleoplasm $\cN/\cD$. The vertical dotted line at $T_\mathrm{growth}$ (cmp.~\Eqref{eq:T_growth}) marks the approximate transition from growth regime to coarsening regime that exhibits a droplet growth with scaling of $V\propto t^1$ (dashed line).
Additional simulation parameters: $\gamma_0\VN=0.1\,a^3$, $\nu=\nuN=\frac13$, $N^\mathrm{init}=100$, $M=10\,\cD a^3$, $V^\mathrm{init}=10^{-3}a^3$, $\sigma_V^\mathrm{init}=10^{-1}V^\mathrm{init}$, $N_\mathrm{sample}=100$.
(B) Shown is the CO assurance of a single chromosome (i.e., probability that there is at least on CO on this chromosome) with an average CO count of $\meanN_i$ in a cell with a total average CO count of $\meanN$ (that is assumed to have a variance of $0.2\meanN$). The dashed lines highlight iso-lines of constant probability.
(C) Shown is the CO assurance of a cell (i.e., the probability that all chromosomes have at least one CO) as a function of the number $n$ of chromosomes (of equal average CO count) and the total average CO count $\meanN$ with the variance used in B.
The dashed lines highlight iso-lines of constant probability, whereas the solid lines are the respective theoretical approximation based on \Eqref{eq:P_CO_assurance} (exact equation in \appref{sub:theoretical_approximations_of_co_assurance}).
}
\label{fig:Figure4}
\end{figure*} 

We next consider the case without SC, but we assume that chromosome pairing still works properly, and droplets are arranged along the axes of the chromosomes, consistent with experiments in \Athaliana{}~\cite{Durand2022}.
In this case, HEI10 droplets are still associated with chromosomes, but they can exchange material only with the nucleoplasm.
Since we assume fast diffusion in nucleoplasm, the kinetics are limited by the exchange rate $\LambdaDN$ between droplets and nucleoplasm.
The resulting growth dynamics exhibit two qualitatively different regimes, which we now discuss before considering CO interference and assurance.

Assuming $N$ initial small droplets form, our model predicts that all droplets grow by depleting the nucleoplasm until their average volume obeys the equilibrium relation~\eqref{eqn:equilibrium}.
During this growth phase, the mean droplet volume $\bar V$ and the corresponding standard deviation $\sigma_V$ grow (\Figref{fig:Figure4}A).
However, in our model, the coefficient of variation, $\sigma_V/\bar{V}$, decreases for $\nuN<1$,
indicating that the size distribution becomes narrower.
After this growth phase, larger droplets grow at the expense of smaller droplets, leading to an increase of $\sigma_V/\bar{V}$ until the first droplet vanishes.
The duration until this happens is approximately given by (\appref{ssub:duration_of_growth_and_coarsening_regime})
\begin{align}
    T_\mathrm{growth} \approx \frac{a^3}{\nu \gamma_0\LambdaDN} 
    \log{\left(\frac{\sigma^\mathrm{init}_V}{V_\mathrm{init}}\right)}
    \left(\frac{M}{\cD a^3 N}\right)^{1+\nu-\nuN}
    \;,
    \label{eq:T_growth}
\end{align}
assuming that most HEI10 ends up in droplets.
This indicates that $T_\mathrm{growth}$ decreases with droplet count $N$.
This expression reveals that the growth duration $T_\mathrm{growth}$ increases with total amount of HEI10 $M$ for $\nuN < 1$, since $\nu>0$.
In any case, the droplet count $N$ remains roughly constant initially and coarsening sets in only after $T_\mathrm{growth}$, consistent with simulations (\Figref{fig:Figure4}A). 

During droplet coarsening, the average droplet size increases whereas the droplet count~$N$ gradually decreases.
After an initial transient, the system approaches an asymptotic behavior (\Figref{fig:Figure4}A), which we describe using a scaling analysis similar to the one in \secref{sec:coarsening_SC} for $0<\nu, \nuN < 1$,
\begin{align}
    N(t)
    &= K_1(\nu, \nuN) \frac{M}{\cD a^3} 
    \left(\frac{1}{\gamma_0} \frac{a^3}{\LambdaDN t}\right)^{\frac{1}{1+\nu - \nuN}}
    \;,
    \label{eq:DN_scaling_dimensional_solution}
\end{align}
where $K_1(\nu, \nuN)$ collects the constant (\appref{ssec:dn_scaling}).
This scaling law is equivalent to \Eqref{eq:scaling_DS_exchange}, essentially since both processes are limited by  material exchange between droplets and their surrounding, while transport in the surrounding is fast.
In particular, we recover the typical Lifshitz--Slyozov--Wagner scaling, $N\sim t^{-1}$, for $\nu=\nuN$.

The absence of the SC also has profound consequences for CO interference and assurance.
Since the spatial positioning of droplets is irrelevant for their dynamics, CO interference is governed by the patterning of initial droplets if only the growth regime is relevant.
Even if coarsening takes place, CO interference will be hardly affected since coarsening corresponds to removal of a random subset of droplets.
Similar considerations hold for CO assurance, which we quantify by the probability of retaining at least one CO per chromosome. 
To derive an analytic estimate, we focus on a single chromosome with average droplet count $\meanN_i$ in a cell with $N$ droplets in total, which are randomly allocated to multiple chromosomes.
The probability that this chromosome has at least one droplet is 
\begin{equation}
    P(N_i\ge1) = 1 - \left(1- \frac{\meanN_i}{N}\right)^N
    \stackrel{\meanN_i\ll N}{\approx} 1 - e^{-\meanN_i}
    \;.\label{eq:P_assurance_single}
\end{equation}
Consequently, CO assurance increases with larger $\meanN_i$ whereas the influence of $N$ is weak (\Figref{fig:Figure4}B).

The fact that each of the chromosomes might lack droplets lowers CO assurance of the entire cell, i.e., the probability that all chromosomes in the cell have at least one droplet.
To estimate this, we consider $n$ chromosomes with same average droplet count $\meanN_i=N/n$, leading to 
\begin{align}
    P(\min(N_i)\geq 1) \approx \exp{\left(-n e^{-N/n}\right)}\;,\label{eq:P_CO_assurance}
\end{align}
which corresponds to the classical occupancy problem~\cite{hald1984moivre,moivre1711mensura,erdHos1961classical} (\appref{sub:theoretical_approximations_of_co_assurance}).
Equation~\eqref{eq:P_CO_assurance} shows that CO assurance approaches $1$ for $N\gg n$, whereas we necessarily have chromosomes with missing COs when $N<n$.
\Figref{fig:Figure4}C shows that these limits are obeyed over a large parameter regime, whereas intermediate values are rare.
This implies that COs are virtually assured for $\meanN/n > 5$, but CO assurance can be strongly suppressed even if the average CO count per chromosome is larger than one.
Moreover, CO assurance is reduced when the average CO count $\meanN_i$ differs between chromosomes (\appref{sub:theoretical_approximations_of_co_assurance}).
In contrast, variations in total CO count $N$ have only a minor effect (\Figref{fig:Figure4}C).
Taken together, absence of  SC implies vanishing CO interference and reduced CO assurance because droplet dynamics are insensitive to allocation of droplets to chromosome and the associated position along the arc.

In summary, we find two qualitatively different regimes for the dynamics of droplets without SC (\Figref{fig:Figure4}).
First, all initiated droplets grow, implying that the droplet count~$\meanN$ is controlled by  details of droplet initiation.
After a duration $T_\mathrm{growth}$, coarsening sets in and $N$ decreases according to \Eqref{eq:DN_scaling_dimensional_solution}.
In the scaling regime, $\meanN$~is then independent of the initial condition, and instead depends on the total amount of material, $M$.
Our analysis of droplet dynamics without SC resembles the \zyp-mutant in \Athaliana{}~\cite{Durand2022,Fozard:2023aa}.
It explains the absence of CO interference and the reduced CO assurance, but we currently cannot decide which of the two regimes is relevant since we lack time-resolved experimental data.

\subsection{Exchange via nucleoplasm reduces crossover interference and assurance}
\label{sec:nucleoplasmic_exchange}

\begin{figure*}[t]
    \centering
\subfloat{
\raisebox{140pt}[0pt]{\makebox[-10pt][l]{\textsf{\textbf{A}}}}
\includegraphics[width=0.3267\textwidth]{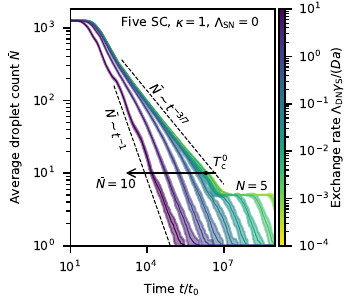}}   \hspace{0.01\textwidth}
\subfloat{
\raisebox{140pt}[0pt]{\makebox[-10pt][l]{\textsf{\textbf{B}}}}
\includegraphics[width=0.3267\textwidth]{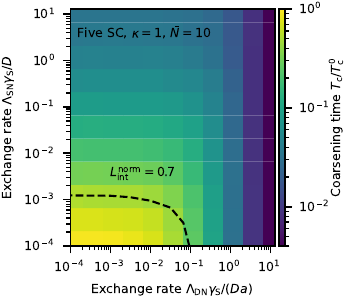}}\hspace{0.01\textwidth}
\subfloat{
\raisebox{140pt}[0pt]{\makebox[-10pt][l]{\textsf{\textbf{C}}}}
\includegraphics[width=0.3267\textwidth]{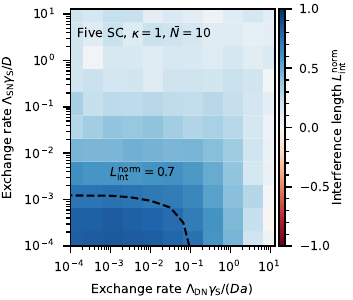}}\\
\subfloat{
\raisebox{140pt}[0pt]{\makebox[-10pt][l]{\textsf{\textbf{D}}}}
\includegraphics[width=0.3267\textwidth]{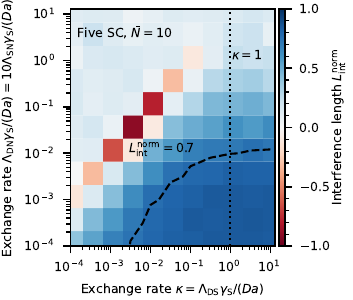}}\hspace{0.01\textwidth}
\subfloat{
\raisebox{140pt}[0pt]{\makebox[-10pt][l]{\textsf{\textbf{E}}}}
\includegraphics[width=0.3267\textwidth]{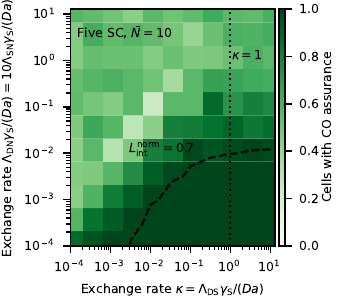}}\hspace{0.01\textwidth}
\subfloat{
\raisebox{140pt}[0pt]{\makebox[-10pt][l]{\textsf{\textbf{F}}}}
\includegraphics[width=0.3267\textwidth]{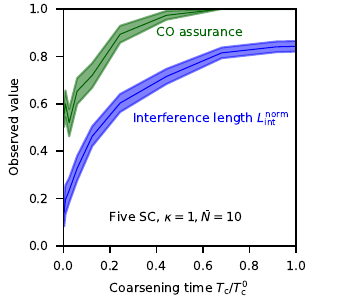}}
\caption{\textbf{Numerical investigation of the extended coarsening model for a cell with five SC.}
(A) Average total droplet count $\meanN$ as a function of time $t/t_0$ for a system of five SC with same length in a nucleoplasm.
The exchange rate $\kappa=\gammaS\LambdaDS/(D a) = 1$ of HEI10 between droplets and SC and the diffusion along the SC $D$ is fixed, the exchane between SC and nucleoplasm is zero, $\LambdaSN=0$, whereas the exchange rate of droplets with nucleoplasm $\LambdaDN$ increases from negligble exchange (yellow) with a slope of $\bar{N}\sim t^{-3/7}$ to strong nucleoplasmic exchange (blue) with a slope of $\bar{N}\sim t^{-1}$.
Increasing the exchange rate, decreases the coarsening time $T_\mathrm{c}$ reaching a certain mean droplet count $\meanN=10$ (along the arrow) compared to the scenario without nucleoplasmic exchange $T_\mathrm{c}^0$ (black dot).
(B) Heatmap showing the parameter dependence of the relative coarsening time $T_\mathrm{c}/T_\mathrm{c}^0$ as a function of the nucleoplasmic exchange rates $\LambdaDN{}$ and $\LambdaSN{}$ for fixed values of diffusion rate $D$, exchange rate $\kappa = 1$, and an average CO count $\meanN = 10$.
(C) Heatmap showing the normalized interference length $\LintNorm$ as a function of the nucleoplasmic exchange rates for the same scenario as in B.
The iso-lines in B--E shows the parameters that exhibit a normalized interference length $\LintNorm=0.7$.
(D) Heatmap showing the normalized interference length $\LintNorm$ as a function of the exchange rate between droplets and SC $\LambdaDS{}$ and the nucleoplasmic exchange rates with $\LambdaDN = 10 \LambdaSN$. 
For $\LambdaDS=0$ we observe $\LintNorm \approx  0.11\pm 0.02$, consistent with the expected CO interference due to the narrower CO count distribution (\appref{ssec:interference_length}).
The observed CO interference for $\LambdaDN=0$ resembles the special case presented in~\Figref{fig:Figure3}C.
(E) Heatmap showing the fraction of cells that do preserve CO assurance as a function of the exchange rate between droplets and SC $\LambdaDS{}$ and the nucleoplasmic exchange rates with $\LambdaDN = 10 \LambdaSN$.
The theoretical value for preserverd CO assurance in the limit of dominating nucleoplasmic exchange for five SC of same length with $\meanN=10$ is $P(\mathrm{min}(N_i)\geq 1 | N=10) \approx 0.52$ (\appref{sub:theoretical_approximations_of_co_assurance}).
For $\LambdaDS=0$ we observe a ratio $0.5\pm 0.02$ which is consistent with the theoretical prediction.
(F) Fraction of cells that preserve CO assurance and normalized interference length $\LintNorm$ as a function of the relative coarsening time $T_\mathrm{c}/T_\mathrm{c}^0$ (where $T_\mathrm{c}^0$ is the time without nucleoplasmic exchange) for a fixed exchange rate $\kappa=1$ (compare vertical dotted line in D--E).
(A--F) Additional parameters: $\nu=\nuN=\frac13$, $\nuS=0$, $\frac{\gamma_0}{\gammaS}=10^{-4}$, $\frac{1}{\gammaS}\frac{V_0}{\VN}=0.1$, $M=10\,\cD a^3$, $L=250\,a$, $N_\mathrm{init} a/L = 1$, $V_\mathrm{init} = 10^{-3}a^3$, $\sigma_V = 10^{-1}V_\mathrm{init}$, $\Delta x_\mathrm{grid}= 0.5\,a$, $N_\mathrm{sample}=50$.
}
\label{fig:Figure5}
\end{figure*}

We finally consider the case where HEI10 can exchange between droplets, SC, and nucleoplasm to describe wild-type behavior.
If exchange via nucleoplasm is negligible, we expect the full model to converge to the simpler case discussed in \secref{sec:coarsening_SC}, which exhibited strong CO interference and assurance.
The exchange between a droplet and the SC together with diffusion along the SC limits competition to droplets in its neighborhood on the same SC, whereas exchange with the nucleoplasm leads to global competition, as well as non-local competition along the same SC.
Since both exchanges might be relevant, we now ask how much exchange can take place via the nucleoplasm so that the predicted behavior is still compatible with wild-type measurements.

We start by investigating the exchange between droplets and nucleoplasm proportional to $\LambdaDN$, while neglecting the exchange between SC and nucleoplasm ($\LambdaSN=0$).
Numerical simulations of a nucleoplasm containing multiple SCs of same length indicate that droplets coarsen faster for larger $\LambdaDN$ (\Figref{fig:Figure5}A).
This is because droplets can now exchange material via an additional pathway.
Note that we recover the scalings derived in \secref{sec:coarsening_SC} and \secref{sec:no_SC} when either pathway dominates.
If competition along the same SC dominates, the system first plateaus when every SC has exactly one droplet.
If nucleoplasmic exchange is non-zero, the system will then exhibit further coarsening and approaches the regime with $\bar{N}\sim t^{-1}$ derived in \secref{sec:no_SC}.
Since the additional exchange via nucleoplasm accelerates coarsening, a comparison of data at the same time is not informative, and we will thus compare at the same mean total droplet count $\meanN$ in the following.

We next ask how exchange between SC and nucleoplasm affects droplet coarsening.
\Figref{fig:Figure5}B shows that increasing either $\LambdaDN$ or $\LambdaSN$ leads to faster coarsening, i.e., the system reaches the same $\meanN$ in shorter times $T_\mathrm{c}$.
However, the two processes differ in the limit of fast exchange:
For large $\LambdaDN$, droplets predominately coarsen via direct exchange through nucleoplasm, approaching the behavior we discussed in \secref{sec:no_SC}, so $T_\mathrm{c} \propto \LambdaDN^{-1}$.
In contrast, increasing $\LambdaSN$ does not decrease $T_\mathrm{c}$ arbitrarily since this pathway is still limited by exchange between droplets and SC.
In any case, \Figref{fig:Figure5}C shows that CO interference, quantified by the normalized interference length $\LintNorm$, tends to decrease if $\LambdaSN$ or $\LambdaDN$ are increased.
Taken together, more exchange via  nucleoplasm thus tends to accelerate coarsening and decrease crossover interference.

Since the two pathways of exchange with nucleoplasm have qualitatively similar effects, we next discuss them together by choosing $\LambdaDN = 10 \LambdaSN$ to compare to the exchange between droplets and SC, which is quantified by $\LambdaDS$.
\Figref{fig:Figure5}D shows that CO interference is strongest when exchange is limited by diffusion along a SC (large $\LambdaDS$, compare to \secref{sec:coarsening_SC}) while exchange via  nucleoplasm is limited (small $\LambdaDN$, $\LambdaSN$, compare to \Figref{fig:Figure5}C).
However, we find a complex behavior in the remaining parameter regime where interference is weaker.
In particular, our model shows highly reduced, and even negative, CO interference for $\LambdaDN \approx 10 \LambdaDS$.
For these parameters, droplets tend to cluster on  SCs, which is a consequence of initial conditions:
If a particular region on a SC starts with relatively few droplets, these droplets have less local competition and can thus grow larger than droplets in denser regions.
This initial bias in size gets amplified by global competition via the nucleoplasm, so that droplets in less dense region have a higher chance of surviving, leading to clustered distributions resulting in negative $\LintNorm$.
To understand why this effect requires that direct exchange via SC is balanced with exchange via nucleoplasm, we next consider various values of $\LambdaDS$:
If $\LambdaDS$ is too large, competition along the individual SC dominates and CO interference is strong.
In contrast, if $\LambdaDS$ is too small, material uptake from the SC, and thus the initial size bias, are negligible.
Taken together, we find a large parameter regime with similarly strong CO interference, although other quantities, particularly the coarsening time $T_\mathrm{c}$ (\Figref{fig:Figure5_SI}), still vary.

The exchange via nucleoplasm also affects CO assurance.
To quantify this, we next measure how often we observe an SC without any droplets.
\Figref{fig:Figure5}E shows that virtually all SCs carry droplets when exchange is limited by diffusion along the SC (large $\LambdaDS$) while exchange via  nucleoplasm is limited (small $\LambdaDN$, $\LambdaSN$).
In contrast, significant exchange via  nucleoplasm leads to a high chance of SCs without droplets (cmp.~\secref{sec:no_SC}), and we observe signatures of the droplet clusters analogous to \Figref{fig:Figure5}D.
Taken together, we expect CO assurance precisely in the parameter regime where CO interference is strong.

In summary, we have shown that exchange of HEI10 via nucleoplasm limits CO interference and assurance.
To obtain an intuitive picture for how much exchange via nucleoplasm is tolerable, we need to quantify the exchange.
This is challenging because the exchange between SC and nucleoplasm is not directly linked to droplet dynamics, so we cannot directly determine how much that nuclear exchange contributes to droplet growth.
To still enable a fair comparison, we instead quantify the coarsening time $T_\mathrm{c}$, relative to the time $T_\mathrm{c}^0$ without exchange via nucleoplasm.
Larger exchange via nucleoplasm reduced $T_\mathrm{c}/T_\mathrm{c}^0$ (\Figref{fig:Figure5}B), so that this quantity serves as a proxy for the exchange via nucleoplasm.
\Figref{fig:Figure5}F shows that CO interference and assurance increase with $T_\mathrm{c}/T_\mathrm{c}^0$, as expected.
However, both quantities are still close to their maximal value even when $T_\mathrm{c} = \frac12 T_\mathrm{c}^0$, suggesting that the material exchange via nucleoplasm can be almost as strong as the direct exchange via the SC.

\section{Comparison with experimental data}

\begin{figure*}[t]
    \centering
\subfloat{
\raisebox{135pt}[0pt]{\makebox[-5pt][l]{\textsf{\textbf{A}}}}
\includegraphics[width=0.333\textwidth]{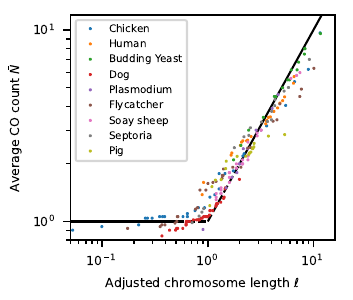}
}    
\subfloat{
\raisebox{135pt}[0pt]{\makebox[-5pt][l]{\textsf{\textbf{B}}}}
\includegraphics[width=0.567\textwidth]{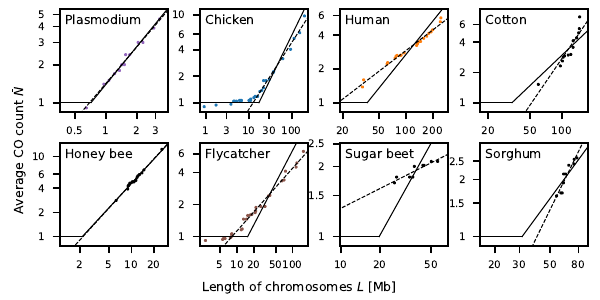}}
\caption{\textbf{Comparison of theoretical scaling prediction and genetic data collected by Fernandes~\cite{Fernandes:2018aa}.}
(A) Shown is the average CO count $\meanN$ as a function of adjusted chromosome length $\ell$ across multiple species using the dataset provided in~\cite{Fernandes:2018aa} that cover at least a factor of $5$ in chromosomes length (measured in base pairs [Mb]).
To determine the CO count we use the given data of the genetic size in centiMorgan [cM] and re-scale these by a factor of $50$ to account for the fact that only half of chromosomes remain in genetic data.
We adjusted the chromosome length so that the root mean square error to the theoretical prediction is minimal: $\meanN=1$ for short chromosomes due to CO assurance and, beyond that, a linear scaling ($\meanN \propto \ell$).
(B) Shown is the average CO count as a function of the chromosome length for some characteristic, individual species.
The black line shows the theoretical predictions and the dashed line the optimal power-law fit for all chromosomes with $N\geq 1.2$ crossover. The respective plots for the remaining data can be found in \Figref{fig:Figure_theory_scaling_SI}.
}
\label{fig:Figure6}
\end{figure*}

Our theoretical model of crossover placement makes detailed predictions that we next compare to published experimental data.

\subsection{Linear scaling with SC length explains CO count across species}
\label{sub:linear_scaling_with_sc_length_explains_co_count}
We start by comparing the average number~$\meanN$ of COs as a function of chromosome length~$L$ across various species using the dataset published by Fernandes \emph{et al.}~\cite{Fernandes:2018aa}.
Our theory predicts CO assurance ($\meanN \geq 1$) when material exchange along the SC dominates.
Beyond that, the theory predicts a linear scaling of class~I COs ($\meanN \propto L$), independent of whether coarsening is limited by exchange or diffusion.
The combined prediction is consistent with most species contained in \cite{Fernandes:2018aa} (\Figref{fig:Figure_theory_scaling_SI}).
This is particularly visible for species whose chromosome lengths vary significantly (\Figref{fig:Figure6}A).
Focusing on individual species, more detailed trends become apparent:
First, some species (e.g., plasmodium and honey bee) exhibit almost perfect, linear scaling, consistent with our theory.
Second, most species (e.g., chicken, human, and flycatcher) exhibit slight sub-linear scaling (with slopes between $0.5$ and $0.8$), and some (e.g., sugar beet) have a much smaller exponent, indicating that long chromosomes have fewer COs than we expect.
Third, some species (e.g., cotton and sorghum) exhibit super-linear scaling, suggesting long chromosomes have more COs than expected in these systems.
Such behavior could originate if shorter chromosomes are more strongly affected by chromosome regions with a lower concentration of the HEI10 analog (e.g., centromers).
Taken together, the scaling relation explains experimental data, and deviations point toward additional processes that affect CO placement.

\subsection{Model parameters determined from \Athaliana{} wild type and \zyp-mutant}
\label{sub:droplets_in_athaliana_}
We next determine parameters of our model that are consistent with empirical data of wild type and \zyp{}-mutants from \Athaliana{}.
We estimate the diffusion constant as $D\approx \SI{1}{\micro\meter\squared\per\second}$, based on the measured diffusion of ZHP3/4 along the SC in \Celegans{}~\cite{Stauffer:2019aa};
the SC diameter is $a\approx \SI{100}{\nano\meter}$~\cite{Capilla-Perez2021}.
We then choose the volume sensitivities as $\nu=\nuN=\frac13$.
To have significant CO interference and assurance in wild type we require $\gammaS\LambdaDS/(Da) > 10^{-1}$ (\Figref{fig:Figure5}D--E), and we choose $\nuS=0$ for simplicity.
The total amount $M$ of HEI10 is limited by the total volume of droplets, assuming that most HEI10 ends up in $N$ droplets of volume~$a^3$ in wild type, $M\approx\cD a^3 N$.
We get $\gamma_0 / \gammaS\approx 10^{-4}$ by applying the analytical scaling relation \eqref{eq:scaling_DS_diffusion} to wild-type CO counts~\cite{Durand2022}, assuming  coarsening approximately lasts for the duration of pachytene, $T_\mathrm{pach}\approx\SI{10}{\hour}$~\cite{Prusicki:2019aa}.
We next estimate the exchange rate $\gammaS\LambdaSN/a$ between SC and nucleoplasm by assuming that HEI10 loads onto  SCs within at most several hours~\cite{prusicki2019live}, implying $\gammaS\LambdaSN / (D a) > 10^{-6}$ using \Eqref{eqn:loading_rate} (\appref{ssub:model_parameters_determined_from_athaliana}).
However, if $\gammaS\LambdaSN$ is too large, CO interference would be impaired (\Figref{fig:Figure5}C), suggesting the upper bound $\gammaS\LambdaSN / (D a) < 10^{-4}$.
Finally, we estimate $\gammaS\LambdaDN$ based on the \zyp-mutant data~\cite{Durand2022}, where we consider both scenarios discussed in \Secref{sec:no_SC}:
If droplets only grow, and do not coarsen, we must have $T_\mathrm{pach} < T_\mathrm{growth}$, implying $\gammaS \LambdaDN/(D a) < 3 \cdot 10^{-2}$ using \Eqref{eq:T_growth}.
In contrast, if droplets coarsen, we get $\gammaS \LambdaDN/(D a) \approx 10^{-1}$ based on CO counts in \zyp-mutants compared to the  scaling relation \eqref{eq:DN_scaling_dimensional_solution}.
Note that we do not have any experimental measurements to estimate $\gammaS$ or $\gamma_0$ directly, so we chose $\gammaS=10\,a^3/V_\mathrm{N}$, which produced reasonable results, and assures that most HEI10 ends up in droplets.

We next check whether our estimate of the magnitude of nucleoplasmic exchange is still compatible with the observed CO interference of $\LintNorm \approx 0.7$~\cite{Morgan2021,Durand2022,Ernst:2024aa} in wild type, which is smaller than the prediction without exchange via nucleoplasm (\Figref{fig:Figure3}C).
One reason for this reduction in genetic data could be non-interfering class II COs, however a similar reduction is also observed in cytological data~\cite{Morgan2021}, suggesting additional effects.
We thus hypothesize that the observed reduction could be a consequence of nucleoplasmic exchange, and \Figref{fig:Figure5}C--D suggests that $\gammaS \LambdaDN/(D a) \approx 10^{-2}\ldots 10^{-1}$ would explain the observations.
We next test whether these parameters are consistent with observed CO assurance.
For wild type data~\cite{Morgan2021}, CO assurance is still close to $1$ and thus consistent with our model (\Figref{fig:Figure5}E).
The \zyp{}-mutant exhibits reduced CO assurance (\Eqref{eq:P_CO_assurance} in \secref{sec:no_SC}), which is consistent with the observation that $3$ of $14$ cells have at least one chromosome that does not carry COs ($p=0.23$)~\cite{Fozard:2023aa}.
Taken together, we find that the coarsening model tolerates substantial exchange via nucleoplasm, essentially because small differences in droplet size suffice to select final droplets.
This implies that there are a range of model parameters that are consistent with observed CO counts, CO assurance, and CO interference in wild type and \zyp{}-mutant.

\subsection{Lack of heterochiasmy in \Athaliana{} \zyp-mutant suggests there is no coarsening}
\label{sub:lack_of_heterochiasmy_in_athaliana_}

We next use our analytical results to investigate heterochiasmy, which refers to differences in CO count between sexes.
For example, \Athaliana{} males exhibit $50\si{\percent}$ more COs than females in  wild type~\cite{Durand2022,Fozard:2023aa}.
Since SCs are roughly $50\si{\percent}$ longer in male~\cite{Durand2022}, the scaling relation~\eqref{eq:scaling_DS_diffusion} directly implies the observed heterochiasmy if the average HEI10 line density~$\lambda$ and the coarsening time $T_\mathrm{c}$ is the same between both sexes in wild type.
However, pachytene is typically longer in female~\cite{hu2025cytological}, suggesting longer coarsening times, which would lead to fewer COs.
Taken together, the scaling relation~\eqref{eq:scaling_DS_diffusion} allows connecting the observed heterochiasmy to measured differences between male and female in wild type.

The observed heterochiasmy vanishes in \Athaliana{} \zyp{}-mutants~\cite{Durand2022,Fozard:2023aa}.
To understand this, we need to discuss the growth and coarsening regime separately:
If droplets only grow, essentially all droplets will become COs, so the absence of heterochiasmy must follow from a mechanism that ensures an equal number of initial droplets in both sexes.
In contrast, if coarsening is relevant, the droplet count reduces through competition for HEI10.
Assuming that physical parameters (like exchange rates and affinities) are identical across sexes, \Eqref{eq:DN_scaling_dimensional_solution} implies
\begin{align}
    \frac{N_\mathrm{male}}{N_\mathrm{female}} \approx \frac{M_\mathrm{male}}{M_\mathrm{female}} \left(\frac{T_\mathrm{female}}{T_\mathrm{male}}\right)^{\frac{1}{1+\nu-\nuN}}
    \;.
\end{align}
If the coarsening times are similar ($T_\mathrm{female}\approx T_\mathrm{male}$), a lack of heterochiasmy implies similar amounts~$M$ of HEI10 in both sexes ($M_\mathrm{male}\approx M_\mathrm{female}$).
Assuming that $M$ is the same in wild type and the \zyp-mutant and that most HEI10 ends up in droplets in wild type ($M\approx \lambda L$), we find $\lambda_\mathrm{female} > \lambda_\mathrm{male}$ since $L_\mathrm{female} < L_\mathrm{male}$.
This increased line density~$\lambda$ in females contradicts the conclusion from wild type in the previous paragraph, suggesting that one of the assumptions is invalid, or droplets in the \Athaliana{} \zyp{}-mutant are indeed not coarsening.

Heterochiasmy is \Athaliana{} is also affected by the overall levels of HEI10.
For instance, the overexpression mutant HEI10\textsuperscript{oe} does increase the CO count in both sexes~\cite{Durand2022}.
This is consistent with the predicted scaling relation~\eqref{eq:scaling_DS_diffusion}, which predicts a sub-linear scaling with total HEI10 line density $\lambda$, i.e. $\meanN\propto \lambda^{4/7}$ for $\nu=\frac13$.
Moreover, heterochiasmy is still present in HEI10\textsuperscript{oe}~\cite{Durand2022}, as expected.
The CO count increases further when additionally the SC is removed (in the \zyp{} HEI10\textsuperscript{oe}-mutant), but heterochiasmy is completely lacking~\cite{Durand2022}.
This observation is difficult to reconcile with coarsening, where we would expect the CO count $\meanN$ to scale with the total amount~$M$ of HEI10; see \Eqref{eq:DN_scaling_dimensional_solution}.
In contrast, if droplets only grow, but do not coarsen, the number of initial droplets $N_\mathrm{init}$ would need to depend on the HEI10 level, consistent with known nucleation mechanisms~\cite{ziethen2023nucleation}.
Taken together, the observations suggest that droplets in the \Athaliana{} \zyp{}-mutant are not coarsening.

\subsection{Coarsening model may explain homeostasis}
We next compare our model to experimentally observed CO homeostasis~\cite{Martini:2006aa,Rosu:2011aa,Cole:2012aa,Xue2018,Diezmann:2021aa, Morgan2021}, i.e., the fact that the observed CO count $\meanN$ depends only weakly on the number~$\meanN_\mathrm{DSB}$ of double-strand breaks (DSBs).
In particular, a \SI{46}{\percent} decrease in DSB count in \Athaliana{} results in only a \SI{17}{\percent} CO count decrease~\cite{Xue2018}, which corresponds to a sublinear scaling of $\meanN\sim \meanN_\mathrm{DSB}^{0.3}$. 
Since DSBs are typically associated with early droplets~\cite{Anderson:2014aa}, the simplest interpretation would be that $\meanN_\mathrm{DSB}$ controls the number $N_\mathrm{init}$ of initial droplets in our coarsening model.
We showed in \secref{sec:droplet_patterning_only_SC} that $N_\mathrm{init}$ does not affect the CO count in the coarsening regime, provided the total amount of HEI10 is unaffected.
The observed weak dependency thus suggest that increasing $\meanN_\mathrm{DSB}$ increases the HEI10 line density $\lambda$, which would affect the CO count sublinearly, $\meanN\propto \lambda^{4/7}$, see \Eqref{eq:scaling_DS_diffusion}.
Taken together, our model predicts that altered DSB counts affect the CO count not by changing the number of initial droplets, but by increasing the HEI10 line density.

\section{Discussion and Conclusion}
We derived a thermodynamically consistent coarsening model with material exchange between nucleoplasm, synaptonemal complex, and HEI10 droplets.
Our model explains key experimental observations, such as CO counts, CO assurance, and CO interference.
The asymptotic scaling relations we derived enable a simple discussion of parameter dependence; For instance, CO counts scales linearly with SC length, which explains the observed heterochiasmy in \Athaliana{} wild type.
More generally, these relations allow us to derive a consistent set of model parameters for describing CO placement in \Athaliana{} wild type and mutants, as well as HEI10 overexpression~\cite{Durand2022}.

Our extended model enables a detailed analysis of the material exchange via nucleoplasm.
Data from \Athaliana{} \zyp{}-mutant suggest that nucleoplasmic exchange is substantial, so it likely cannot be neglected in wild type.
This is consistent with reduced CO interference in wild type compared to a model without nucleoplasmic exchange~\cite{Ernst:2024aa}.
In particular, the observed CO assurance and reduction of CO interference in wild type is consistent with an acceleration of coarsening due to nucleoplasmic exchange of up to a factor of $2$, suggesting that up to half of the HEI10 exchange could be mediated by the nucleoplasm.
However, it remains unclear whether droplets simply grow, and the lack of heterochiasmy is a consequence of initial droplet formation, or whether also coarsen in the \Athaliana{} \zyp{}-mutant (cmp.~\cite{Durand2022,Fozard:2023aa,Girard2023}).
Taken together, more detailed cytological observations of \Athaliana{} \zyp{}-mutant could help to answer this question and refine our model.

Our model suggests that CO patterning is insensitive to details of initial conditions, provided the total initial HEI10 density is uniform along the SC.
This is a direct consequence of the rapid initial coarsening of small droplets in close vicinity.
If the initial HEI10 density is uniform along the SC, our model predicts that chromosome end regions (telomeres) exhibit fewer COs (\Figref{fig:Figure3}B).
In contrast, some experiments register elevated CO frequencies at telomeres~\cite{Morgan2021,Durand2022, Fozard:2023aa,Jing:2025aa}, which points at one or more heterogenities along the SC:
(i) Initial HEI10 loading could be heterogeneous~\cite{Morgan2021,Fozard:2023aa}, e.g., because SC might assemble from telomeres, enabling more loading of HEI10.
(ii) Earlier SC formation might induce earlier droplet growth and coarsening, allowing these droplets to outcompete others.
(iii) A heterogeneous density of DSBs could affect HEI10 loading~\cite{Lloyd2022}, e.g., DSB hotspots~\cite{borde2013programmed,Diezmann:2021aa}.
(iv) HEI10 affinity, captured by $\omegaS$ in our model, might depend on position along the SC.
Such heterogeneities, or combinations, could promote COs at telomeres, but also explain why centromeric regions typically contain no crossovers~\cite{Berchowitz:2010aa,smith2020new}.
An extreme form of heterogeneity are partial SC mutants, where chromosome zipping is not complete and multiple intact SC fragments exist along the same chromosome pair.
More generally, to understand these heterogeneities in more detail, a better understanding of SC assembly, initial HEI10 loading onto the SC, and dynamics of early HEI10 droplets is needed.
Our current model cannot explain the initial formation of droplets, it does not explicitly model active modifications (e.g., via post-translational modifications) of the involved proteins~\cite{kim2024control,Zhang:2025aa}, and it currently does not capture the physical interaction between droplets and SCs, e.g., due to wetting~\cite{zwicker2025physics}.
However, the model provides a solid framework and can be extended to study these details in the future.

\begin{acknowledgments}
We thank Abby Dernburg, Chloé Girard, Simone Köhler, Raphaël Mercier, Oliver Paulin, Gerrit Wellecke, and Noah Ziethen for fruitful discussions.
We thank Simone Köhler for a careful reading of the manuscript.
We gratefully acknowledge funding from the Max Planck Society and the European Research Council (ERC, EmulSim, 101044662).
\end{acknowledgments}
\FloatBarrier
\newpage
\onecolumngrid
\renewcommand\thefigure{\thesection.\arabic{figure}}    
\appendix
\setcounter{figure}{0}   
\section{General investigations}
This section contains general considerations and calculations, which are in principle valid in a larger class of models, beyond the coarsening model discussed in the main text.

\subsection{Relaxation of concentrations in a linear compartment coupled to a bulk}
\label{ssec:theory_relaxation}
We here investigate the relaxation behaviour of an effectively one-dimensional compartment coupled to a large bulk.
The one-dimensional compartment has length $L$ and an initial line concentration $c(x)$.
We consider diffusion along this compartment with diffusion constant $D$.
Assuming that diffusion in the bulk of volume $V$ is fast, we consider the concentration in the bulk to be spatially invariant and describe it by  the average concentration $c_\mathrm{b}$.
We further consider simple mass-action kinetics with off-rate $k_\mathrm{off}$, describing the unbinding from the linear compartment to the bulk, and on-rate $k_\mathrm{on}$ for the other direction.
The system can be described by the following dynamical equation and conservation law:
\begin{subequations}
\begin{align}
    \dif{c(x)}{t} &= D \nabla_x^2 c(x) +  \left(k_\mathrm{on} c_\mathrm{b} - k_\mathrm{off} c(x)\right)\;, \\ 
    c^\mathrm{tot}_\mathrm{b} V &= c_\mathrm{b} V + \int_{0}^{L} c(x, t) \diff x\;,
\end{align}
\end{subequations}
where the second equation describes mass conservation and $c^\mathrm{tot}_\mathrm{b}$ is the effective concentration if all material is in the bulk.
We can use mass conservation, to obtain the single equation
\begin{equation}
    \dif{c(x)}{t} = D \nabla_x^2 c(x, t) + k_\mathrm{on} \left(c^\mathrm{tot}_\mathrm{b} - \frac{1}{V}\int_{0}^{L} c(x, t) \diff x\right) -  k_\mathrm{off} c(x, t)\;,
\end{equation}
and impose no-flux boundary conditions at both ends.
We now mirror $c(x, t)$ at $x=0$ to get a $2L$-periodic function $c(x, t)=c(-x, t)$ and apply a Fourier series,
\begin{equation}
    c(x, t) = \sum_{q \in \mathbb{Z}} c_q(t) e^{i \pi q x/L} =  \sum_{n=1}^\infty 2 c_n(t) \cos{\left(\frac{\pi q x}{L}\right)} + c_0(t)\;,
\end{equation}
with $c_q(t)= c_{-q}(t)$ to ensure $c(x, t)\in \mathbb{R}$, yielding
\begin{equation}
    \sum_{q \in \mathbb{Z}} \dif{c_q(t)}{t} e^{i \pi q x/L} = k_\mathrm{on} c^\mathrm{tot}_\mathrm{b} -\sum_{q \in \mathbb{Z}} \left(\left[ D \left(\frac{\pi q}{L}\right)^2 + k_\mathrm{off} \right] c_q(t) e^{i \pi q x/L} + \frac{k_\mathrm{on}}{2 V}\int_{-L}^{L} e^{i \pi q x/L} c_q(t) \diff x \right)\;.
\end{equation}
We now observe that
\begin{equation}
    \frac12 \int_{-L}^{L} e^{i \pi q x/L} \diff x = 
    \begin{cases}
      L & \text{if $q=0$}\\
      \frac{L}{\pi q}\sin{\left(\pi q\right)}=0 & \text{if $q\in\mathbb{Z}\backslash\{0\}$}
    \end{cases}\;.
\end{equation}
\begin{subequations}
Since the Fourier modes are orthogonal, we obtain the following result for $q=0$:
\begin{equation}
    \dif{c_0(t)}{t} =  k_\mathrm{on} c^\mathrm{tot}_\mathrm{b} -\left(k_\mathrm{off}  + k_\mathrm{on} \eta\right) c_0(t)\;,
\end{equation}
with $\eta = L/V$.
This yields the relaxation time  $t^\mathrm{relax}(0)$ of the constant mode  ($q=0$),
\begin{equation}
    t^\mathrm{relax}(0) = \frac{1}{k_\mathrm{off}  + \eta k_\mathrm{on}}\;.
    \label{eq:t_relax_0}
\end{equation}
\end{subequations}
Notably, the relaxation time does depend on the size ratio of the compartments $\eta$: If the one-dimensional compartment increases in size, the relaxation time decreases. This is because the bulk depletes faster, implying the compartment and the bulk will reach equilibrium earlier.
\begin{subequations}
We now conclude the following for $q\in\mathbb{Z}\backslash\{0\}$
\begin{equation}
    \dif{c_q(t)}{t}  = - \left[D \left(\frac{\pi q}{L}\right)^2 + k_\mathrm{off} \right] c_q(t) = D_q \left(\frac{\pi q}{L}\right)^2\;,
\end{equation}
which yields the relaxation time $t^\mathrm{relax}_q$ of the $q$-th mode for $q\neq 0$,
\begin{equation}
    t^\mathrm{relax}(q) = \frac{1}{k_\mathrm{off} +  D \left(\frac{\pi q}{L}\right)^2}
    \label{eq:t_relax_q}\;.
\end{equation}
We observe that the relaxation time does depend on the wavelength of the modes and, in particular, increases for larger wavelengths $\tilde{\lambda}=2L/q$.
This is due to the non-local, fast transport through the bulk.
We can now describe the behaviour of the non-constant modes as effective non-local diffusion, with an effective diffusion constant, $D_q$, that depends on the mode $q$:
\begin{equation}
    D_\mathrm{q} = D + k_\mathrm{off} \left(\frac{L}{\pi q}\right)^2\;.
    \label{eq:theory_D_q}
\end{equation}
\end{subequations}
This converges to the normal diffusion constant $D$ for modes with small wavelengths, whereas we get increased effective diffusion by increased transport via the bulk (where we assumed fast diffusion)  for modes of large wavelength.

\subsection{Stationary profiles between droplets along one-dimensional compartment with bulk-exchange}
\label{ssec:theory_concentration_profile}
\begin{figure}[tp]
\centering
\includegraphics[width=0.6\textwidth]{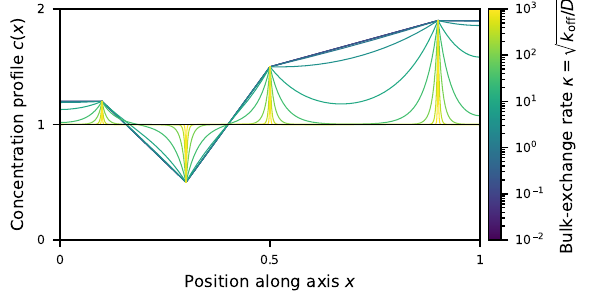}
\caption{
\textbf{Concentration profile along one-dimensional compartment with bulk-exchange} as given by \Eqref{eq:theory_concentration_profile} for varying bulk-exchange rates $\kappa = \sqrt{k_\mathrm{off}/D}$ along the axis with some exemplary droplets at $x_i$, which impose $c(x_i)$.}
\label{fig:theory_concentration_profile}
\end{figure}

We here analyze the diffusion-limited concentration profile in a one-dimensional compartment with bulk-exchange, where droplets are described by the effective droplet model.
Specifically, we consider a bulk with a homogeneous mean-field concentration $c_b$, a one-dimensional compartment of length $L$ with linear concentration profile $c(x)$, and $N$ droplets with point-like exchanges at $x_i$ for $i\in1, \ldots, N$.
We further consider diffusion along the one-dimensional compartment and exchange between compartment and bulk via simple mass-action kinetics, as well as exchange between compartment and droplets.
We assume that the size of the droplets, the concentration profile in the compartment, and the concentration in the bulk change on a slow timescale, and that the dynamics of the compartment is diffusion-limited.
We also assume that the exchange between the droplets and the compartment is fast, implying that the local concentration in the compartment is equal to the outside equilibrium concentration of the droplets.
The dynamical equation describing the concentration profile along the compartment is given by
\begin{subequations}
\begin{align}
    \dif{c(x)}{t} &= D \nabla_x^2 c(x) + \left(k_\mathrm{on} c_\mathrm{b} - k_\mathrm{off} c(x)\right)\;, \\ 
    c(x_i, t) &= c_i\;, 
\end{align}
\end{subequations}
where $D$ is the diffusion constant along the compartment, $k_\mathrm{on}$ and $k_\mathrm{off}$ are the kinetic exchange rates, and droplets placed at $x_i$ enforce the respective equilibrium concentrations $c_i$.
Consequently, the concentration profile $c(x)$ only depends on the values of the adjacent droplets and the kinetic exchange rates.
Since we assume a slow change of concentration profile within the compartment, we set $\dif{c(x)}{t} = 0$, and find
\begin{subequations}
\begin{align}
    \nabla_x^2 c(x) + \frac{k_\mathrm{off}}{D} \left(\frac{k_\mathrm{on}}{k_\mathrm{off}} c_\mathrm{b}  - c(x) \right) &= 0\;, \\ 
\Rightarrow \nabla_x^2 c(x) + \kappa^2 \left(c_{0}  - c(x) \right) &= 0 \label{eq:theory_laplace_equation}\;,
\end{align}
\end{subequations}
where $\kappa = \sqrt{k_\mathrm{off}/D}$ and $c_0 = k_\mathrm{on}/k_\mathrm{off} c_\mathrm{b}$.
The general solution to this equation is given by (\Figref{fig:theory_concentration_profile})
\begin{subequations}
\begin{equation}
    c(x) = \tilde{c}_\mathrm{b} + k_1 e^{\kappa x} + k_2 e^{-\kappa x}\;.
    \label{eq:theory_concentration_profile}
\end{equation}
By applying the boundary conditions for each piecewise interval $[x_i, x_{i+1}]$, we get
\begin{align}
    k_1 &= \frac{\Delta c_{i+1} e^{\kappa x_i} - \Delta c_{i}e^{\kappa x_{i+1}}}{e^{\kappa \Delta x} - e^{-\kappa \Delta x}}\;,\\ 
k_2 &= \frac{\Delta c_{i}e^{\kappa x_{i+1}} - \Delta c_{i+1} e^{\kappa x_{i}}}{e^{\kappa \Delta x_i} - e^{-\kappa \Delta x_i}}\;,
\end{align}
\end{subequations}
where $\Delta x_{i} = x_{i+1}- x_{i}$ with concentrations $\Delta c_{i} = c_{i} - c_0$.
Hence, the concentration profile for $x_i < x < x_{i+1}$ reads
\begin{subequations}
\begin{equation}
    c(x) = c_0 + \frac{\Delta c_{i+1} \sinh{\kappa (x - x_i)} -  \Delta c_{i} \sinh{\kappa (x - x_{i+1})}}{\sinh{\kappa \Delta x_i}}\;.
\end{equation}
Meanwhile, for $x = x_{i}$ where $i\in [1, N]$, we obtain
\begin{equation}
    \int_{x_{i}-\epsilon}^{x_{i}+\epsilon} \left[D {\nabla_x}^2 c(x) - \mathcal{S}_i \delta{\left(x-x_i\right)}\right]\diff x= 0
\end{equation}
for sufficiently small $\epsilon >0$, so that $x_{i}+\epsilon < x_{i+1}$ etc., and a diffusion-limited flux of $\mathcal{S}_i$ into the droplet
\begin{align}
    \mathcal{S}_i &= D\left.{\nabla_x} c(x)\right|_{x_{i}-\epsilon}^{x_{i}+\epsilon},\\ 
    &=  D\kappa \left(\frac{\Delta c_{i+1}}{\sinh{\kappa \Delta x_i}} + \frac{\Delta c_{i}}{\sinh{\kappa \Delta x_{i+1}}}  - \Delta c_{i} \coth{\kappa \Delta x_{i}} - \Delta c_{i+1} \coth{\kappa \Delta x_{i+1}}  \right)\;, \\ 
    &\approx D \left[\frac{\Delta c_{i+1} - \Delta c_{i}}{\Delta x_{i}} + \frac{\Delta c_{i-1} - \Delta c_{i}}{\Delta x_{i-1}} - \kappa^2 \left(\frac{\Delta c_{i} \left(\Delta x_{i} + \Delta x_{i-1}\right)}{3} - \frac{\Delta c_{i+1} \Delta x_{i} + \Delta c_{i-1} \Delta x_{i-1}}{6}\right)\right]\;, \label{SI_theory:full_sds}
\end{align}
\label{eq:theory_concentration_with_bulk_exchange}
\end{subequations}
where the last line provides a second-order approximation for small $\kappa$. 
Notably, exchange between the one-dimensional compartment and the bulk introduces a non-local component, increasing diffusional transport between droplets.
The first term describes diffusion along the one-dimensional compartment. For $\kappa = 0$, this describes the profile without bulk exchange, yielding a piecewise linear profile between the droplets.
The second term is a second-order approximation of the bulk exchange, leading to equilibration of the concentration with the surrounding bulk, steeper concentration profiles near the droplets, and thus larger diffusive fluxes between the droplets.

\subsection{Scaling law for diffusion-limited droplet coarsening along one dimension}
\label{ssec:theory_scaling_1D}
We here derive an asymptotic solution to diffusion-limited coarsening along an infinite one-dimensional compartment.
Following the approach of Lifshitz, Slyozov, and Wagner~\cite{Lifshitz1961,Wagner:1961aa}, we derive scaling relations for mean droplet growth and the number of remaining droplets as a function of time, as well as the asymptotic droplet size distribution.

We consider a one-dimensional compartment with $N(t)$ droplets per unit length with volume size distribution $n(V, t)$.
We describe the size-dependent equilibrium concentration $c_\mathrm{eq}$ of the droplets by
\begin{equation}
    c_\mathrm{eq}(V) = c_0 + c_1 \left(\frac{V}{V_0}\right)^{-\alpha}\;,
\end{equation}
with droplet volumes $V$, volume scale $V_0$, concentration scales $c_0$ and $c_1$, and size-sensitivity exponent $\alpha>0$, which ensures that larger droplets have lower equilibrium concentration to drive coarsening.
The particular case $\alpha = \frac13$ recovers the equilibrium concentration corresponding to typical surface tension effects~\cite{Weber2019}.
We assume that exchange between droplets and compartment occurs on a fast timescale, whereas droplet size $V$ and the concentration profile $c(x)$ in the compartment change on a slow timescale, and that the dynamics is diffusion-limited.
It follows (\secref{ssec:theory_concentration_profile}) that the concentration profile $c(x)$ is piecewise linear between droplets (since we do not consider exchange with other compartments) and the compartment is in equilibrium with the droplets, i.e., for droplet $i$ with position $x_i$ and size $V_i$, we have $c(x_i) = c_{\mathrm{eq}}(V_i)$.

In our mean-field-like approach we assume that the distance between all adjacent droplets is equal at all times, and thus given by the inverse of the number of droplets per unit length,
\begin{equation}
    \Delta x(t) = \frac{1}{N(t)}\;.
\end{equation}
Using this in \Eqref{SI_theory:full_sds} yields the dynamical equation for the growth of the $i$-th droplet,
\begin{subequations}
\begin{align}
    \dif{V_{i}}{t} &= D \left[\frac{c_{i+1} - c_{i}}{\Delta x(t)} + \frac{c_{i-1} - c_{i}}{\Delta x(t)}\right]\;,\\
    &= \frac{D c_1}{\Delta x(t)} \left[V_{i+1}^{-\alpha} + V_{i-1}^{-\alpha}- 2V_i^{-\alpha}\right]\;,
    \label{eq:theory_dV_1}
\end{align}
\end{subequations}
with diffusion coefficient $D$. 
Note that vanishing droplets, i.e. $V_i=0$, are excluded from subsequent dynamics, and thus indirectly affect the effective droplet distance $\Delta x(t)$.
To investigate the growth of a particular droplet $i$, we assume that the size of the two adjacent droplets is given by the stationary droplet size $V_\mathrm{eq}$, implying their equilibrium concentrations can be approximated by the average concentration $c_{\mathrm{eq}, i\pm 1}  = \bar{c}(t) = c_0+ c_1  V_\mathrm{eq}^{-\alpha}(t)$.
We thus have
\begin{equation}
    \dif{V_i}{t} = 2 D c_1 \frac{V_\mathrm{eq}^{-\alpha}(t) - V_i^{-\alpha}(t)}{\Delta x(t)}\;.
\end{equation}
To derive the solution in the asymptotic limit, we assume that this limit exists and the asymptotic relative droplet size distribution follows a universal distribution.
The average droplet size $\bar{V}(t)$ in that limit is proportional to the stationary droplet size
\begin{equation}
    \bar{V}(t) = V_\mathrm{eq}(t) k^\infty \label{eq:theory_barVt}\;.    
\end{equation}
The conservation law per unit length reads
\begin{subequations}
\begin{align}
    \textrm{const} &= \cD N(t) \bar{V}(t) + \bar{c}(t) \;, \\
    \Rightarrow M &= N(t) \bar{V}(t) + \mu V_\mathrm{eq}^{-\alpha}(t) \;,\\
    \Rightarrow N(t) &= \frac{M - \mu V_\mathrm{eq}^{-\alpha}(t)}{k^\infty V_\mathrm{eq}(t)} = \frac{1}{\Delta x(t)}\
    \label{theory:conservation_N}
\end{align}
\end{subequations}
where $\cD$ is the average concentration in the droplets, $\mu = c_1/\cD$, and $M$ is a constant.
\Eqref{eq:theory_dV_1} then yields
\begin{equation}
    \dif{V_i}{t} = \frac{2 D M c_1}{k^\infty}  V_\mathrm{eq}^{-1-\alpha}(t)  \left[1 - \left(\frac{V_i(t)}{V_\mathrm{eq}(t)}\right)^{-\alpha} \right] \left(1 - \frac{\mu V_\mathrm{eq}^{-\alpha}(t)}{M}\right)\;. 
    \label{eq:theory_dV_i_dt}
\end{equation}
Using $V_{\mathrm{eq}, 0} = V_\mathrm{eq}(t_0)$, we define $\rho = V_i(t)/V_{\mathrm{eq}, 0}$, $x = V_\mathrm{eq}(t)/V_{\mathrm{eq}, 0}$, $z = \rho/x = V_i(t)/V_\mathrm{eq}(t)$, $\mu' = \mu V_{\mathrm{eq}, 0}^{-\alpha}/M$, $\kappa = M/V_{\mathrm{eq}, 0}$, and $T_0 = k^\infty V_{\mathrm{eq}, 0}^{2+\alpha}/(2 D M c_1)$ (to rescale the time $\tau = t/T_0$).
We then get
\begin{subequations}
\begin{equation}
    \dif{\rho}{\tau} = x^{-1-\alpha} \left(1 - z^{-\alpha}\right)  \left(1 - \mu x^{-\alpha}\right)\;.\label{eq:theory_scaling_drhodt}
\end{equation}
We re-write the droplet size distribution $f(\rho, \tau)$,
\begin{equation}
    \int n(V, t) \diff \rho = \int f(\rho, \tau) \diff \rho = N(\tau T_0)\label{eq:theory_scaling_frhotau2}\;.
\end{equation}
\end{subequations}
By integrating $f(\rho, \tau)$ and applying \Eqref{theory:conservation_N}, we write the conservation law as
\begin{subequations}
\begin{align}
    \int_0^\infty f(\rho, \tau) \rho \diff \rho &\hat{=}  \sum_{i=1}^{N(\tau T_0)} \rho_i= \sum_{i=1}^{N(\tau T_0)} \frac{V_i(\tau T_0)}{V_{\mathrm{eq}, 0} } 
    =  \frac{N(\tau T_0) \bar{V}(\tau T_0)}{V_{\mathrm{eq}, 0}};,\\
    &= \kappa - \mu x^{-\alpha}\;. \label{eq:theory_conservation_frhotau}
\end{align}
\end{subequations}
We rescale time $\tau' = \ln{(x(\tau))}$ and re-write the dynamical equation for the relative droplet growth \Eqref{eq:theory_scaling_drhodt} as
\begin{subequations}
\begin{align}
 v(z,\tau') := \dif{z}{\tau'} &= x^{-1-\alpha}\left(1 - z^{-\alpha}\right)  \left(1 - \mu x^{-\alpha}\right) \left[\dif{x}{\tau} \right]^{-1} - z\;, \label{eq:theory_dzt_v}\\
    &= e^{-(1+\alpha) \tau'}\left(1 - z^{-\alpha}\right)\left(1 - \mu e^{-\alpha \tau'}\right) \left[\dif{x}{\tau} \right]^{-1} -z\;, \\ 
    &= \left(1 - z^{-\alpha}\right)\left(1 - \mu e^{-\alpha \tau'}\right) \gamma{(\tau')} -z \label{eq:theory_scaling_drhodt2}\;,
\end{align}
by applying
\begin{align}
    \dif{\rho}{\tau} &= \dif{(z x)}{\tau}
    = z \dif{x}{\tau} + x \dif{z}{\tau'} \dif{\tau'}{\tau}\
    = \left(z + \dif{z}{\tau'}\right) \dif{x}{\tau}\;, \label{eq:theory_scaling_drhodtau}\\
    \gamma(\tau') &= e^{-(1+\alpha) \tau'} \left[\dif{x}{\tau} \right]^{-1} = x^{-(1+\alpha)} \left[\dif{x}{\tau} \right]^{-1}\;. \label{eq:theory_scaling_gamma}
\end{align}
\end{subequations}
In the asymptotic limit, $\tau'\rightarrow \infty$, we get
\begin{equation}
 \dif{z}{\tau'} =  \left(1 - z^{-\alpha}\right) \gamma{(\tau')} -z \label{eq:theory_scaling_drhodt3}\;.
\end{equation}
Following the approach in~\cite{Lifshitz1961} $\gamma(\tau')$ does need to converge to constant $\gamma$ for $\tau'\rightarrow \infty$ to allow for an asymptotically stationary state.
For $\gamma(\tau') \rightarrow \gamma > 0$ we define the relative growth rate
\begin{equation}
v(z, \gamma) = \dif{z}{\tau'}  = \left(1 - z^{-\alpha}\right) \gamma -z \label{eq:theory_scaling_vzgamma}\;.
\end{equation}
This relative growth rate asymptotically must obtain a maximum at $z=z_0>0$ for $\gamma=\gamma_0>0$ with $v(z_0, \gamma_0)\rightarrow 0$ (compare~\cite{Lifshitz1961}).
If, on the other hand $\gamma > \gamma_0$ or $\gamma \rightarrow \infty$, the total droplet volume diverges, and if $\gamma < \gamma_0$, then $\dif{z}{\tau'} < 0$ for all $z>0$, and all the droplets will disappear.
Consequently, the condition for the maximum $z_0, \gamma_0$ follows as
\begin{subequations}
\begin{align}
    \left.v{(z, \gamma_0)}\right|_{z=z_0} &=0\;,\\
    \left.\dif{}{z} v{(z, \gamma_0)} \right|_{z=z_0}&=\left.\alpha z^{-1-\alpha} \gamma_0 - 1\right|_{z=z_0} = 0\;,
\end{align}
\label{eq:theory_conditions_gamma}
\end{subequations}
which indeed corresponds to a maximum, since $\frac{\diff^2}{\diff z^2} v(z, \gamma_0) = -\alpha (1+\alpha) z^{-2-\alpha} \gamma_0 < 0$ for $\alpha>0$, yielding
\begin{subequations}
\begin{align}
z_0 &= (1+\alpha)^{1/\alpha}\;, \label{eq:theory_z0}\\
    \gamma_0 &= \frac{1}{\alpha}  (1+\alpha)^{\frac{1+\alpha}{\alpha}} \label{eq:theory_gamma0}\;.
\end{align}
\end{subequations}
In particular, for $\alpha=\frac13$ we have $\gamma_0=\frac{256}{27}$ and $z_0=\frac{64}{27}$, and for $\alpha=\frac14$ we have $\gamma_0=\frac{3125}{256}$ and $z_0=\frac{625}{256}$.

\begin{figure*}[t]
    \centering
\subfloat{
\raisebox{135pt}[0pt]{\makebox[-10pt][l]{\textsf{\textbf{A}}}}
\includegraphics[width=0.333\textwidth]{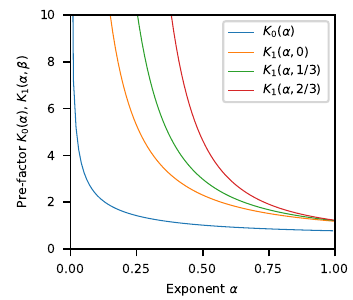}}
\subfloat{
\raisebox{135pt}[0pt]{\makebox[-5pt][l]{\textsf{\textbf{B}}}}
\includegraphics[width=0.333\textwidth]{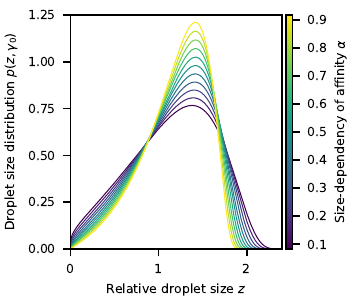}}
\subfloat{
\raisebox{135pt}[0pt]{\makebox[-5pt][l]{\textsf{\textbf{C}}}}
\includegraphics[width=0.333\textwidth]{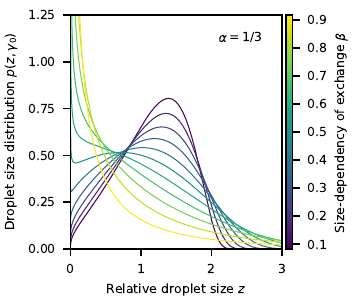}}
\caption{
\textbf{Pre-factors $K_0(\alpha)$, and $K_1(\alpha, \beta)$, and droplet size distributions from asymptotical scaling analysis.}
(A) Shown is the pre-factor $K_0(\alpha)$ for the scaling relation of diffusion-limited coarsening along one dimension (\Eqref{eq:theory_K0} in \secref{ssec:theory_scaling_1D}), and the pre-factor $K_1(\alpha, \beta)$ for exemplary size-depencies of the exchange rate $\beta$ for coarsening in the generalized case (\Eqref{eq:theory2_K1} in \secref{ssec:theory_scaling_generalized}).
(B) Universal droplet size distribution $p(z, \gamma_0)$ for varying $\alpha$ for diffusion-limited coarsening along one dimension (\Eqref{eq:theory_pzgamma0}).
(C) Universal droplet size distribution $p(z, \gamma_0)$ for $\alpha=1/3$, and varying $\beta$ for generalized coarsening (\Eqref{eq:theory2_pzgamma0}).}
\label{fig:Figure_theory_distributions}
\end{figure*}

The relative droplet size distribution, that we write in terms of $z$ and $\tau'$,
\begin{equation}
    \phi(z,\tau') \diff z = f(\rho, \tau) \diff\rho\;, \label{eq:theory_phiztau}
\end{equation}
fulfils the continuity equation
\begin{equation}
    \frac{\partial \phi(z,\tau')}{\partial \tau'} = \frac{\partial}{\partial z}\left(-\dif{z}{\tau'} \phi(z,\tau')\right)\;. \label{eq:theory_scaling_continuity2}
\end{equation}
Applying $\diff z/\diff \tau' = v{(z, \gamma)}$ (\Eqref{eq:theory_scaling_vzgamma}) this gives the general solution~\cite{Lifshitz1961} (for $\gamma{(\tau')}\rightarrow \gamma_0$)
\begin{subequations}
\begin{equation}
    \phi{(z, \tau')} = \chi{(\tau+\psi(z))}\frac{-1}{v{(z,\gamma_0)}}\;, \label{eq:theory_scaling_phiztau2}
\end{equation}
whereas $\chi$ is an arbitrary function still to be determined, and
\begin{equation}
    \psi(z) = \int_0^z \frac{-1}{v(z, \gamma)} \diff z\;. \label{eq:theory_scaling_psi}
\end{equation}
\end{subequations}
We re-write the droplet size distribution in the asymptotic limit as
\begin{equation}
    \phi{(z, \tau')} = e^{-\tau'} \Phi{(z, \gamma_0)}\;,
\end{equation}
which asymptotically fulfils the conservation \Eqref{eq:theory_conservation_frhotau},
\begin{equation}
     \int_0^\infty f(z, \tau) \rho \diff \rho = e^{\tau'} \int_0^\infty \phi(z, \tau') z \diff z = \int_0^\infty \Phi{(z, \gamma_0)} z \diff z = \kappa \;,  \label{eq:theory_scaling_conservation2}
\end{equation}
since the second term vanishes for $\tau'\rightarrow \infty$.
To allow consistency with the general solution given by \Eqref{eq:theory_scaling_phiztau2}, it follows that the function $\chi$ has to have the form
\begin{equation}
    \chi{(\tau' + \psi)} = \chi_0 e^{-(\tau' + \psi)}\;.
\end{equation}
In particular, the distributions $\Phi{(z, \gamma_0)}$ and $\phi{(z, \gamma_0)}$ have to vanish for $z > z_0$ or otherwise these droplets would all converge to a size of $z_0$ and the total volume diverges,
\begin{equation}
    \phi{(z, \gamma_0)} = \begin{cases}
        \chi_0 e^{-\tau'} \frac{-e^{-\psi}}{v(z, \gamma_0)} & z\leq z_0\;, \\
        0 & z>z_0\;, \end{cases} \label{eq:theory_scaling_phi}
\end{equation}
where $\chi_0$ can be derived by applying \Eqref{eq:theory_scaling_conservation2},
\begin{equation}
    \chi_0 = \kappa \left[ \int_0^{z_0} \frac{-e^{-\psi} z}{v{(z, \gamma_0)}} \diff z \right]^{-1}\;. \label{eq:theory_scaling_A}
\end{equation}
We then derive the relative universal size distribution (\Figref{fig:Figure_theory_distributions}B)
\begin{subequations}
\begin{align}
p{(z, \gamma_0)} = \left[\int_0^\infty \phi{(z, \gamma_0)} \diff z\right]^{-1} \phi(z, \gamma_0) &= \left[\chi_0 e^{-\tau'}\right]^{-1} \phi(z, \gamma_0) =\begin{cases}\frac{-e^{-\psi}}{v(z, \gamma_0)} & z\leq z_0, \\ 0 & z>z_0 \end{cases} \label{eq:theory_pzgamma0}  \;,\\
    \textrm{with } \int_0^\infty p{(z, \gamma_0)} \diff z &= 1\;, \label{eq:theory_scaling_int_pzy0}
\end{align}
\end{subequations}
and the number of remaining droplets $n'(\tau')$,
\begin{equation}
    n'{(\tau')} = \int_0^\infty \phi{(z, \gamma_0)} \diff z = \chi_0 e^{-\tau'} = n(\tau) = \frac{\chi_0}{x(\tau)}=N{(\tau T_0)}\;.
\end{equation}
Applying \Eqref{eq:theory_scaling_gamma} and assuming a power law relation in the asymptotic limit $x(\tau) = x_0 {\tau}^{\beta}$ yields
\begin{subequations}
\begin{align}
    \gamma{(\tau)} \rightarrow \gamma_0 &= {x(\tau)}^{-(1+\alpha)} \left[ \dif{x(\tau)}{\tau}\right]^{-1}\;,\\
    \Rightarrow \beta &= \frac{1}{2+\alpha}\;, \\ 
    \Rightarrow x(\tau) &= \left(\frac{2+\alpha}{\gamma_0}\right)^{\beta} {\tau}^{\beta}\;.
\end{align}
\end{subequations}
Consequently, the number of droplets per unit length in the asymptotic limits scales with
\begin{subequations}
\begin{align}
    N(t) &= \frac{\chi_0}{x(t/T_0)} = \frac{\chi_0}{x_0} \left(\frac{t}{T_0}\right)^{-\frac{1}{2+\alpha}}\;,\\
    &= \frac{M}{V_{\mathrm{eq}, 0}} \left[ \int_0^{z_0} \frac{-e^{-\psi} z}{v{(z, \gamma_0)}} \diff z \right]^{-1} \left(\frac{\gamma_0}{2+\alpha}\right)^{\frac{1}{2+\alpha}} \left(\frac{k^\infty V_{\mathrm{eq}, 0}^{2+\alpha}}{2 D M c_1 t }\right)^{\frac{1}{2+\alpha}}\;, \\ 
    &= K_0(\alpha) M^{\frac{1+\alpha}{2+\alpha}} \left(\frac{1}{D c_1 t }\right)^{\frac{1}{2+\alpha}}\;,
\end{align}
where the pre-factor
\begin{align}
    K_0(\alpha) &= \left[\frac{k^\infty}{2}\right]^{\frac{1}{2+\alpha}} \left[ \int_0^{z_0} \frac{-e^{-\psi(z)} z}{v{(z, \gamma_0)}} \diff z \right]^{-1} \left(\frac{\gamma_0}{2+\alpha}\right)^{\frac{1}{2+\alpha}}\;,\label{eq:theory_K0}\\
    k^\infty &= \bar{z} = \int_0^{z_0} p{(z, \gamma_0)}z dz \label{eq:theory_kinfty}\;,
\end{align}
collects all constants that only depend on the exponent $\alpha$ (\Figref{fig:Figure_theory_distributions}A).
\end{subequations}

The equilibrium droplet volume and radius scales with
\begin{subequations}
\begin{align}
    V_\mathrm{eq}(t) &= \frac{\bar{V}(t)}{k^\infty} = \frac{M}{k^\infty N(t)} 
    = \frac{1}{k^\infty K_0(\alpha)} \left(2 D c_1 M t\right)^{\frac{1}{2+\alpha}}\;,\\
    R_\mathrm{eq}(t) &\propto \left(D  M t\right)^{\frac{1}{3(2+\alpha)}}\;.
\end{align}
\end{subequations}

For $\alpha=\frac13$, the number of droplets per unit length $N(t)$ scales with exponent $-\frac37$ and the droplet radius with exponent $\frac17$.
This scaling is much weaker than the typical scalings with respective exponents $-1$ and $\frac13$ for Ostwald ripening in diffusion-limited coarsening~\cite{Lifshitz1961,Wagner:1961aa}.

\subsection{Scaling law of droplet coarsening with generalized exponents}
\label{ssec:theory_scaling_generalized}
In this section we derive the asymptotic scaling relations of a generalized version of the classical diffusion-mediated coarsening~\cite{Lifshitz1961,Wagner:1961aa}.
In particular, we consider an infinite system with $N(t)$ droplets per unit-volume, a droplet size distribution $n(V, t)$, and a bulk that is described by a mean-field concentration $c(t)$.
We generalize the equilibrium concentration of the droplet and the droplet growth rate as
\begin{subequations}
\begin{align}
    c_\mathrm{eq}(V) &= c_0 + c_1 \left(\frac{V}{V_0}\right)^{-\alpha}\;,\\ 
    \dif{V}{t} &\propto V^\beta \left(c(t) - c_\mathrm{eq}(V)\right)\;,
\end{align}
\end{subequations}
with droplet volumes $V$, bulk concentration $c(t)$, volume scale $V_0$, and concentration scales $c_0$, $c_1$.
Moreover, $\alpha>0$ ensures larger droplets have lower equilibrium concentration to drive coarsening, and  $0<\beta < 1$ describes the size-dependence of the exchange between droplets and bulk.
We require $\beta<1$ since the subsequent derivation does otherwise not yield an asymptotic solution.
With the stationary droplet size defined by $c(t) = c_0 + c_1\left(\frac{V_\mathrm{eq}}{V_0}\right)^{-\alpha} $, and exchange rate $\Lambda$, we obtain the dynamics and conservation law,
\begin{subequations}
\begin{align}
    \dif{V}{t} &= \Lambda V^\beta \left(V_\mathrm{eq}^{-\alpha} - V^{-\alpha}\right)\;,\\
    V_\mathrm{tot} &= \int_0^\infty n(V, t) V \diff V + \mu_0 c(t) = N(t) \bar{V} + \mu_0 c_1 \left(\frac{V_\mathrm{eq}}{V_0}\right)^{-\alpha}\;,
    \label{eq:theory2_dynamics_equation}    
\end{align}
with total material per unit volume $V_\mathrm{tot}$ and constant $\mu_0$.
\end{subequations}
Note that we recover diffusion-limited coarsening for $\alpha=\frac13$ and $\beta = \frac13$~\cite{Lifshitz1961,Wagner:1961aa} and conversion-limited coarsening for $\alpha=\frac13$ and $\beta=\frac23$~\cite{Wagner:1961aa}.

We closely follow the approach of the previous section, inspired by Lifshitz, Slyozov, and Wagner~\cite{Lifshitz1961,Wagner:1961aa}.
With $V_{\mathrm{eq}, 0}=V_\mathrm{eq}(t_0)$ at time $t_0$, we define $\rho = V/V_{\mathrm{eq}, 0}$, $x = V_{\mathrm{eq}}(t)/ V_{\mathrm{eq}, 0}$ and $z = \rho/x$, $T_0 = V_{\mathrm{eq}, 0}^{1+\alpha-\beta}/\Lambda$, rescale time $\tau=t/T_0$ to get the dynamics (compare \Eqref{eq:theory_scaling_drhodt}),
\begin{subequations}
\begin{equation}
    \dif{\rho}{\tau} = x^{\beta-\alpha} \left(z^\beta - z^{\beta - \alpha}\right)\;.\label{eq:theory_2_drhotau}
\end{equation}
We re-write the droplet size distribution in terms of $\rho$ and $\tau$
\begin{equation}
    \int_0^\infty n(V, t) \diff V = \int_0^\infty f(\rho, \tau) \diff\rho = N(t) = N(\tau T_0)\;,
\end{equation}
and write material conservation in the asymptotic limit (compare \Eqref{eq:theory_conservation_frhotau}), as
\begin{equation}
    \int_0^\infty f(\rho, \tau) \rho \diff\rho  + \mu x^{-\alpha}= \kappa \;,
\end{equation}
with constants $\kappa=V_\mathrm{tot}/V_{\mathrm{eq}, 0}$ and $\mu = \mu_0 c_1 V_0^{-\alpha}/V_{\mathrm{eq}, 0}^{1+\alpha}$.
\end{subequations}

We rescale time with $\tau'=\ln{(x(\tau))}$ and analogously re-write the dynamical equation for the relative droplet growth \Eqref{eq:theory_2_drhotau} (compare \Eqref{eq:theory_scaling_drhodt2}),
\begin{subequations}
\begin{align}
    v(z, \tau') := \dif{z}{\tau'} &= x^{\beta-\alpha} \left(z^\beta - z^{\beta - \alpha}\right)\;,\\
    &= z^{\beta} \left(1 - z^{-\alpha}\right) \gamma{(\tau')} -z\;, \label{eq:theory2_scaling_vztau}
\end{align}
where
\begin{equation}
    \gamma(\tau') = e^{(\beta-\alpha) \tau'} \left[\dif{x}{\tau} \right]^{-1} = x^{(\beta-\alpha)} \left[\dif{x}{\tau} \right]^{-1}\;. \label{eq:theory2_scaling_gamma}
\end{equation} 
\end{subequations}
Following the approach of the previous section, we only get an asymptotically stationary solution if $\gamma(\tau') \rightarrow \gamma>0$ for $\tau'\rightarrow \infty$.
Analogously, we define \Eqref{eq:theory2_scaling_vztau} (compare \Eqref{eq:theory_scaling_vzgamma}) as
\begin{subequations}
\begin{equation}
v(z, \gamma) = \dif{z}{\tau'} = z^{\beta} \left(1 - z^{-\alpha}\right) \gamma -z\;,\label{eq:theory2_scaling_vzgamma}
\end{equation}
and, when applying the conditions for the maximum $z_0, \gamma_0$ (compare \Eqref{eq:theory_conditions_gamma}), we get
\begin{align}
z_0 &= \left(\frac{1+\alpha - \beta}{1- \beta}\right)^{\frac{1}{\alpha}}\;, \\
\gamma_0 &= \frac{1+\alpha-\beta}{\alpha} \left(\frac{1+\alpha-\beta}{1-\beta}\right)^{\frac{1-\beta}{\alpha}}\;,
\end{align}
which has positive solutions for $0 \leq \alpha, \beta < 1$.
In particular, for $\alpha=\beta=\frac13$ we have $\gamma_0=\frac{27}{4}$ and $z_0=\frac{27}{8}$, and for $\alpha=\frac13$ and $\beta=\frac23$ we have $\gamma_0=4$ and $z_0=8$.
\end{subequations}
Closely following the derviation in \Eqsref{eq:theory_phiztau} to \eqref{eq:theory_scaling_int_pzy0}, we can derive the relative droplet size distribution (\Figref{fig:Figure_theory_distributions}C),
\begin{subequations}
\begin{align}
p{(z, \gamma_0)} = \left[\int_0^\infty \phi{(z, \gamma_0)} \diff z\right]^{-1} \phi(z, \gamma_0) &= \left[\chi_0 e^{-\tau'}\right]^{-1} \phi(z, \gamma_0) =\begin{cases}\frac{-e^{-\psi}}{v(z, \gamma_0)} & z\leq z_0, \\ 0 & z>z_0 \end{cases} \label{eq:theory2_pzgamma0}  \;,\\
    \textrm{with } \int_0^\infty p{(z, \gamma_0)} \diff z &= 1\;, \label{eq:theory2_scaling_int_pzy0}
\end{align}
\end{subequations}
and the number of remaining droplets,
\begin{equation}
    N{(\tau T_0)} = \frac{\chi_0}{x(\tau)}\;.
\end{equation}
Analogously, applying \Eqref{eq:theory2_scaling_gamma} and assuming a power law relation in the asymptotic limit $x(\tau) = x_0 {\tau}^{\nu}$, we get
\begin{subequations}
\begin{align}
    \gamma{(\tau)} \rightarrow \gamma_0 & = x(\tau)^{(\beta-\alpha)} \left[\dif{x}{\tau} \right]^{-1}\;,\\
    \Rightarrow \nu &= \frac{1}{1+\alpha - \beta}, \\ 
    \Rightarrow x(\tau) &= \left(\frac{1+\alpha- \beta}{\gamma_0}\right)^{\nu} {\tau}^{\nu}\;.
\end{align}
\end{subequations}

Consequently, the number of droplets per unit volume in the asymptotic limits scales as
\begin{subequations}
\begin{align}
    N(t) &= \frac{\chi_0}{x(t/T_0)} = \frac{\chi_0}{x_0} \left(\frac{t}{T_0}\right)^{\frac{-1}{1+\alpha - \beta}}\;,\\
     &= \kappa \left[ \int_0^{z_0} \frac{-e^{-\psi} z}{v{(z, \gamma_0)}} \diff z \right]^{-1} \left(\frac{\gamma_0}{1+\alpha- \beta}\right)^{\frac{1}{1+\alpha - \beta}} V_{\mathrm{eq}, 0} \left[\Lambda t\right]^{-\frac{1}{1+\alpha-\beta}}\;, \\ 
     &= K_1(\alpha, \beta)\, V_\mathrm{tot} \left[\Lambda t\right]^{-\frac{1}{1+\alpha-\beta}}\;.
\end{align}
where the pre-factor (\Figref{fig:Figure_theory_distributions}A)
\begin{equation}
    K_1(\alpha, \beta) = \left[ \int_0^{z_0} \frac{-e^{-\psi} z}{v{(z, \gamma_0)}} \diff z \right]^{-1} \left(\frac{\gamma_0}{1+\alpha- \beta}\right)^{\frac{1}{1+\alpha - \beta}} \label{eq:theory2_K1}\;,
\end{equation}
collects all constants that only depend on the exponents $\alpha$ and $\beta$.
\end{subequations}

The equilibrium droplet volume and radius in the asymptotic limit scale as
\begin{subequations}
\begin{align}
    V_\mathrm{eq}(t) &= \frac{V_\mathrm{tot}}{k^\infty N(t)}= \frac{V_\mathrm{tot}}{k^\infty K_1(\alpha, \beta)} \left(\Lambda t\right)^{\frac{1}{1+\alpha-\beta}}\;,\\
    R_\mathrm{eq}(t) &\propto \left(\Lambda t\right)^{\frac{1}{3(1+\alpha-\beta})}\;,
\end{align}
where (compare \Eqsref{eq:theory_barVt} and \eqref{eq:theory_kinfty})
\begin{align}
    k^\infty = \frac{\bar{V}(t)}{V_\mathrm{eq}(t)} = \bar{z} = \int_0^{z_0} p{(z, \gamma_0)}z \diff z \label{eq:theory2_Kinfty} \;.
\end{align}
\end{subequations}
Consistent with literature, this reproduces the scalings for special cases:
For classical diffusion-limited coarsening, $\alpha=\frac13$ and $\beta = \frac13$, this recovers $R\propto t^{1/3}$~\cite{Lifshitz1961,Wagner:1961aa}, whereas for conversion-limited coarsening, $\alpha=\frac13$ and $\beta=\frac23$, it recovers $R\propto t^{1/2}$~\cite{Wagner:1961aa}.

\subsection{Theoretical approximations of CO assurance}
\label{sub:theoretical_approximations_of_co_assurance}
In this section we quantify the reduction of CO assurance in mutants with abolished SC.
To derive an analytic estimate of CO assurance, we focus on a single chromosome with average droplet count $\meanN_i$ in a cell with $N$ droplets in total, which are randomly allocated to multiple chromosomes.
The probability that this chromosome has at least one droplet, i.e. the CO assurance, then reads
\begin{subequations}
\begin{equation}
    P(N_i\ge1) = 1 - \left(1- \frac{\meanN_i}{N}\right)^N
    \stackrel{\meanN_i\ll N}{\approx} 1 - e^{-\meanN_i}
    \;.\label{eq:P_assurance_single_appendix}
\end{equation}
\end{subequations}
since the allocation of individual droplets to chromosomes is independent for abolished SC, and 
the CO count distribution on the considered chromosome follows a Poisson distribution in the limit of small $\bar{N}_i/N$.
The estimation in this limit does not depend on $N$, and thus also holds for the generalized case of a non-constant, average total average CO count in the cell.
However, the precise result depends on the distribution of the CO count in the full cell.

We next consider the probability that all chromosomes in the cell have at least one COs, i.e., the CO assurance of the cell.
For simplicity, we consider a cell with exactly $N$ droplets and $n$ chromosomes of equal average CO count $N/n$.
An equivalent problem is known as the occupancy problem~\cite{hald1984moivre,moivre1711mensura}, which asks for the probability that a certain number $k$ of $n$ boxes have at least one ball, if $N$ balls are randomly distribited into these boxes.
The general solution to the occupancy problem yields
\begin{subequations}
\begin{equation}
    P(\min{(N_i)}\geq 1) = \sum_{i=0}^n (-1)^{i} \binom{n}{i} \left(1-\frac{i}{n}\right)^N\;.
\end{equation}
Exemplarily, this simplifies to
\begin{equation}
    P(\min{(N_i)}\geq 1) \approx 1- \frac{1}{2^{N-1}}\;.
\end{equation}
for $n=2$ chromosomes, and for $n=5$ chromosomes and $N=10$ droplets, this results in a CO assurance of $P(\min{(N_i)}\geq 1) \approx 0.52$.
\end{subequations}

We provide a simpler estimate of this probability based on the solution of the classical coupon collector's problem~\cite{erdHos1961classical}, which solves the problem of how many independent draws $N$ one needs to collect all of $n$ different coupons (with equal probability) at least once.
This is equivalent to our problem and allows us to estimate the number of necessary COs $N$, so that each of the $n$ chromosomes have at least one CO,
\begin{subequations}
\begin{equation}
    N \approx n \log{n} - \log{\log{\frac{1}{P}}}\;.
\end{equation}
Consequently, we get the estimated probability of cells with CO assurance,
\begin{equation}
    P(\min{(N_i)}\geq 1) \approx \exp{\left(-n \exp{-N/n}\right)}\;.
\end{equation}
To generalize this solution to the case that individual chromosome do not have equal average CO count $\meanN_i$, we can use the solution to the generalized coupon collector's problem with unequal probabilities $p_i$~\cite{FLAJOLET1992207}.
In particular, the number of expected draws to have each coupon at least once is described by
\begin{align}
    E(T, \{p_1, \ldots, p_n\}) &= \int_0^\infty \left(1- \prod_{i=1}^n \left(1-e^{-p_i t}\right) \right) \diff t &\geq \int_0^\infty \left(1- \prod_{i=1}^n \left(1-e^{-p t}\right) \right) \diff t = E(T, \{p, \ldots, p\})\;,
\end{align}
\end{subequations}
which by inspection is larger than the expected number of draws for equal probabilities.
This strongly suggests that the estimated probability of cells with CO assurance for a constant total CO count $N$ is lower for chromosomes with non-equal average CO counts $\meanN_i$.

\subsection{Interference length for non-Poissonian total CO count distribution}
\label{ssec:interference_length}

In our recent publication~\cite{Ernst:2024aa}, we introduced the interference length $\Lint$ and the normalized interference length $\LintNorm$ to quantify CO interference.
To do this, we assumed a null hypothesis (absent CO interference) where COs were independently designated and placed on chromosomes.
In this null hypothesis, the CO count~$N$ per chromosome follows a Poisson distribution with the observed average CO count $\meanN$ on the respective chromosome.
In contrast, systems that follow coarsening dynamics exhibit a distribution $P_\mathrm{cell}(N)$ of the total CO count that is typically narrow and does not follow a Poisson distribution.
This is even the case if the allocation of COs to chromosomes and their spatial positions is completely random.
Since these distributions deviate from the null hypothesis, $\Lint$ would still report positive values.
To estimate this value, we next consider two scenarios; (i) exchange-limited coarsening along the SC without nucleoplasmic exchange, and (ii) purely nucleoplasmic exchange between multiple SC in a cell.

In the first case of exchange-limited coarsening along the SC without nucleoplasmic exchange,  chromosomes are independent, and the CO count distribution on chromosomes, $P(N_i)$, is typically relatively narrow.
For simplicity, we assume a sharp CO count distribution with $N_i$ COs per chromosome, which gives us also the expected ratio $\phi$~\cite{Ernst:2024aa},
\begin{subequations}
\begin{align}
    \phi = \frac{\bar{N}^\mathrm{pair}_\mathrm{obs}}{\bar{N}^\mathrm{pair}_\mathrm{noInt}} &= \frac{N_i (N_i-1)/2}{N_i^2/2} = 1-\frac{1}{N_i}\;.
\end{align}
For random placement of these COs, i.e., $d_\mathrm{obs} = d_\mathrm{noInt}$~\cite{Ernst:2024aa}, the residual interference length reads
\begin{align}
 \Lint &= \frac{1}{N_i} \left(L - d_\mathrm{noInt}\right)\;,\\
 \LintNorm &= \left(1 - \frac{d_\mathrm{noInt}}{L}\right)\approx \frac13\;,
\end{align}
\end{subequations}
for uniform CO distributions.

In the second case of purely nucleoplasmic exchange between multiple chromosomes and total CO count distribution $P_\text{cell}(N)$, the
CO count distribution $P(N_i)$ on a chromosome with average CO count $\bar{N}_i$ is given by a weighted sum of Binomial distributions,
\begin{subequations}
\begin{equation}
P(N_i) = \sum_{N=0}^\infty \binom{N}{N_i} \left(\frac{\bar{N}_i}{N}\right)^{k} \left(1 -\frac{\bar{N}_i}{N}\right)^{N-N_i} P_\text{cell}(N)\;.
\end{equation}
If cells have always the same total CO count $N$, this simplifies to
\begin{equation}
    P(N_i) = \binom{N}{N_i} \left(\frac{\bar{N}_i}{N}\right)^{N_i} \left(\frac{N-\bar{N}_i}{N}\right)^{N-N_i}\;,
\end{equation}
for a constant total CO count $N$. The distribution converges to a Poisson distribution with mean $\bar{N}_i$ for large total CO count, i.e., $N\rightarrow \infty$, yielding us also the expected ratio $\phi$~\cite{Ernst:2024aa},
\begin{align}
    \bar{N}^\mathrm{pair}_\mathrm{obs} &= \sum_{N_i=0}^\infty P(N_i) \frac{N_i(N_i-1)}{2} = \left(\frac{\bar{N}_i}{N}\right)^2 \frac{N(N-1)}{2}\;,\\ 
    \Rightarrow \phi &= \frac{\bar{N}^\mathrm{pair}_\mathrm{obs}}{\bar{N}^\mathrm{pair}_\mathrm{noInt}} = \frac{2 \bar{N}^\mathrm{pair}_\mathrm{obs}}{\bar{N}_i^2} = 1-\frac{1}{N}\;.
\end{align}
Consequently for random placement of COs, i.e. $d_\mathrm{obs} = d_\mathrm{noInt}$~\cite{Ernst:2024aa}, we get
\begin{align}
 \Lint &= \frac{1}{N} \left(L - d_\mathrm{noInt}\right)\;,\\
 \LintNorm &= \frac{\bar{N}_i}{N} \left(1 - \frac{d_\mathrm{noInt}}{L}\right)\propto \frac{1}{N}\;,
\end{align}
\end{subequations}
which scales inversely with the \emph{total} droplet count $N$ in the cell.

Taken together, $\Lint$ and $\LintNorm$ yield a residual CO interference in case of non-Poissonian total CO count distribution, despite random spatial placement and allocation of COs toward chromosomes.

\section{Detailed analysis of the model of the main text}

The following sections contain detailed analysis of the coarsening model presented in the main text.

\subsection{Non-dimensionalization}
We simulate the system in non-dimensional units, choosing appropritate length and time scales.
Since we discuss various limits of the full model, we cannot use one non-dimensionalization for all cases, and instead introduce two in the following, depening on whether the synaptonemal complex is present or not.

\subsubsection{Non-dimensionalization of a system with SC}
\label{ssec:non_dim}
To set-up a non-dimensional version of the model described in the main text (\secref{sub:thermodynamically_consistent_coarsening_model}), we first need to define length scale, time scale, energy scale, and HEI10 concentration scale.
For the latter, we use the constant concentration $\cD$ in HEI10 droplets.
We use the effective cross-section $a^2$ of the SC to define a natural length scale $a$.
This defines the non-dimensional position $\xi$ along the SC and SC lengths $\ell$ as 
\begin{subequations}
\begin{align}
\xi &= \frac{x}{a}\;,\\ 
\ell &= \frac{L}{a} \;.
\end{align}
\end{subequations}
The length scales also defines a natural volume $a^3$, which is comparable to the size of real HEI10 droplets since their radius is comparable to the SC width $a$.
We normalize  time $t$ using the time scale of diffusion across the SC,
\begin{equation}
    T=\frac{{a}^2}{D}\;, \label{eq:non_dim_time_scale}
\end{equation}
defining $\tau=\frac{t}{T}$.
Consequently, the normalized state variables are given as
\begin{subequations}
\begin{align}
\mD_{j,i} &= \frac{V_{j,i}}{a^3}\;, \\ 
\mS_j &= \frac{c_j a^2 a}{\cD a^3} = \frac{c_j}{\cD}\;, \\ 
\mN &=\frac{\cN \VN}{\cD a^3}\;,
\end{align}
\end{subequations}
for the individual droplet sizes $\mD_{j,i}$, the HEI10 concentration $\mS_j$ along the $j$-th SC arc, and the amount of HEI10 in the nucleoplasm $\mN$.
We now define the coefficients that are associated with  differences in affinity of the compartments,
\begin{subequations}
\begin{align}
\phiD &= \frac{\gamma_0}{\gammaS}\;, \\
\phiN &= \frac{1}{\gammaS}\frac{a^3}{\VN}\;,
\end{align}
\end{subequations}
as well as the non-dimensional exchange rates,
\begin{subequations}
\begin{align}
\gammaDS &= \frac{\gammaS \LambdaDS }{D a}\;, \\
\gammaDN &= \frac{\gammaS \LambdaDN }{D a}\;, \\
\gammaSN &= \frac{\gammaS \LambdaSN }{D a}\;.
\end{align}
\end{subequations}
This yields the following non-dimensional exchange rates
\begin{subequations}
\begin{align}
    \mathcal{S}^\mathrm{DS}_{j, i} &= \gammaDS \left(\mS(\xi_{j,i}) - \phiD\left[\mD_{j,i}\right]^{-\nu}\right) \left[\mD_{j,i}\right]^{\nuS}\;, \\  
    \mathcal{S}^\mathrm{DN}_{j,i} &= \gammaDN \left(\phiN \mN - \phiD\left[\mD_{j,i}\right]^{-\nu}\right) \left[\mD_{j,i}\right]^{\nuN}\;,\\ 
    \mathcal{S}^\mathrm{SN}_{j}(\xi) &= \gammaSN \left(\phiN \mN -   \mS(\xi)\right)\;,
\end{align}
\end{subequations}
and the dynamical equations
\begin{subequations}
\begin{align}
     \dif{\mD_{j,i}(\tau)}{\tau} &= \mathcal{S}^{\mathrm{DN}}_{j,i}(\tau) + \mathcal{S}^{\mathrm{DS}}_{j,i}(\tau)\;, \label{eq:dMDdt_nd}  \\ 
\dif{\mS_{j}(\xi, \tau)}{\tau} &= \nabla_\xi^2 \mS_{j}(\xi, \tau) + \mathcal{S}^{\mathrm{SN}}_{j}(\xi, \tau) - \sum_{i=1}^{N_j}  \mathcal{S}^{\mathrm{DS}}_{j,i}(\tau) \delta{(\xi-\xi_{j,i})}\;, \label{eq:dlambdaSdt_nd}\\ 
    \dif{\mN(\tau)}{\tau} &= -\sum_{j=1}^{N_\mathrm{S}} \left[ \int_0^{\ell_j} \mathcal{S}^{\mathrm{SN}}_{j}(\xi, \tau) \diff \xi+   \sum_{i=1}^{N_j}  \mathcal{S}^{\mathrm{DN}}_{j,i}(\tau) \right] \label{eq:dMNdt_nd}\;.
\end{align}
\end{subequations}
Material conservation reads
\begin{align}
\frac{M}{\cD a^3} = m = \mN(\tau) + \sum_{j=1}^{N_\mathrm{S}} \left( \int_0^{\ell_j} \mS_j(\xi,\tau) \diff \xi +  \sum_{i=1}^{N_i} \mD_{j, i}(\tau)\right) \;.
    \label{eq:M_total_nondim}
\end{align}
Consequently, the dynamics are determined by the following non-dimensional parameters: the total amount of HEI10~$m$, the sensitivity exponents $\nu$, $\nuN$ and $\nuS$, the ratios $\phiD$ and $\phiN$, and the exchange rates $\gammaDS$, $\gammaDN$ and $\gammaSN$.
The state of the system is defined by the number of SCs $\NS$, the number of initial droplets $N_j$, the SC lengths $\ell_j$, the initial droplet positions $\xi_{j,i}$, as well as the initial state variables $\mN$, $\mS_j$ and $\mD_{j, i}$.

If exchange between different compartments is permitted, all chemical potentials are balanced in thermodynamic equilibrium. Consequently, the concentrations in the compartments will equilibrate,
\begin{equation}
    \label{eqn:equilibrium_nondim}
    \phiN m^{\mathrm{N}, *}
    = m^{\mathrm{S}, *}_j
    = \phiD {m^{\mathrm{D}, *}}^{-\nu}
    \;.
\end{equation}
After initial loading of SCs without considering droplets, the SCs are homogeneous with concentration
\begin{equation}
    m^{\mathrm{S}, *}_j = \frac{m}{\ell + 1/ \phiN}
    \;,
\end{equation}
which follows from combining \Eqsref{eqn:equilibrium_nondim} and \eqref{eq:M_total_nondim}.
Consequently, SCs receive more HEI10 per unit length if the overall amount $m$ is higher, if the partition coefficient $\phiN$ is larger, and if the normalized SC lengths are smaller.

\subsubsection{Non-dimensionalization for system without a SC}
\label{ssec:non_dim_dn}
In the scenario where the system does not have an SC, we can simplify the equations by only considering the exchange rate~${S}^\mathrm{DN}$.
We thus choose another non-dimensionalization, which helps us to reduce the number of non-dimensional parameters by choosing the time-scale as
\begin{equation}
    T_\mathrm{noSC}=\frac{\VN}{\LambdaDN}\;,\label{eq:non_dim_time_scale_zyp}
\end{equation}
while using the same length, energy, and concentration scale.
We define the same normalized state variables 
\begin{subequations}
\begin{align}
\mD_{i} &= \frac{V_{i}}{a^3}\;, \\ 
\mN &=\frac{c_\mathrm{N} \VN}{\cD a^3}\;,
\end{align}
and now choose
\begin{equation}
    \phiD_\mathrm{noSC} = \gamma_0\frac{\VN}{a^3} = \frac{\phiD}{\phiN}\;,
\end{equation}
\end{subequations}
eliminating a non-dimensional variable to get the simplified dynamical equations
\begin{subequations}
\begin{align}
    \dif{\mD_{i}}{\tau} &= \left( \mN - \phiD_\mathrm{noSC} \left[\mD_{i}\right]^{-\nu}\right) \left[\mD_{i}\right]^{\nuN}\;, \label{eq:dMDdt_nd2} \\
    m &= \mN(\tau) + \sum_{i=1}^{N}  \mD_{i}(\tau)= \mN(\tau) + N(\tau)  \overline{\mD}(\tau)\;, \label{eq:non_dim_DN_conservation}\\
    \Rightarrow \dif{\mN}{\tau} &= - \sum_{i=1}^{N^\mathrm{D}} \dif{\mD_{i}}{\tau}\;, \label{eq:dMNdt_nd2} 
\end{align}
\label{eq:non_dim_DN_system}
\end{subequations}
where $N^\mathrm{D}$ is the respective number of droplets in the system, and $m$ is the total amount of HEI10.

The two respective time scales in \Eqsref{eq:non_dim_time_scale} and \eqref{eq:non_dim_time_scale_zyp} yield a ratio of
\begin{equation}
    \phi_\mathrm{full/noSC} = \frac{T}{T_\mathrm{noSC}} = \frac{\VN D}{a^2 \LambdaDN}\;.
\end{equation}

\subsection{Relaxation dynamics of HEI10 concentration on a SC in nucleoplasm without droplets}
\label{app:relaxation_no_droplets}
We first analyze our model without droplets and for simplicity consider one SC with length $L = \ell a$ to investigate the time scale of HEI10 loading onto the SC and the relaxation dynamics of non-homogeneous concentration profiles along the SC which are also affected by the exchanges with the nucleoplasm.
The system can be described by the following non-dimensional, dynamical equation and conservation law
\begin{subequations}
\begin{align}
    \dif{\mS(\xi, \tau)}{\tau} &= \nabla_\xi^2 \mS(\xi, \tau) + \gammaSN \left(\phiN \mN(\tau) -  \mS(\xi, \tau)\right)\;, \\ 
    m &=  \mN(\tau) +  \int_{0}^{\ell} \mS(\xi, \tau) \diff\xi\;.
\end{align}
\end{subequations}
Applying the calculation from \secref{ssec:theory_relaxation}, we get the non-dimensional relaxation times $\tau^\mathrm{SN}(q)$ as
\begin{subequations}
\begin{equation}
    \tau^\mathrm{SN}(q) = \begin{cases}
      \left[\gammaSN \left(1  + \ell \phiN\right)\right]^{-1} & \text{if $q=0$}\;,\\
      \left[\gammaSN +  \left(\frac{\pi q}{\ell}\right)^2\right]^{-1} & \text{if $q\in\mathbb{Z}\backslash\{0\}$}\;,
    \end{cases} 
\end{equation}
yielding a relaxatation time for a pertubation with wavelength $\tilde{\lambda} = 2\ell/q$ ($q\neq 0$) as
\begin{equation}
    \tau^\mathrm{SN}(\tilde{\lambda}) = \left[\gammaSN +  \left(\frac{2 \pi}{\tilde{\lambda}}\right)^2\right]^{-1}\;,
\end{equation}
and an effective diffusion constant $\mathcal{D}_\mathrm{q}$ of 
\begin{equation}
    \mathcal{D}_\mathrm{q} = 1 + \gammaSN \left(\frac{\ell}{\pi q}\right)^2\;.
\end{equation}
\end{subequations}
\begin{subequations}
This yields the dimensional relations for $T^\mathrm{SN}(q)$
\begin{equation}
     T^\mathrm{SN}(q) = 
    \begin{cases}
      \frac{a^3}{\gammaS \LambdaSN}\left(1  + \frac{L a^2}{\gammaS \VN}\right)^{-1} & \text{if $q=0$}\;,\\
      \left( \frac{\gammaS \LambdaSN}{a^3} + D\left(\frac{\pi q}{L}\right)^2 \right)^{-1} & \text{if $q\in\mathbb{Z}\backslash\{0\}$}\;,
    \end{cases} 
\end{equation}
which corresponds to the loading rate of HEI10 onto the SC,
\begin{equation}
     k_\mathrm{load} = \frac{1}{T^\mathrm{SN}(0)} =  \frac{\gammaS \LambdaSN}{a^3} +  \frac{\LambdaSN L}{\VN a}
     =  \frac{\LambdaSN}{a^3} \left(\gammaS +  \frac{V_\mathrm{S}}{\VN}\right)\;,\label{eq:k_load}
\end{equation}
where $V_\mathrm{S}=L a^2$ is the effective volume of the SC.
The relaxation rate of the $q$-th mode with wavelength $\lambda = 2L/q$ is
\begin{align}
    k^\mathrm{SN}(q) &=  \frac{\gammaS \LambdaSN}{a^3} + D\left(\frac{\pi q}{L}\right)^2\;,\\ 
    k^\mathrm{SN}(\lambda) &= \frac{\gammaS \LambdaSN}{a^3} + D \left(\frac{2\pi}{\lambda}\right)^2 \;.
\end{align}
The effective diffusion constant $D_\mathrm{q}$ reads
\begin{equation}
    D_\mathrm{q} = D + D \frac{\gammaS \LambdaSN}{D a} \left(\frac{L}{\pi a q}\right)^2
     = D +   \frac{\gammaS \LambdaSN }{a^3}\ \left(\frac{L}{\pi  q}\right)^2\;. 
\end{equation}
\end{subequations}
\subsection{Investigation of a coarsening of droplets on a SC without nucleoplasm}
\label{ssec:no_nucleoplasm}
In this section we consider a system of one SC with droplets that does not exchange HEI10 with the surrounding nucleoplasm, using the non-dimensionalisation description (\secref{ssec:non_dim}).

\subsubsection{SC profile in diffusion-limited case}
\label{ssec:sc_profile_ds}

In this section, we derive the concentration profile on a single SC with multiple droplets in the diffusion-limited case without nucleoplasmic exchange.
In particular, we consider a SC of non-dimensional length $\ell$ with droplets of sizes $\mD_{i}(\tau)$ at positions $\xi_{i}$.
In order to calculate the profile $\mS(\xi, \tau)$, we assume that changes of the SC concentration profile are slow (\secref{ssec:theory_concentration_profile}).
Thus, for $\xi \neq \xi_{i}$ \Eqref{eq:dlambdaSdt_nd} implies
\begin{subequations}
\begin{equation}
    {\nabla_\xi}^2 \mS(\xi, \tau) = 0\;.
\end{equation}
This results in piecewise linear profiles $\mS(\xi, \tau)$ between droplets.
Conversely, for $\xi = \xi_{i}$ for $i=1,\ldots, N$, we get
\begin{align}
    \int_{\xi_{i}-\epsilon}^{\xi_{i}+\epsilon} \left[{\nabla_\xi}^2 \mS(\xi, \tau) - \dif{\mD_{i}(\tau)}{\tau} \delta{\left(\xi-\xi_i\right)}\right]\diff \xi = 0\;, \label{eq:sc_profile_integral_droplet}
\end{align}
\end{subequations}
for a infinitesimal $\epsilon >0$. Thus,
\begin{subequations}
\begin{equation}
    \left.{\nabla_\xi} \mS(\xi, \tau)\right|_{\xi_{i}-\epsilon}^{\xi_{i}+\epsilon} = \dif{\mD_{i}(\tau)}{\tau}\;.
\end{equation}
For simplicity, we consider a droplet that is not subject to boundary effects at the end of the SC,
\begin{align}
    \frac{\mS(\xi_{i+1}, \tau) - \mS(\xi_{i}, \tau)}{\xi_{i+1} - \xi_{i}} - \frac{\mS(\xi_{i-1}, \tau) - \mS(\xi_{i}, \tau)}{\xi_{i-1} - \xi_{i}} = \gammaDS \left( \mS(\xi_{i}, \tau) - \phiD \left[\mD_{i}(\tau)\right]^{-\nu}\right) \left[\mD_{i}(\tau)\right]^{\nuS}\;,\label{eq:sc_profile_ds}
\end{align}
\end{subequations}
which can be solved to compute $\mS(\xi_{i}, \tau)$.
This equation describes that the flux of material between droplets and SC is balanced by diffusion of material along the SC towards neighbouring droplets.
In the case of a boundary droplet $i_0$, i.e., one that is not surrounded by two adjacent droplets, we consider a no-flux boundary condition on the SC.
This implies that the profile of $\mS(\xi, \tau)$ is constant between the droplet and the boundary, and is given by $\mS(\xi_{i_0}, \tau)$.
Assuming that the exchange between droplets and SC is fast ($\gammaDS\rightarrow \infty$) and that the dynamics are diffusion-limited, this corresponds the the solution from \secref{ssec:theory_concentration_profile} for the case of negligble bulk-exchange ($\kappa=0$).

\subsubsection{Asymptotic scaling for diffusion-limited coarsening}
\label{ssec:scaling_ds}
In order to study the simplest case of droplet coarsening, we focus on the dynamics of droplets on a single SC, neglecting the nucleoplasm.
The asymptotic scaling of coarsening depends on the ratio of the exchange rate, $\LambdaDS$, and the diffusion, $D$, quantified by the non-dimensional ratio, $\kappa = \LambdaDS/(a D) = \gammaDS/\gammaS$ (We use the non-dimensional quantity $\kappa$ in the main text for simplicity.)
The aim of this section is to derive the coarsening behaviour in the limit of fast exchange between  SC and droplets compared to the diffusion of HEI10 along the SC ($\gammaDS,\kappa \rightarrow \infty$), i.e., the diffusion-limited scenario.
Following the approach of Lifshitz, Slyozov, and Wagner (\secref{ssec:theory_scaling_1D}), we estimate the asymptotic scaling of droplet sizes $V(t)$ and the number of remaining droplets $N(t)=n(t/T)$, as well as the universal size distribution. 
This approach assumes an infinite system, but we nevertheless apply it to our system, assuming that the SC initially has sufficiently many droplets, to estimate the scalings.
To validate these scalings, we later compare these to numerical simulation.

We assume that we have a sufficiently long SC of length $L=\ell a$ with periodic boundary conditions. For simplicity, we conisder a uniform distribution of initial droplets along the SC.
Analogously to \secref{ssec:theory_scaling_1D}, we assume that changes to the SC concentration profile are slow, implying that it is piecewise linear between droplets.
Since we assume fast exchange between droplets and SC, the concentration $\mS(x)$ is equal to the equilibrium concentration of the droplets at their position (\secref{ssec:sc_profile_ds}),
\begin{equation}
    \mS(x_i) = m_\mathrm{eq, i} = \phiD \left[\mD_i\right]^{-\nu}\;.
\end{equation}
Consequently, we obtain the concentration profile by applying \Eqref{eq:sc_profile_ds} 
\begin{equation}
    \dif{\mD_{i}(\tau)}{\tau} = 2  \frac{\phiD \left[\mD_\mathrm{eq}(\tau)\right]^{-\nu} - \mS(\xi_{i}, \tau)}{\Delta \xi(\tau)} = \gammaDS \left( \mS(\xi_{i}, \tau) - \phiD \left[\mD_{i}(\tau)\right]^{-\nu}\right) \left[\mD_{i}(\tau)\right]^{\nuS}\;,
\end{equation}
which yields the concentration profile
\begin{subequations}
\begin{align}
\mS(\xi_{i}, \tau)  &= 
    \frac{2  \left[\mD_\mathrm{eq}(\tau)\right]^{-\nu}  + \Delta \xi(\tau)\gammaDS  \left[\mD_{i}(\tau)\right]^{\nuS-\nu}}{2+\Delta \xi(\tau) \gammaDS\left[\mD_{i}(\tau)\right]^{\nuS}} \phiD \;,\\
\Rightarrow \dif{\mD_{i}(\tau)}{\tau} &= \frac{2 \phiD}{\Delta \xi(\tau)}  \left(\left[\mD_\mathrm{eq}(\tau)\right]^{-\nu} - \frac{2  \left[\mD_\mathrm{eq}(\tau)\right]^{-\nu}  + \Delta \xi(\tau)\gammaDS  \left[\mD_{i}(\tau)\right]^{\nuS-\nu}}{2+\Delta \xi(\tau) \gammaDS\left[\mD_{i}(\tau)\right]^{\nuS}} \right)\;, \\ 
 &= \frac{2 \phiD}{\Delta \xi(\tau)}  \left(\left[\mD_\mathrm{eq}(\tau)\right]^{-\nu} - \left[\mD_i(\tau)\right]^{-\nu}\right) \frac{1}{1 + 2\left(\gammaDS \Delta \xi(\tau) \left[\mD_{i}(\tau)\right]^{\nuS}\right)^{-1}}\;,\\ 
&= \frac{2 \phiD N(\tau)}{\ell}  \left(\left[\mD_\mathrm{eq}(\tau)\right]^{-\nu} - \left[\mD_i(\tau)\right]^{-\nu}\right) \frac{1}{1 + 2\left(\gammaDS \Delta \xi(\tau) \left[\mD_{i}(\tau)\right]^{\nuS}\right)^{-1}}\;.
\end{align}
\end{subequations}
For large  exchange rates $\gammaDS, \kappa \rightarrow \infty$, the last term converges to $1$, since both $\Delta \xi(\tau)$ and $\left[\mD_{i}(\tau)\right]^{\nuS}$ (for $\nuS\leq 0$) do not decrease over time, and  we thus have
\begin{subequations}
\begin{align}
    \dif{\mD_{i}(\tau)}{\tau}  &= \frac{2 \phiD N(\tau)}{\ell}  \left(\left[\mD_\mathrm{eq}(\tau)\right]^{-\nu} - \left[\mD_i(\tau)\right]^{-\nu}\right) \;.\label{eq:ds_scaling_md_dtau}
\end{align}
Consequently, we can ignore the last factor in the asymptotic analysis.
We now consider the non-dimensional conservation law given by
\begin{equation}
    n(\tau) \overline{\mD}(\tau) + \ell \overline{\mS}(\tau) = m\;.\label{eq:ds_conservation_n}
\end{equation}
\end{subequations}
Inspecting these equations, and comparing with \Eqsref{theory:conservation_N} and \eqref{eq:theory_dV_i_dt}, allows us to use the result from \secref{ssec:theory_scaling_1D}.
Using $T_0 = k_\infty \ell \left[\mD_{\mathrm{eq}, 0}\right]^{2+\nu}/(2 m \phiD)$, we get
\begin{subequations}
\begin{align}
    n(\tau) &= \frac{\chi_0}{x_0} \left(\frac{\tau}{T}\right)^{-\frac{1}{2+\nu}}
    = K_0(\nu) \ell\left(\frac{m}{\ell}\right)^{\frac{1+\nu}{2+\nu}} \left(\phiD\tau \right)^{-\frac{1}{2+\nu}}\;,
\end{align}
where the pre-factor $K_0(\nu)$ and constant $k^\infty$ are defined in \Eqsref{eq:theory_K0} and \eqref{eq:theory_kinfty}. The universal distribution function is given in \Eqref{eq:theory_pzgamma0}. 
\end{subequations}
We now write the solution in dimensional variables, with $M = m \cD a^3$, to get
\begin{subequations}
\begin{equation}
    N(t) = K_0(\nu) \frac{L}{a} \left(\frac{M}{\cD a^2 L}\right)^{\frac{1+\nu}{2+\nu}} \left(\frac{a^2\gammaS}{D \gamma_0 t} \right)^{\frac{1}{2+\nu}}\;, \label{eq:scaling_diffusion_limited}
\end{equation}
while the droplet volume and radius scales as follows,
\begin{align}
    V_\mathrm{eq}(t) &= \frac{\bar{V}(t)}{k^\infty} = \frac{M}{k^\infty \cD N(t)}
 = \frac{1}{k^\infty K_0(\nu)}  a^3 \left(\frac{\gamma_0 D M t}{\cD \gammaS a^3 a L }\ \right)^{\frac{1}{2+\nu}}\;,\\
R^\mathrm{D}_\mathrm{eq}(t) &\propto a \left(\frac{\gamma_0 D M t}{\cD \gammaS a^3 a L } \right)^{\frac{1}{3(2+\nu)}}\;.
\end{align}
\end{subequations}
For $\nu=\frac13$ this results in slowed coarsening, $R\propto t^{1/7}$.
The universal size distribution in the asymptotic limit is given by \Eqref{eq:theory_pzgamma0} and shown in \Figref{fig:Figure_theory_distributions}B.

\subsubsection{Asymptotic scaling for exchange-limited coarsening}
\label{ssec:scaling_exchange_limited}
In this section, we consider exchange-limited coarsening, i.e., the case where exchange between droplets and SC is slow ($\kappa\rightarrow 0$), for a single SC of length $\ell$ and $N(t_0)=n(t_0/T)$ droplets at time $t_0$.
Consequently, diffusive transports acts on a faster time-scale and we can assume that concentrations along the entire SC are equilibrated.
In this limit the concentration profile $\mS(\xi) = \bar(\mS)$, which sets the size of the stationary droplet by $\mS = \phiD \left[\mD_\mathrm{eq}\right]^{-\nu}$.
The dynamics following from \Eqref{eq:dMDdt_nd} and conservation law \Eqref{eq:M_total_nondim} in non-dimensional units then read
\begin{subequations}
\begin{align}
    \dif{\mD_{i}(\tau)}{\tau} 
&= \gammaDS \phiD  \left(\left[\mD_\mathrm{eq}(\tau)\right]^{-\nu} - \left[\mD_i(\tau)\right]^{-\nu}\right) \left[\mD_{i}(\tau)\right]^{\nuS}\;,\\
m &= \ell \phiD \left[\mD_\mathrm{eq}(\tau)\right]^{-\nu} +  \sum_{i=1}^{N_i} \mD_{j, i}(\tau) = \ell \phiD \left[\mD_\mathrm{eq}(\tau)\right]^{-\nu} +  n(\tau) \overline{\mD}(\tau)\;.
\end{align} 
\end{subequations}
To determine the scaling relations, we can use the results from \secref{ssec:theory_scaling_generalized}.
Inspecting \Eqref{eq:theory2_dynamics_equation} 
yields the number of droplets per unit volume, and non-dimensional droplet size in the asymptotic limit,
\begin{subequations}
\begin{align}
    N(\tau) &= K_1(\nu, \nuS)\, m \left[\gammaDS \phiD \tau\right]^{-\frac{1}{1+\nu-\nuS}}\;,\\
    \mD_\mathrm{eq}(\tau) &= \frac{m}{k^\infty N(t)}= \frac{1}{k^\infty K_1(\nu, \nuS)} \left(\gammaDS \phiD \tau\right)^{\frac{1}{1+\nu-\nuS}}\;,
\end{align}
where the pre-factors $K_1(\nu, \nuS)$ and $k^\infty$ are defined in \Eqsref{eq:theory2_K1} and \eqref{eq:theory2_Kinfty}.
In dimensional units,
\begin{align}
    N(t) &= K_1(\nu, \nuS)\, \frac{M}{\cD a^3} \left[\frac{1}{\gamma_0} \frac{a^3}{\LambdaDS t}\right]^{\frac{1}{1+\nu-\nuS}}\;,\\
    V_\mathrm{eq}(t) &= \frac{M}{k^\infty N(t)}= \frac{\cD a^3}{k^\infty K_1(\nu, \nuS)} \left[\gamma_0 \frac{\LambdaDS t}{a^3}\right]^{\frac{1}{1+\nu-\nuS}}\;,\\
    R_\mathrm{eq}(t) &\propto \left[\gamma_0 \frac{\LambdaDS t}{a^3}\right]^{\frac{1}{3(1+\nu-\nuS})}\;.
\end{align}
\end{subequations}
The universal distribution function is given in \Eqref{eq:theory2_pzgamma0} and shown in \Figref{fig:Figure_theory_distributions}C.

\subsubsection{Model parameters based on previous coarsening model}
\label{ssec:parameter_ds}
In this section, we estimate key model parameters based on our earlier model for \Athaliana{} described in~\cite{Durand2022}.
In that work, we used a Hill function to describe the equilibrium concentration outside droplets, $\cD^\mathrm{eq} \propto V_i/[1 + (V_i/a^3)^{1+\alpha}]$ with $\alpha=\frac14$, which yields a stable droplet size below the critical radius. For large enough droplets the chosen function relates to an exponents of $\nu=\frac14$.
We do not consider a Hill function in the present paper, and insteaed focus on the dynamics of larger droplets, assume the power-law expression given by \Eqref{eqn:droplet_affinity}.
The size sensitivity of the exchange rates were also chosen as $\nuS=0$.
In this publication, we choose $\nu=\frac13$ and $\nuS=0$ and assume that small droplets simply vanish.
We measure length scales in terms of the SC diameter, $a\approx \SI{100}{\nano \metre}$~\cite{Capilla-Perez2021}, so that the SC lengths, which are approximateley $20$ to $\SI{50}{\micro \metre}$, correspond to $\ell\approx 200\ldots 500$ in non-dimensional units.
Estimating the diffusion constant as $D\approx \SI{1}{\micro\meter\squared\per\second}$, based on the measured diffusion of ZHP3/4 along the SC in \Celegans{}~\cite{Stauffer:2019aa}, we estimate the time-scale as
\begin{subequations}
\begin{equation}
    T=\frac{{a}^2}{D} \approx 10^{-2} \si{s}\;.
\end{equation}
In \refcite{Durand2022}, we assumed that the total coarsening time in wild type, $T_\mathrm{wt}=T_\mathrm{pach}\approx \SI{10}{h}$, corresponds to the duration of pachytene.
In non-dimensional units, $\tau_\mathrm{wt} = \SI{10}{h}/T \approx 4\cdot 10^6$.

Comparing Eq.~2 in \refcite{Durand2022} (and Eq.~2 in Fig.~1 in \refcite{Morgan2021a}) with \Eqref{eq:dMDdt_nd}, we get
\begin{equation}
    \Lambda=\SI{2.1}{\micro\meter\per\second} = \frac{\gammaDS a}{T}\;,
\end{equation}
and thus $\gammaDS \approx 0.2$.
Inspecting these equations yields
\begin{equation}
    \phiD = \frac{c^\mathrm{eq}_0 a}{M_0} \approx \frac{\SI{1.35}{au\per\micro\meter} \cdot \SI{0.1}{\micro\meter}}{\SI{1000}{au}} \approx 10^{-4}\;,
\end{equation}
\end{subequations}
implying that large droplets have a size of approximateley $\SI{1000}{au}$ in \refcite{Durand2022}.

\subsubsection{Numerical simulations}
\label{ssec:DS_numerical_investigations}
\begin{figure*}[t]
    \centering
\subfloat{
\raisebox{100pt}[0pt]{\makebox[-5pt][l]{\textsf{\textbf{A}}}}
\includegraphics[width=0.25\textwidth]{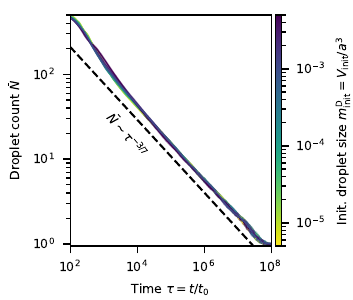}}
\subfloat{
\raisebox{100pt}[0pt]{\makebox[-5pt][l]{\textsf{\textbf{B}}}}
\includegraphics[width=0.25\textwidth]{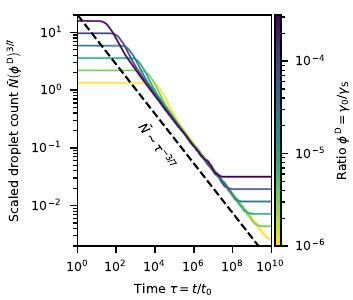}}
\subfloat{
\raisebox{100pt}[0pt]{\makebox[-5pt][l]{\textsf{\textbf{C}}}}
\includegraphics[width=0.25\textwidth]{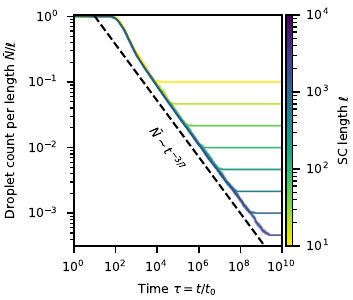}}
\subfloat{
\raisebox{100pt}[0pt]{\makebox[-5pt][l]{\textsf{\textbf{D}}}}
\includegraphics[width=0.25\textwidth]{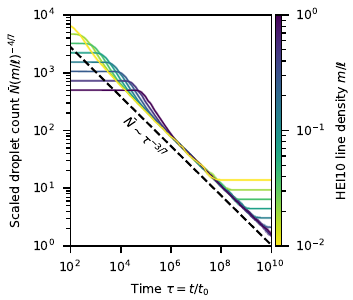}}
\caption{
\textbf{Numerical investigation of the coarsening model without nucleoplasmic exchange.}
Shown is the the (rescaled) average number of droplets as a function of time $\tau = t/t_0$.
(A) Time development of the average number of droplets $\meanN$ for varying initial droplet sizes $\mD_\mathrm{init}=V_\mathrm{init}/a^3$.
(B) Time development of the rescaled average number of droplets $\meanN \left(\phiD\right)^{3/7}$  for varying ratios $\phiD=\gamma_0/\gammaS$.
(C) Time development of the rescaled average number of droplets $\meanN/\ell$ for varying SC lengths $\ell=L/a$.
(D) Time development of the rescaled average number of droplets $\meanN \left(m/\ell\right)^{-4/7}$ for varying HEI10 line density $m/\ell=M/(\cD L a^2)$.
(A-C) The overall HEI10 line density $m/\ell = M/(\cD L a^2)$ per unit length along the SC is kept constant.
(B-D) The re-scaling is based on the scaling relation \Eqref{eq:scaling_diffusion_limited}.
(A-D) The dashed line gives the theoretical asymptotic scaling for the diffusion-limited scenario with $\bar{N}\sim \tau^{-\frac{1}{2+\nu}} \hat{=} \tau^{-3/7}$.
Additional parameters from \Figref{fig:Figure2}.}
\label{fig:Figure_numerics_without_nucleoplasm}
\end{figure*}

We here present some additional numerical simulations.
\Figref{fig:Figure_numerics_without_nucleoplasm} shows the dynamics of the (rescaled) average number of droplets as a function of multiple parameters, such as the SC length $\ell$, the ratio $\phiD$, the initial droplet size $\mD_\mathrm{init}$, or the total HEI10 line density.
The curves collapse in the asymptotic scaling regime if we rescale the droplet count $\meanN$ according to the scaling relation~\eqref{eq:scaling_diffusion_limited}.
This also underlines the interpretation in the main text (\secref{sec:droplet_patterning_only_SC}), that details of the initial conditions do not affect the asymptotic behavior, when accounting for the scaling relation.
Meanwhile the scaling with time is also consistent with the predicted scaling of $\bar{N}\sim \tau^{-\frac{1}{2+\nu}}$.

\begin{figure*}[t]
    \centering
\subfloat{
\raisebox{125pt}[0pt]{\makebox[-10pt][l]{\textsf{\textbf{A}}}}
\includegraphics[width=0.32\textwidth]{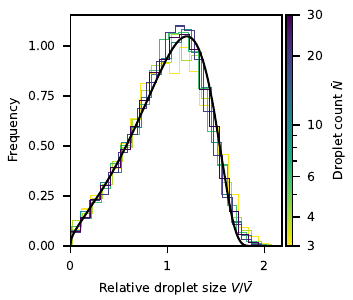}}
\hspace{0.01\textwidth}
\subfloat{
\raisebox{125pt}[0pt]{\makebox[-10pt][l]{\textsf{\textbf{B}}}}
\includegraphics[width=0.32\textwidth]{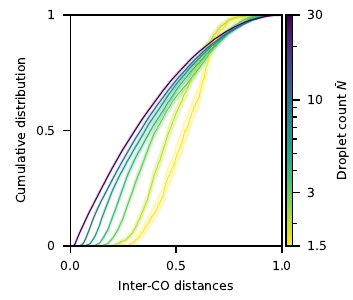}}
\hspace{0.01\textwidth}
\subfloat{
\raisebox{125pt}[0pt]{\makebox[-10pt][l]{\textsf{\textbf{C}}}}
\includegraphics[width=0.32\textwidth]{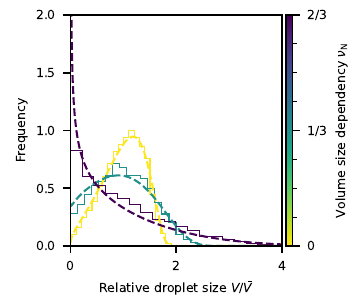}}
\caption{
\textbf{Droplet size distribution and inter-CO distances resulting from the coarsening model.}
The theoretical universal size distributions $p(z, \gamma_0)$ are compared to numerical data.
(A) $p(z, \gamma_0)$ given by \Eqref{eq:theory_pzgamma0} as a function of $z=V/\bar V$ for $\alpha=\frac13$ and various average droplet counts $\meanN$ for diffusion-limited coarsening along one dimension.
(B) Cumulative distribution of the observed inter-CO distances $d/L$  as a function of the average droplet count $\meanN$.
(A-B) Additional simulation parameters are given in \Figref{fig:Figure3}B.
(C) $p(z, \gamma_0)$ given by \Eqref{eq:theory2_pzgamma0} as a function of $V/\bar V$ for $\alpha=\frac13$ and various $\nuN$ for nucleoplasmic exchange coarsening.
Model parameters are $\meanN=100$, $N^\mathrm{init}=1000$, $M=100\,\cD a^3$, and $N_\mathrm{sample}=50$ ($N_\mathrm{sample}=100$ for $\nuN=\frac13$).}
\label{fig:Figure_numerics_distributions}
\end{figure*}

\Figref{fig:Figure_numerics_distributions}A shows that the theoretical droplet size distributions (see \Figref{fig:Figure_theory_distributions}B) is consistent with numerical simulations.

\subsection{Investigation of droplet growth and coarsening without SC}
In this section, we consider a system without SC which resembles the \textit{zyp1}-mutant in \Athaliana{}~\cite{Durand2022,Fozard2022}.
Because we assumed fast diffusion in the nucleoplasm, the allocation and positions of individual droplet along chromosomes is no factor in the dynamics. 
Consequently, CO interference vanishes, and CO assurance is reduced.

The simplified dynamical equations in this section are given by \Eqsref{eq:non_dim_DN_system}.
These equations imply that droplets ranked by their size will remain in the same order, and the development of each individual droplet only depends on $\nu$, $\nuN$, $\phiD$, its size, and the total remaining material in nucleoplasm.

\subsubsection{Asymptotic scaling for exchange-rate-limited coarsening}
\label{ssec:dn_scaling}
In this section, we consider coarsening in a system without SC, so the dynamics are limited by the exchange rate $\LambdaDN$ between droplets and nucleoplasm.
We assume we have $N(t_0)=n(t_0/T)$ droplets at time $t_0$ and in the coarsening regime, the concentration in the nucleoplasm defines the stationary droplet size by $\mN = \phiD_\mathrm{noSC} \left[\mD_\mathrm{eq}\right]^{-\nu}$.
The dynamics and conservation law given by \Eqsref{eq:non_dim_DN_system} in non-dimensional units then read
\begin{subequations}
\begin{align}
    \dif{\mD_{i}(\tau)}{\tau} 
&= \phiD_\mathrm{noSC}  \left(\left[\mD_\mathrm{eq}(\tau)\right]^{-\nu} - \left[\mD_i(\tau)\right]^{-\nu}\right) \left[\mD_{i}(\tau)\right]^{\nuN}\;,\\
m &= \phiD_\mathrm{noSC} \left[\mD_\mathrm{eq}(\tau)\right]^{-\nu} +  \sum_{i=1}^{N_i} \mD_{j, i}(\tau) = \phiD_\mathrm{noSC} \left[\mD_\mathrm{eq}(\tau)\right]^{-\nu} +  n(\tau) \overline{\mD}(\tau)\;.
\end{align} 
\end{subequations}
To determine the scaling relations we use the result from \secref{ssec:theory_scaling_generalized}.
Inspecting \Eqref{eq:theory2_dynamics_equation} 
yields the number of droplets per unit volume and the non-dimensional droplet size in the asymptotic limit,
\begin{subequations}
\begin{align}
    N(\tau) &= K_1(\nu, \nuS)\, m \left[\phiD_\mathrm{noSC} \tau\right]^{-\frac{1}{1+\nu-\nuN}}\;,\\
    \mD_\mathrm{eq}(\tau) &= \frac{m}{k^\infty N(t)}= \frac{1}{k^\infty K_1(\nu, \nuN)} \left(\phiD_\mathrm{noSC} \tau\right)^{\frac{1}{1+\nu-\nuN}}\;,
\end{align}
where the pre-factor $K_1(\nu, \nuN)$ and $k^\infty$ is given by \Eqsref{eq:theory2_K1} and \eqref{eq:theory2_Kinfty}.
In dimensional units,
\begin{align}
    N(t) &= K_1(\nu, \nuN)\, \frac{M}{\cD a^3} \left[\frac{1}{\gamma_0} \frac{a^3}{\LambdaDN t}\right]^{\frac{1}{1+\nu-\nuN}}\;,\label{eq:DN_scaling}\\
    V_\mathrm{eq}(t) &= \frac{M}{k^\infty N(t)}= \frac{\cD a^3}{k^\infty K_1(\nu, \nuS)} \left[\gamma_0 \frac{\LambdaDN t}{a^3}\right]^{\frac{1}{1+\nu-\nuN}}\;,\\
    R_\mathrm{eq}(t) &\propto \left[\gamma_0 \frac{\LambdaDN t}{a^3}\right]^{\frac{1}{3(1+\nu-\nuN})}\;.
\end{align}
\end{subequations}
The universal distribution function is given by \Eqref{eq:theory2_pzgamma0} and shown in \Figref{fig:Figure_theory_distributions}C.

\subsubsection{Duration of growth and coarsening regime}
\label{ssub:duration_of_growth_and_coarsening_regime}
In the numerical simulation for the mutant without SC (main text \secref{sec:no_SC}), we observe three distinct dynamical regimes.
In the first regime all $N$ initial droplets grow by depleting the nucleoplasm until their average volume is in equilibrium with the nucleoplasm.
The second regime is characterized by growing variation of droplet sizes, whereas the average droplet size does not change significantly.
Together, we denote these two phases as the \emph{growth regime}, which ends after some duration $T_\mathrm{growth}$ when droplets start vanishing and the system transitions to the \emph{coarsening regime}, which last a duration $T_\mathrm{coarsen}$.
In this section, we estimate the duration of these regimes, and compare the theoretical results with numerical simulations as a function of system parameters.

To estimate the duration of the growth regime, we start by estimating the change of droplet sizes in the first regime, i.e., when growing $N$ droplets from the initial size $\mD_\mathrm{init}$ to the stationary size $\mD_\mathrm{eq}$.
In particular, the average, stationary droplet size is described by \Eqref{eq:non_dim_DN_conservation},
\begin{subequations}
\begin{equation}
    m = \phiD \left(\mD_\mathrm{eq}\right)^{-\nu} + N \mD_\mathrm{eq} \approx N \mD_\mathrm{eq}\;,
\end{equation}
\end{subequations}
assuming most material ends up in droplets.
We simplify the droplet growth dynamics in the first regime \eqref{eq:dMDdt_nd2} as
\begin{subequations}
\begin{align}
    \dif{\mD}{\tau} &= \left(\mD\right)^{\nuN} \left[\mN - \phiD_\mathrm{noSC} \left(\mD\right)^{-\nu}\right]\;, \\
    &= \left(\mD\right)^{\nuN} \left[m  - N \mD - \phiD_\mathrm{noSC} \left(\mD\right)^{-\nu}\right] \;,
    \label{eqn:tgrowth_intermediate_calculation}
    \\
    &\approx \left(\mD\right)^{\nuN} m \;,
\end{align}
neglecting the last two terms in \Eqref{eqn:tgrowth_intermediate_calculation}.
Integrating this equation from $\mD_0$ to $\mD_1$ for $\nuN \neq 1$,
\begin{align}
    \tau_1 &= \frac{1}{m} \int_{\mD_0}^{\mD_1}\frac{\diff \mD}{\left(\mD\right)^{\nuN}} 
    = \frac{1}{m} \frac{\left(\mD_1\right)^{1-\nuN} - \left(\mD_0\right)^{1-\nuN}}{1-\nuN} \;.
\end{align}
Using $\mD_0=\mD_\mathrm{init}$ and $\mD_0=\mD_\mathrm{eq}\approx m/N$,
\begin{align}
    \tau_1 \approx \frac{1}{m} \frac{\left(\frac{m}{N}\right)^{1-\nuN} - \left(\mD_\mathrm{init}\right)^{1-\nuN}}{1-\nuN}\;.
    \label{eqn:tgrowth_final_tau_1}
\end{align}
This time estimates the duration until the mean intial droplet reaches its equilibrium size.
We next ask how the size of a slightly smaller droplet, with volume $\mD_\mathrm{init} (1-\sigma_\mathrm{init})$, change during $\tau_1$.
Using \Eqref{eqn:tgrowth_final_tau_1}, we find
\begin{align}
    \tau_1 (1-\nuN) m= \left(\mD_\mathrm{eq} (1+\sigma_\mathrm{eq})\right)^{1-\nuN} - \left(\mD_\mathrm{init} (1+\sigma_\mathrm{init})\right)^{1-\nuN} = \left(\mD_\mathrm{eq}\right)^{1-\nuN} - \left(\mD_\mathrm{init}\right)^{1-\nuN}\;,
\end{align}
and consequently
\begin{align}
    \frac{\sigma_\mathrm{eq}}{\sigma_\mathrm{init}} \approx \left(\frac{\mD_\mathrm{eq}}{\mD_\mathrm{init}}\right)^{\nuN-1}\;.
\end{align}
Notably, the relative droplet variation reduces over time in this regime.
\end{subequations}

To estimate the duration $T_\mathrm{growth}$ of the growth regime, we neglect the time it takes to deplete the nucleoplasm and focus on the second part of the growth regime:
Assuming $N$ droplets with average droplet size $\mD_\mathrm{eq}$ in equilibrium with  nucleoplasm $\mN$, and a droplet size distribution that follows a Gaussian distribution with standard deviation $\sigma_\mathrm{eq}$, we estimate the time until the first droplets disappear.
In particular, we perform a linear stability analysis around the stationary state in order to estimate the dominant rate of this regime, which will allow us to determine the time it takes for initial deviations to grow and droplets that are approximately one standard deviation smaller than the mean droplet to start vanishing.
Applying the dynamics given by \Eqref{eq:dMDdt_nd2} with $\mD = \mD_\mathrm{eq} + \Delta \mD$ and $\mN = \phiD_\mathrm{noSC} \left(\mD_\mathrm{eq}\right)^{-\nu}$, we get (to linear order in the deviation $\Delta \mD$)
\begin{subequations}
\begin{align}
    \dif{\mD}{\tau} = \dif{\Delta \mD}{\tau} &\approx \left(\mD_\mathrm{eq}\right)^{\nuN} \left(1 + \nuN \frac{\Delta \mD}{\mD_\mathrm{eq}}\right) \left[\mN - \phiD_\mathrm{noSC} \left(\mD_\mathrm{eq}\right)^{-\nu} \left(1 -\nu \frac{\Delta \mD}{\mD_\mathrm{eq}} \right)\right] \;,\\ 
    &\approx  \nu \left(\mD_\mathrm{stat}\right)^{\nuN-\nu - 1}  \phiD_\mathrm{noSC} \Delta \mD\;.
\end{align}
\end{subequations}
We now integrate this equation for $\Delta \mD_0 / \mD_\mathrm{eq} = -\sigma_\mathrm{eq} \approx -\sigma_\mathrm{init} \left(\frac{m}{N \mD_\mathrm{init}}\right)^{\nuN-1}$ to determine the time it takes from the initial deviation until the droplet vanishes, at $\Delta \mD/\mD_\mathrm{eq} = -1$, and get
\begin{subequations}
\begin{align}
    \tau_\mathrm{growth} &\approx \frac{1}{\nu \phiD_\mathrm{noSC}}  \left(\mD_\mathrm{eq}\right)^{1 + \nu -\nuN}  \int_{\sigma_\mathrm{eq}}^{1}\frac{\diff V}{V} \;,\\
    &= \frac{1}{\nu \phiD_\mathrm{noSC}} \left(\mD_\mathrm{eq}\right)^{1 + \nu - \nuN}  \log{\left(\frac{1}{\sigma_\mathrm{eq}}\right)}\;,\\
    &\approx \frac{1}{\nu \phiD_\mathrm{noSC}} \left(\frac{m}{N}\right)^{1 + \nu - \nuN} \left[\log{\left(\frac{1}{\sigma_\mathrm{init}}\right)} + (1-\nuN) \log{\left(\frac{m}{N \mD_\mathrm{init}}\right)}\right] \;.
\end{align}
Notably, this transition time does depend logarithmically on the relative initial variance of the droplet size distribution.
In dimensional units,
\begin{align}
    T_\mathrm{growth} \approx& \frac{a^3}{\nu \gamma_0 \LambdaDN} \left(\frac{V_\mathrm{eq}}{a^3}\right)^{1 + \nu - \nuN} \log{\left(\frac{1}{\sigma_\mathrm{eq}}\right)}\;,\\
    \approx& \frac{a^3}{\nu \gamma_0 \LambdaDN} \left(\frac{M}{N \cD a^3}\right)^{1 + \nu - \nuN} \log{\left(\frac{1}{\sigma_\mathrm{eq}}\right)}\;. \label{eq:T_growth_appendix}
\end{align}
For $\nu=\nuN$, this simplifies to
\begin{equation}
    T_\mathrm{growth} \approx \frac{M}{\nu \gamma_0 \cD \LambdaDN} \frac{1}{N} \log{\left(\frac{1}{\sigma_\mathrm{eq}}\right)}\;.
\end{equation}
\end{subequations}
Finally, we estimate the duration $T_\mathrm{coarsen}$ of the coarsening regime by applying the scaling relation \eqref{eq:DN_scaling}.
Assuming we start with $N_0$ droplets at time $t_0=\tau_0 T_\mathrm{noSC}$ and coarsen until $N_1$ droplets remain at time $t_1=\tau_1 T_\mathrm{noSC}$,
\begin{subequations}
\begin{align}
\tau_i &= \frac{1}{\phiD_\mathrm{noSC}} \left(\frac{N_i}{m K_1(\nu, \nuN)}\right)^{-(1+\nu - \nuN)}\;,\\
t_i &= \frac{a^3}{\gamma_0 \LambdaDN} \left(\frac{N_i \cD a^3}{M K_1(\nu, \nuN)}\right)^{-(1+\nu - \nuN)}\;,
\end{align}
yielding the duration
\begin{align}
\tau_\mathrm{coarsen}(N_1) &= \left(N_1^{-(1+\nu - \nuN)} - N_0^{-(1+\nu - \nuN)}\right)\frac{1}{\phiD_\mathrm{noSC}} \left[K_1(\nu, \nuN) m \right]^{-(1+\nu - \nuN)} \\ 
&\approx \frac{1}{\phiD_\mathrm{noSC}} \left[K_1(\nu, \nuN) m \right]^{-(1+\nu - \nuN)}\;,\\
T_\mathrm{coarsen}(N_1) 
&\approx \frac{a^3}{\gamma_0 \LambdaDN} \left(\frac{M K_1(\nu, \nuN)}{\cD a^3 N_1}\right)^{-(1+\nu - \nuN)}
\end{align}
for $N_1\ll N_0$. 
For $\nu=\nuN$, the equation simplifies to
\begin{equation}
T_\mathrm{coarsen} \approx K_1(\nu, \nuN) \frac{M}{\gamma_0 \cD \LambdaDN} \frac{1}{N_1}\;.
\end{equation}
\end{subequations}

\subsubsection{Numerical simulations}
\begin{figure*}[t]
    \centering
\subfloat{
\raisebox{100pt}[0pt]{\makebox[-5pt][l]{\textsf{\textbf{A}}}}
\includegraphics[width=0.23\textwidth]{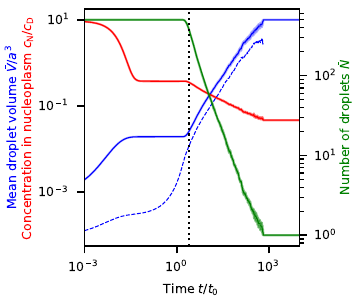}}\hspace{0.01\textwidth}
\subfloat{
\raisebox{100pt}[0pt]{\makebox[-5pt][l]{\textsf{\textbf{B}}}}
\includegraphics[width=0.23\textwidth]{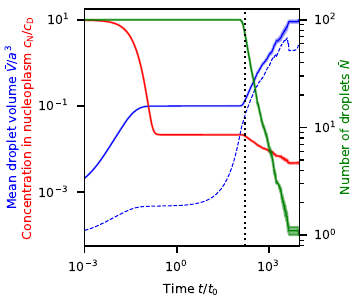}}\hspace{0.01\textwidth}
\subfloat{
\raisebox{100pt}[0pt]{\makebox[-5pt][l]{\textsf{\textbf{C}}}}
\includegraphics[width=0.23\textwidth]{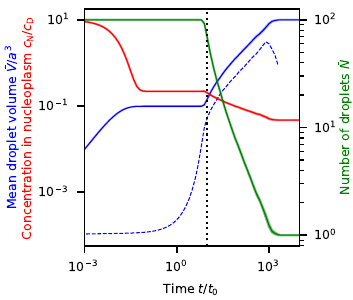}}\hspace{0.01\textwidth}
\subfloat{
\raisebox{100pt}[0pt]{\makebox[-5pt][l]{\textsf{\textbf{D}}}}
\includegraphics[width=0.23\textwidth]{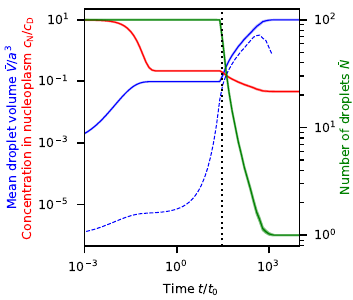}}
\caption{\textbf{Numerical investigation of the coarsening model without synaptonemal complex.} 
Average droplet volume $\bar{V}/a^3$ (blue solid lines), average variance of droplets sizes $\sigma_V/a^3$ (blue dashed lines), HEI10 concentration in nucleoplasm $\cN/\cD$ (red lines), and average number of droplets (green lines) as a function of time $t$ for 
(A) $N^\mathrm{init}=500$, (B) $\phiD_\mathrm{noSC}=\gamma_0 \frac{\VN}{a^3} = 10^{-2}$, (C) $\nuN=0$, and (D) $\sigma_V^\mathrm{init}=10^{-3}V^\mathrm{init}$.
The vertical dotted line marks $T_\mathrm{growth}$ given by \Eqref{eq:T_growth_appendix}, which indicates the transition from growth regime to coarsening regime.
(A--D) Additional parameters taken from \Figref{fig:Figure4} and $N_\mathrm{sample}=10$.
}
\label{fig:Figure4_SI}
\end{figure*}

\Figref{fig:Figure_numerics_distributions}C shows that the theoretical droplet size distributions (\Figref{fig:Figure_theory_distributions}C) are consistent with numerical simulations for various volume-size dependencies $\nuN$.
Moreover, \Figref{fig:Figure4_SI} shows additional examples of the time-development of the average droplet volume $\bar{V}/a^3$ and average variance of droplets sizes $\sigma_V/a^3$, as well as the  HEI10 concentration in nucleoplasm $\cN/\cD$ (cmp.~\Figref{fig:Figure4}A). The estimation for the duration of the growth regime (vertical dashed line) is consistent with these simulations.

\subsection{Numerical analysis of full model  and comparison to experiments}
In the following, we consider the full system with all three compartments (the nucleoplasm, the SCs, and droplets) and analyze teh dynamics numerically.

\subsubsection{HEI10 concentration profile along the SC}
\label{ssec:full_sc_profile}
We determine the concentration profile along the SC, extend the analysis from \secref{ssec:sc_profile_ds} while also considering the exchange between SC and nucleoplasm.
We assume that changes in the concentration profile $\mS(\xi)$ are slow, whereas exchanges between compartments are sufficiently fast so the concentration $\mN$ in the nucleoplasm is in equilibrium with the average SC concentration, $\overline{\mS}(\xi) =\phiN \mN$.
The general solution for the concentration profile, described by \Eqref{eq:theory_laplace_equation}, is given in \secref{ssec:theory_concentration_profile}, assuming $\kappa = \sqrt{\gammaSN}$.

\subsubsection{Model parameters determined from \Athaliana{} wild type and \zyp-mutant}
\label{ssub:model_parameters_determined_from_athaliana}
In \secref{sub:droplets_in_athaliana_} of the main text, we estimate the order of magnitude of some model parameters that are consistent with empirical data of wild type and \zyp{}-mutants from \Athaliana{}~\cite{Durand2022,Morgan2021,Fozard:2023aa}.
For simplicity, we describe the \Athaliana{} system by assuming that we have $n=5$ SCs of equal length $L=\SI{30}{\micro\meter}$ with an average droplet count of $N_\mathrm{wt}=2$ per SC in male meiosis and $N_\mathrm{zyp}=3$ in the \zyp{}-mutant~\cite{Durand2022}.
In the following, we describe further details of the approach:

First, in wild type, we assume that coarsening times correspond to the duration of pachytene, $T_\mathrm{wt} = T_\mathrm{pach}\approx\SI{10}{\hour}$~\cite{Prusicki:2019aa}. 
We define the effective coarsening time to reach the same average CO count $\meanN$ of the system assuming we have no exchange via the nucleoplasm as $T_\mathrm{SC}\geq T_\mathrm{wt}$.
We further estimate the total amount of HEI10 in the system as $M_\mathrm{wt}=\alpha_\mathrm{wt} N_\mathrm{wt} n \cD a^3$, with some pre-factor $\alpha_\mathrm{wt}$, which is $1$ if all HEI10 finally ends up in droplets of size $a^3$, which corresponds to the observation that late recombination nodules are of order SC size~\cite{Carpenter:2003aa}.
For the \zyp{}-mutant, we define the duration $T_\mathrm{zyp}$, and use $M_\mathrm{zyp}=\alpha_\mathrm{zyp}/\alpha_\mathrm{wt} M_\mathrm{wt}$, since the total amount of HEI10 is not necessarily the same.

We start by estimating a lower bound for the exchange rate between SC and nucleoplasm from the initial HEI10 loading time $T_\mathrm{load}\approx \SI{1}{h}$ onto the SC. 
Applying \Eqref{eq:k_load}  gives
\begin{subequations}
\begin{align}
\frac{a^3}{\LambdaSN}\frac{1}{\gammaS + \frac{V_\mathrm{S}}{\VN}} &<T_\mathrm{load} \;, \\ 
\frac{a^2}{D}\frac{1}{T_\mathrm{load}} \approx \frac{a^2}{D}\frac{1}{T_\mathrm{load}} \frac{1}{1+ \frac{V_\mathrm{S}}{\VN} \frac{1}{\gammaS}} &< \frac{\LambdaSN \gammaS}{D a}= \gammaSN\;, \\
3\cdot 10^{-6} &< \gammaSN\;.
\end{align}
\end{subequations}
To estimate $\phiD$, we consider the scaling relation \eqref{eq:scaling_diffusion_limited} and approximate
\begin{subequations}
\begin{equation}
    N_\mathrm{wt}(t_\mathrm{SC}) = K_0(\nu) \frac{L}{a} \left(\frac{M_\mathrm{wt}}{\cD a^2 L}\right)^{\frac{1+\nu}{2+\nu}} \left(\frac{a^2}{D \phiD t_\mathrm{SC}} \right)^{\frac{1}{2+\nu}}\;,
\end{equation}
which yields
\begin{align}
    \phiD &= \left(\frac{K_0(\nu) L}{N_\mathrm{wt} a}\right)^{2+\nu}  \left(\frac{\alpha_\mathrm{wt} \cD a^3 n N_\mathrm{wt}}{\cD a^2 n L}\right)^{1+\nu} \frac{a^2}{D t_\mathrm{SC}}\;, \\ 
     &= K_0^{2+\nu}  \alpha_\mathrm{wt}^{1+\nu} \frac{a L }{D t_\mathrm{SC} N_\mathrm{wt}}\label{eq:approx_phiD} \approx 10^{-4}\;,
\end{align}
\end{subequations}
when applying the numerical values from above (also in \secref{sub:droplets_in_athaliana_} of main text), and $K_0(1/3)\approx 1.2$.
Moreover, we assumed that nucleoplasmic exchange does not affect coarsening times ($T_\mathrm{SC}\approx T_\mathrm{wt}$), and $\alpha_\mathrm{wt} \approx 1$.

To estimate the exchange rate $\gammaDN$ between droplets and nucleoplasm, we have to separately consider the two scenarios for the \zyp{}-mutant discussed in the main text \secref{sec:no_SC}.
We start with the first case where all initial droplets still grow without apparent coarsening.
In this case, the number of initial droplets corresponds to the number of final droplets, $N_\mathrm{init}=n N_\mathrm{zyp}$, and it thus  follows from \Eqref{eq:T_growth_appendix} that
\begin{subequations}
\begin{align}
    T_\mathrm{zyp} &< T_\mathrm{growth} = \frac{a^3}{\nu \gamma_0 \LambdaDN} \left(\frac{M_\mathrm{zyp}}{\cD n N_\mathrm{zyp} a^3}\right)^{1 + \nu - \nuN} \log{\left(\frac{V_\mathrm{stat}}{\Delta V}\right)}\;,\\
    &< \frac{a^2}{\nu \phiD \gammaDN D} \left(\frac{\alpha_\mathrm{zyp} N_\mathrm{wt}}{N_\mathrm{zyp}}\right)^{1 + \nu - \nuN} \log{\left(\frac{V_\mathrm{stat}}{\Delta V}\right)}\;.
\end{align}
Applying \Eqref{eq:approx_phiD} then yields
\begin{align}
    \gammaDN &< \frac{a^2}{\nu D} \left(\frac{\alpha_\mathrm{zyp} N_\mathrm{wt}}{N_\mathrm{zyp}}\right)^{1 + \nu - \nuN} \log{\left(\frac{V_\mathrm{stat}}{\Delta V}\right)} \frac{D T_\mathrm{SC} N_\mathrm{wt}}{a L K_0^{2+\nu} \alpha_\mathrm{wt}^{1+\nu}} \frac{1}{T_\mathrm{zyp}}\;,\\
 &\approx \frac{1}{\nu K_0^{2+\nu}}\frac{a}{L}\frac{N_\mathrm{wt}}{\alpha_\mathrm{wt}^{\nuN}} \left(\frac{\alpha_\mathrm{zyp} N_\mathrm{wt}}{\alpha_\mathrm{wt} N_\mathrm{zyp}}\right)^{1 + \nu - \nuN} \log{\left(\frac{V_\mathrm{stat}}{\Delta V}\right)} \frac{T_\mathrm{SC}}{T_\mathrm{zyp}}\;,
\end{align}
which simplifies to
\begin{align}
    \gammaDN &< \frac{1}{\nu K_0^{2+\nu}}\frac{a}{L}\frac{\alpha_\mathrm{zyp} N_\mathrm{wt}^2}{\alpha_\mathrm{wt}^{1+\nu} N_\mathrm{zyp}} \log{\left(\frac{V_\mathrm{stat}}{\Delta V}\right)} \frac{T_\mathrm{SC}}{T_\mathrm{zyp}}\;,
\end{align}
\end{subequations}
for $\nu=\nuN$.
When applying $\alpha_\mathrm{wt} \approx \alpha_\mathrm{zyp} \approx 1$ and approximating $T_\mathrm{zyp} \approx T_\mathrm{wt} > \frac12 T_\mathrm{SC}$ (to get an upper bound), and $\log{\left(\frac{V_\mathrm{stat}}{\Delta V}\right)} < 3$, we estimate $\gammaDN < 3\cdot 10^{-2}$.

Next, we consider the case where droplets are coarsening in the \zyp-mutant.
\Eqref{eq:DN_scaling} then implies
\begin{subequations}
\begin{align}
    N_\mathrm{zyp}(t_\mathrm{zyp}) &\approx K_1(\nu, \nuN) \frac{M_\mathrm{zyp}}{\cD a^3} \left(\frac{1}{\gamma_0} \frac{a^3}{\LambdaDN t_\mathrm{zyp}}\right)^{\frac{1}{1+\nu-\nuN}}\;, \\ 
    &\approx K_1(\nu, \nuN) \alpha_\mathrm{zyp} n N_\mathrm{wt} \left(\frac{a^2}{\phiD \gammaDN D} \frac{1}{t_\mathrm{zyp}}\right)^{\frac{1}{1+\nu-\nuN}}\;,
\end{align}
and with applying \Eqref{eq:approx_phiD} and $\nu=\nuN$ we get
\begin{align}
N_\mathrm{zyp}(t_\mathrm{zyp}) &\approx \frac{K_1 \alpha_\mathrm{zyp}}{K_0^{2+\nu} \alpha_\mathrm{wt}^{1+\nu} } \frac{a}{L} \frac{n N_\mathrm{wt}^2}{\gammaDN} \frac{T_\mathrm{SC}}{T_\mathrm{zyp}}\;,\\ 
    \Rightarrow \gammaDN &\approx\frac{K_1 \alpha_\mathrm{zyp}}{K_0^{2+\nu} \alpha_\mathrm{wt}^{1+\nu}} \frac{a}{L}  \frac{n N_\mathrm{wt}^2}{N_\mathrm{zyp}}  \frac{T_\mathrm{SC}}{T_\mathrm{zyp}} \;.
    \end{align}
\end{subequations}
Using numerical values from above (also in \secref{sec:droplet_patterning_only_SC} of main text), and $K_1(1/3, 1/3)\approx 5.98$, we get $\gammaDN \approx 10^{-1}$.

\subsubsection{Numerical simulations}
\begin{figure*}[t]
    \centering
\subfloat{
\raisebox{135pt}[0pt]{\makebox[5pt][l]{\textsf{\textbf{A}}}}
\includegraphics[width=0.3267\textwidth]{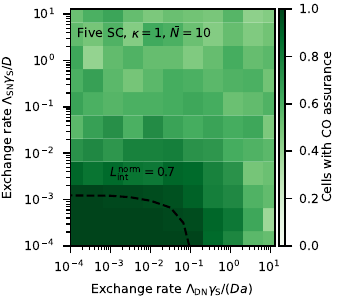}}\hspace{0.05\textwidth}
\subfloat{
\raisebox{135pt}[0pt]{\makebox[5pt][l]{\textsf{\textbf{B}}}}
\includegraphics[width=0.3267\textwidth]{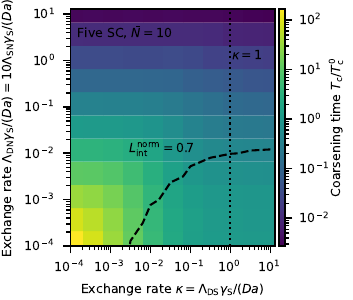}}

\caption{\textbf{Numerical investigation of the extended coarsening model for a cell with five SC.}
(A) Heatmap showing the fraction of cells that do preserve CO assurance as a function of the nucleoplasmic exchange rates $\LambdaDN{}$ and $\LambdaSN{}$ for fixed values of diffusion rate $D$, exchange rate $\kappa = 1$, and an average CO count $\meanN = 10$. 
(B) Heatmap showing the parameter dependence of the relative coarsening time $T_\mathrm{c}/T_\mathrm{c}^0$ as a function of the exchange rate between droplets and SC $\LambdaDS{}$ and the nucleoplasmic exchange rates with $\LambdaDN = 10 \LambdaSN$. 
(A--B) Additional parameters taken from \Figref{fig:Figure5}.
}
\label{fig:Figure5_SI}
\end{figure*}

In \Figref{fig:Figure5_SI} we show additional plots for the numerical investigation of the extended coarsening model for a cell with five SC presented in \Figref{fig:Figure5}.

\subsubsection{Empirical investigations}
\begin{figure*}[t]
    \centering
\includegraphics[width=\textwidth]{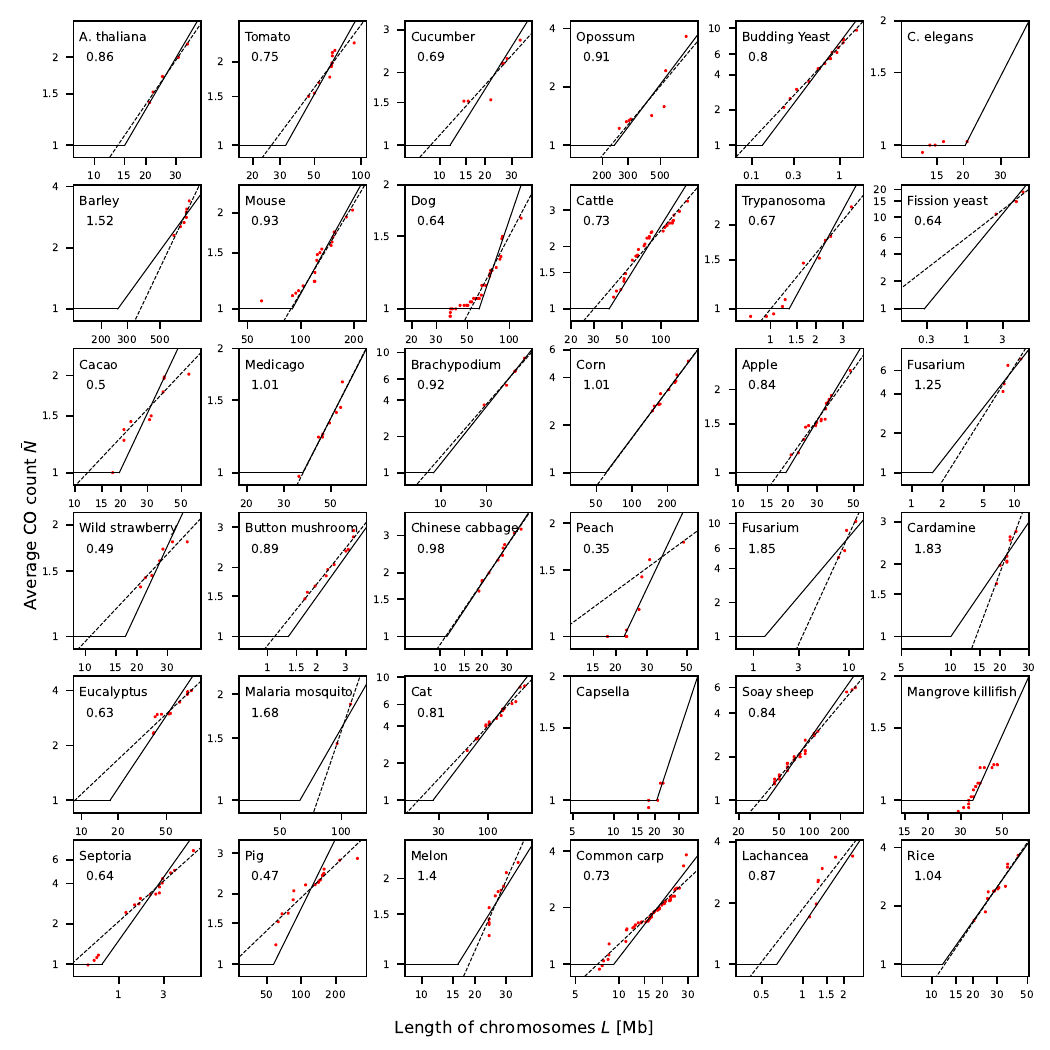}
\caption{\textbf{Comparison of theoretical scaling prediction and genetic data collected by Fernandes~\cite{Fernandes:2018aa}.}
Shown is the average CO count $\meanN$ as a function of the chromosome length  (measured in base pairs [Mb]) across multiple species using the dataset provided in~\cite{Fernandes:2018aa} for those species not shown in \Figref{fig:Figure6}.
To determine the CO count we use the given data of the genetic size in centiMorgan [cM] and rescale these by a factor of $50$ to account for the fact that only half of chromosomes remain in genetic data.
The black line shows the theoretical predictions and the dashed line the optimal power-law fit for all chromosomes with $N\geq 1.2$ crossover. The observed slope is given in the respective Figure.}
\label{fig:Figure_theory_scaling_SI}
\end{figure*}

In \Figref{fig:Figure_theory_scaling_SI} we show the average CO count $\meanN$ as a function of the chromosome length using the dataset provided in~\cite{Fernandes:2018aa} for those species not shown in \Figref{fig:Figure6}.
\FloatBarrier
\subsection{Implementation details of numerical simulations}
The implementation of our model is based on the non-dimensional versions of the equations, which are described in \secref{ssec:non_dim} for the model without nucleoplasm and the full model, and \secref{ssec:non_dim_dn} for the system without SC.
We discretized the concentration field on the SC using equidistant grid points separated by $\Delta x_\mathrm{grid}/a$ and approximated spatial derivatives using finite differences.
We choose $\Delta x_\mathrm{grid}$ such that a further decrease does not significantly alter the dynamics anymore.

We initialized droplets along all chromosomes, assuming a uniform droplet density of $N^\mathrm{init}/\ell$.
Droplet sizes are chosen independently from a normal distribution with mean $\bar{V}^\mathrm{init}=\mD_\mathrm{init} a^3$ and standard deviation $\sigma_V^\mathrm{init}/\bar{V}^\mathrm{init}$, truncated to $\left[0.1 \mD_\mathrm{init}, 10 \mD_\mathrm{init}\right]$.
The total amount of HEI10 $M=m \cD a^3$, that remains after initializing the droplets, is distributed between nucleoplasm and uniformly along all SCs (insofar these compartments are included in the simulation), so that both compartments are in equilibrium.
The diffusional flux along the SC is computed by imposing no-flux boundary conditions at the ends of the SC.
If a droplets vanishes, i.e., its size would be smaller or equal to zero in the next time step, we set its size to zero and neglect it subsequently.
To ensure material conservation, fluxes of vanishing droplets are reduced accordingly.

We solved the temporal dynamics using an explicit fist-order Euler scheme and we use Richardson extrapolation~\cite{ilie2015adaptive} for adaptive time-stepping.
We compare the maximal absolute error $\Delta_i$ of all apparent fluxes between the dynamics with time-step $\Delta t^{(i)}$ and two half-steps $\frac12 \Delta t^{(i)}$ with a certain threshold ($\text{tol}=10^{-7}$), to calculate the next time step,
\begin{equation}
    \Delta t^{i+1} = 0.9 \Delta t \min{\left(\max{\left(\sqrt{\frac{\text{tol}}{2 \Delta_i}}, 0.3\right)}, 2\right)}\;.
\end{equation}
All relevant simulation parameters are given in the captions of the  respective figures.
\FloatBarrier
\bibliography{literature.bib}

\end{document}